\documentclass[iop]{emulateapj}

\usepackage{txfonts}
\usepackage{graphicx}
\usepackage{subfigure}
\usepackage{multirow}
\usepackage{verbatim}
\usepackage{longtable}
\usepackage{url}
\usepackage{float}
\usepackage{ esint }
\usepackage{dcolumn}
\usepackage{hyperref}
\usepackage[utf8]{inputenc}
\slugcomment{Revision draft}

\shorttitle{Robo-AO \textit{Kepler} Planetary Candidate Survey III}
\shortauthors{Ziegler et al.}

\begin{document}

\title{Robo-AO Kepler Planetary Candidate Survey III: Adaptive Optics Imaging of 1629 Kepler Exoplanet Candidate Host Stars}

\author{Carl Ziegler\altaffilmark{1}, Nicholas M. Law\altaffilmark{1}, Tim Morton\altaffilmark{2}, Christoph Baranec\altaffilmark{3}, Reed Riddle\altaffilmark{4}, Dani Atkinson\altaffilmark{3}, Anna Baker\altaffilmark{5},\\ Sarah Roberts\altaffilmark{6}, and David R. Ciardi\altaffilmark{7}}

\email{carlziegler@unc.edu}
\altaffiltext{1}{Department of Physics and Astronomy, University of North Carolina at Chapel Hill, Chapel Hill, NC 27599-3255, USA}
\altaffiltext{2}{Department of Astrophysical Sciences, Princeton University, Princeton, NJ 08544, USA}
\altaffiltext{3}{Institute for Astronomy, University of Hawai`i at M\={a}noa, Hilo, HI 96720-2700, USA}
\altaffiltext{4}{Division of Physics, Mathematics, and Astronomy, California Institute of Technology, Pasadena, CA 91125, USA}
\altaffiltext{5}{Durham Academy Upper School, 3601 Ridge Road, Durham, NC 27705, USA}
\altaffiltext{6}{Juniata College, 1700 Moore St, Huntingdon, PA 16652, USA}
\altaffiltext{7}{NASA Exoplanet Science Institute, California Institute of Technology, Pasadena, CA 91125, USA}

\begin{abstract}
The Robo-AO \textit{Kepler} Planetary Candidate Survey is observing every \textit{Kepler} planet candidate host star with laser adaptive optics imaging to search for blended nearby stars, which may be physically associated companions and/or responsible for transit false positives.  We present in this paper the results of our search for stars nearby 1629 \textit{Kepler} planet candidate hosts. With survey sensitivity to objects as close as $\sim$0$\farcs$15 and magnitude differences $\Delta$m$\le$6, we find 223 stars in the vicinity of 206 target KOIs; 209 of these nearby stars have not previously been imaged in high resolution.  We measure an overall nearby-star probability for \textit{Kepler} planet candidates of 12.6\%$\pm$0.9\% at separations between 0$\farcs$15 and 4$\farcs$0.  Particularly interesting KOI systems are discussed, including 26 stars with detected companions which host rocky, habitable zone candidates, and five new candidate planet-hosting quadruple star systems.  We explore the broad correlations between planetary systems and stellar binarity using the combined dataset of \citet{baranec16} and this paper.  Our previous 2$\sigma$ result of a low detected nearby star fraction of KOIs hosting close-in giant planets is less apparent in this larger dataset.  We also find a significant correlation between detected nearby star fraction and KOI number, suggesting possible variation between early and late \textit{Kepler} data releases.

\end{abstract}


\section{Introduction}

The primary \textit{Kepler} mission vastly increased the tally of known extrasolar planets, discovering over 2300 confirmed planets and approximately 4700 planet candidates \citep{borucki10, borucki11a, borucki11b, batalha13, burke14, rowe14, coughlin15, morton16}.  Using high-precision photometry to detect the periodic dip in stellar brightness consistent with a transiting planet, \textit{Kepler} exoplanet candidates (\textit{Kepler} Objects of Interests, or KOIs) require follow-up observations to rule out astrophysical false positives and for host star characterization \citep{brown11}.

\begin{figure*}
\centering
\includegraphics*[width=500pt]{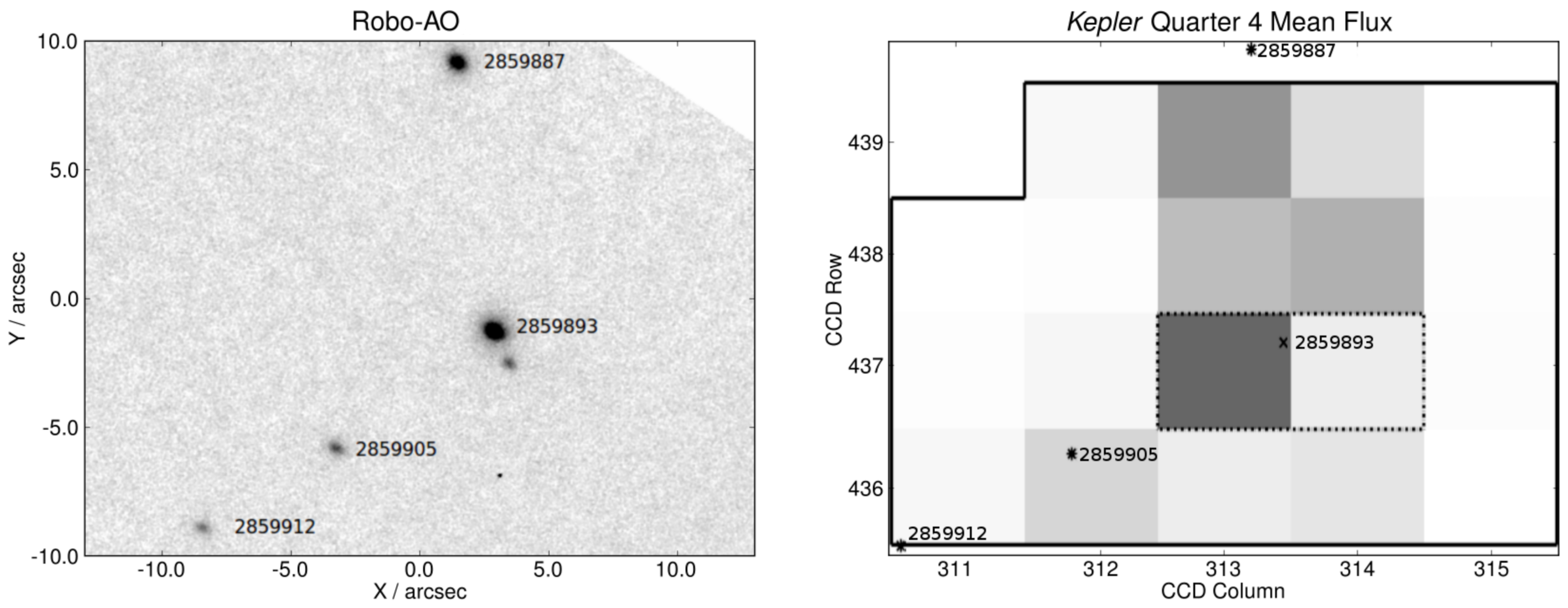}
\caption{On the left, the full-frame Robo-AO reduced image of KOI-4418 (KIC2859893) rotated and scaled to match the \textit{Kepler} view of the same field, displayed on the right, each pixel colored by the mean flux in Quarter 4.  KICs in the field are marked on both images, as well as K$_{P}$ magnitude in the \textit{Kepler} image.  The 1$\farcs$41 binary to KOI-4418 is not visible in the $\sim$4$\arcsec$ pixels of \textit{Kepler}, illustrating how real companions and background stars can blend with the KOIs, resulting in astrophysical false positives or inaccurate planetary characteristics.  High resolution follow-ups are a crucial step in the validation and characterization of \textit{Kepler} planetary systems.}
\label{fig:kepler_field}
\end{figure*}

Most solar-type stars, which comprise the majority of \textit{Kepler} targets \citep{batalha13}, form with at least one companion star \citep{duquennoy91, raghavan10}.  The large effective point-spread function (6-10\arcsec) and coarse resolution (pixel size of $\sim$4\arcsec) \citep{haas10} of \textit{Kepler} allow these companion stars and background objects to be blended with the host candidate, illustrated in Figure$~\ref{fig:kepler_field}$.  High-angular-resolution follow-up imaging is crucial to distinguish these blended multiple stellar systems and identify false transit signals.  Even when the candidates are bona fide planets, the planet radius measurements based on the diluted transit signal are underestimated due to the presence of multiple stars in the system or unbounded stars within the \textit{Kepler} aperture \citep{ciardi15}.

Before being elevated to planet candidate status, each KOI is vetted for clear signatures of being an astrophysical false positive: center-of-light shifts during transit, an identifiable secondary eclipse signal indicating the eclipsing object is self-luminous, or sharing the exact ephemeris as another KOI.  While these vetting efforts on early catalogs were largely based on human inspection \citep{batalha10}, the most recent DR24 catalog has fully automated this process \citep{coughlin15}.  Notably, the candidate status of a KOI is \textit{not} a function of its depth or shape (i.e., whether it is V-shaped), which means that a large fraction of the deeper signals ($\sim$50\%) can be expected to be false positives \citep{santerne12, santerne15}.  Shallower candidates have a much lower predicted false positive rate ($\sim$10\%) \citep{morton11, fressin13}, a prediction that has been confirmed by follow-up observations from the \textit{Spitzer} space telescope \citep{desert15}.  Nevertheless, even if a large fraction of the candidate signals are real planets, many of the inferred properties of these planets are affected by the presence of blended sources \citep{dressing13,santerne13}.  Therefore to fully characterize individual \textit{Kepler} planets and to measure any possible biasing effects of stellar multiplicity on the planetary populations, every KOI needs to be searched for stellar companions\footnote{For brevity we denote stars which we found within our detection radius of KOIs as ``companions,'' in the sense that they are asterisms associated on the sky.}.

There has been considerable effort by the community to perform high-resolution imaging surveys of the KOIs \citep{howell11, adams12, adams13, lillo12, lillo14, horch12, marcy14, dressing14, horch14, wang15a, wang15b, torres15, everett15, kraus16}.  These surveys, however, have combined covered approximately 30\% of the full set of \textit{Kepler} planetary candidates.  This piecemeal approach leads to inconsistent vetting, while limiting the comprehensive statistics and correlations that can be derived from a large dataset of high resolution images of multiple stellar systems hosting planetary systems.  In addition, target lists of past surveys are often biased towards brighter targets, possibly skewing any interpretations drawn from the data.

A complete, consistent high-resolution survey of all the the KOIs with ground-based adaptive optics (AO) is limited by the typical overheads required with traditional systems.  Taking advantage of the order-of-magnitude increase in time-efficiency provided by Robo-AO, the first robotic laser adaptive optics system, we are performing high-resolution imaging of every KOI system.  The first paper in this survey, \citet[hereafter Paper I]{law14}, observed 715 \textit{Kepler} planetary candidates, identifying 53 companions, with 43 new discoveries, for a detected companion fraction of 7.4\%$\pm$1.0\% within separations of 0$\farcs$15 to 2$\farcs$5.  The second paper in this survey, \citet[hereafter Paper II]{baranec16}, observed 969 \textit{Kepler} planetary candidates, identifying 202 companions, with 139 new discoveries, for a detected companion fraction of 11.0\%$\pm$1.1\% within separations of 0$\farcs$15 to 2$\farcs$5., and 18.1\%$\pm$1.3\% within separations of 0$\farcs$15 to 4$\farcs$0.

This paper presents a total of 1629 targets observed, around which we find 223 companions around 206 KOIs, 209 of which have not been previously imaged in high resolution, for a detected companion fraction of 12.6\%$\pm$0.9$\%$ within 4$\farcs$0 of planetary candidate hosting stars.

We begin in Section \ref{sec:targetselection} by describing our target selection, the Robo-AO system, and follow-up observations. In Section \ref{sec:datareduction} we describe the Robo-AO data reduction and the companion detection and analysis.  In Section \ref{sec:Discoveries} we describe the results of this survey, including discovered companions, and compare to other KOI surveys.  We discuss the results in Section \ref{sec:Discussion}, detailing the effects on the planetary characteristics of the survey's discoveries and looking at the overall binarity statistics of the \textit{Kepler} planet candidates.  We conclude in Section \ref{sec:conclusion}.

\begin{figure*}
\centering
\includegraphics[width=0.83\paperwidth]{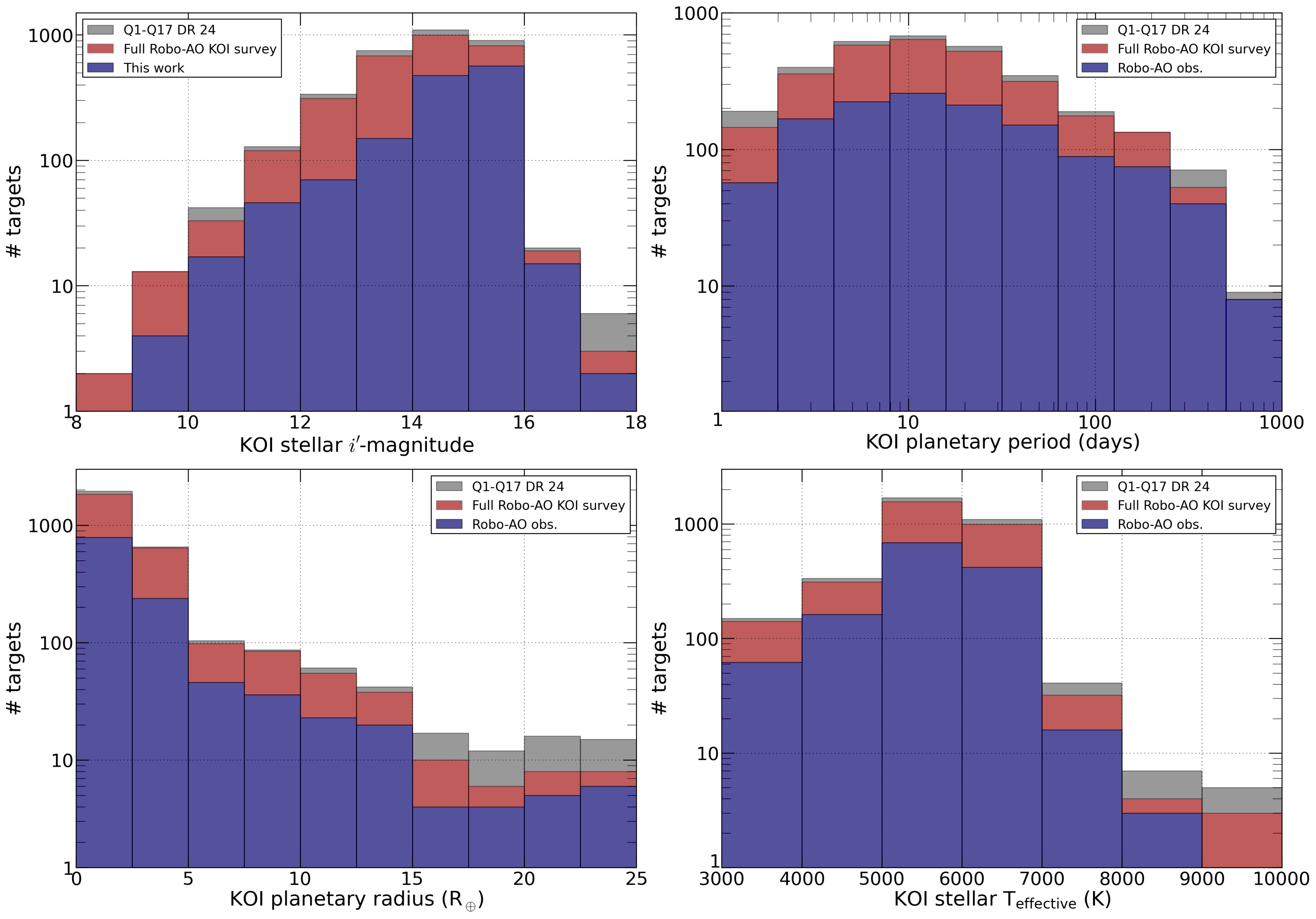}
\caption{Comparison of the distribution of the Robo-AO sample in this paper as well as the combined Robo-AO survey (Paper I, Paper II, and this work) to the complete set of KOIs from Q1-Q17 \citep{borucki10, borucki11a, borucki11b, batalha13, burke14, rowe14, coughlin15}.}
\label{fig:histograms}
\end{figure*}

\section{Survey Targets and Observations}
\label{sec:targetselection}

\subsection{Target Selection}
KOI targets were selected from the KOI catalog based on Q1-Q17 \textit{Kepler} data \citep{borucki10, borucki11a, borucki11b, batalha13, burke14, rowe14, coughlin15}.  We selected targets not observed in Paper I and Paper II, with the objective of completing the Robo-AO survey of all KOIs, including those with already detected companions.  Observations in this paper are primarily from the 2014-15 observing seasons; residual observations of dim targets from 2012-13 are also included, their analysis now possible using our improved binary detection and characterization pipeline.  KOIs flagged as false positives using \textit{Kepler} data were removed.  In Figure$~\ref{fig:histograms}$ the properties of the targeted KOIs in this work as well as for the full Robo-AO survey as of the end of the 2015 observing season are compared to the set of all KOIs from Q1-Q17 with CANDIDATE dispositions based on \textit{Kepler} data.  The Robo-AO target distribution closely matches the full KOI list in magnitude, planetary radius, planetary orbital period, and stellar temperature.  On-sky positions of all targeted KOIs in the complete survey are displayed in Figure$~\ref{fig:fovplot}$.

\subsection{Observations}
\subsubsection{Robo-AO}
We obtained high-angular-resolution images of the 1629 KOIs during 55 separate nights of observations between 2012 July 16 and 2015 June 12 (UT), detailed in Table$~\ref{tab:whitelist}$ in the Appendix.  The observations were performed using the Robo-AO laser adaptive optics system \citep{baranec13, baranec14, riddle12} mounted on the Palomar 1.5-m telescope.  The first robotic laser guide star adaptive optics system, the automatic Robo-AO system can efficiently perform large, high angular resolution surveys. The AO system runs at a loop rate of 1.2 kHz to correct high-order wavefront aberrations, delivering a median Strehl ratio of 9\% in the \textit{i}\textsuperscript{$\prime$}-band.  Observations were taken in either a \textit{i}\textsuperscript{$\prime$}-band filter or a long-pass filter cutting on at 600 nm (LP600 hereafter).  The LP600 filter approximates the \textit{Kepler} passband at redder wavelengths, while also suppressing blue wavelengths that reduce adaptive optics performance.

Typical seeing at the Palomar Observatory is between 0$\farcs$8 and 1$\farcs$8, with median around 1$\farcs$1 \citep{baranec14}. The typical FWHM (diffraction limited) resolution of the Robo-AO system is 0$\farcs$15. Images are recorded on an electron-multiplying CCD (EMCCD), allowing short frame rates for tip and tilt correction in software using a natural guide star ($m_V < 16$) in the field of view.  Specifications of the Robo-AO KOI survey are summarized in Table$~\ref{tab:specs}$.

\subsubsection{Keck LGS-AO}
\label{sec:keckao}
Eight candidate multiple systems were selected for re-imaging by the NIRC2 camera behind the Keck-II laser guide star adaptive optics system \citep{KeckLGS1, KeckLGS2}, on 2015 July 25 (UT) to confirm possible companions.  The targets were selected for their low significance of detectability, either because of low contrast ratio or small angular separation. Observations were performed in the K$_{prime}$ filter using the narrow mode of NIRC2 (9.952 mas pixel$^{-1}$; \citealt{Yelda10}), dithering the primary target at intervals of 30 s into the 3 lowest noise quadrants, for a total exposure time of 90 s. The images were corrected for geometric distortion using the NIRC2 distortion solution of \citet{Yelda10}.    Targets observed with Keck are detailed in Table$~\ref{tab:keck}$.  Further follow-up observations of low-significance companion detections are ongoing and will appear in future papers in this survey.

\subsubsection{Gemini LGS-AO}
\label{sec:gemini}
Seven candidate multiple systems from this work and three from Paper I and Paper II, again selected for their low detection significance, were re-imaged with the adaptive optics assisted NIRI instrument \citep{hodapp03} on the Gemini North telescope.  Three targets were observed on 2015 July 31 (UT) and seven targets were observed on 2015 August 27, using Band 3 allocated time. Targets observed with Gemini are detailed in Table$~\ref{tab:gemini}$.  Observations were performed with the F/32 camera, providing resolution of 21.9 mas pixel$^{-1}$ across a field of view of 22$\arcsec \times 22\arcsec$.  Total integration times were 90 s in the K$_{prime}$ band across three dithered images, used to increase dynamic range and allow sky subtraction.  The common striping pattern found in NIRI images was removed using the \textit{cleanir.py} script provided by the Gemini staff. The images were flat fielded, bad pixel corrected, and sky subtracted.   The distortion solution provided by the Gemini staff was used to correct the images for distortion.

\begin{table}
\renewcommand{\arraystretch}{1.3}
\begin{longtable}{ll}
\caption{\label{tab:survey_specs}The specifications of the Robo-AO KOI survey}
\\
\hline
KOI targets    	& 1629 \\
FWHM resolution   	& $\sim$0$\farcs$15 (@600-750 nm) \\
Observation wavelengths & 600-950 nm\\
Field size & 44\arcsec $\times$ 44\arcsec\\
Detector format & 1024$^2$ pixels\\
Pixel scale & 43.1 mas / pix\\
Exposure time & 90 seconds \\
Targets observed / hour & 20\\
Observation dates & 2012 July 16 --\\ &  2015 June 12\\
\hline
\label{tab:specs}
\end{longtable}
\end{table}

\begin{figure}
\centering
\includegraphics[width=245pt]{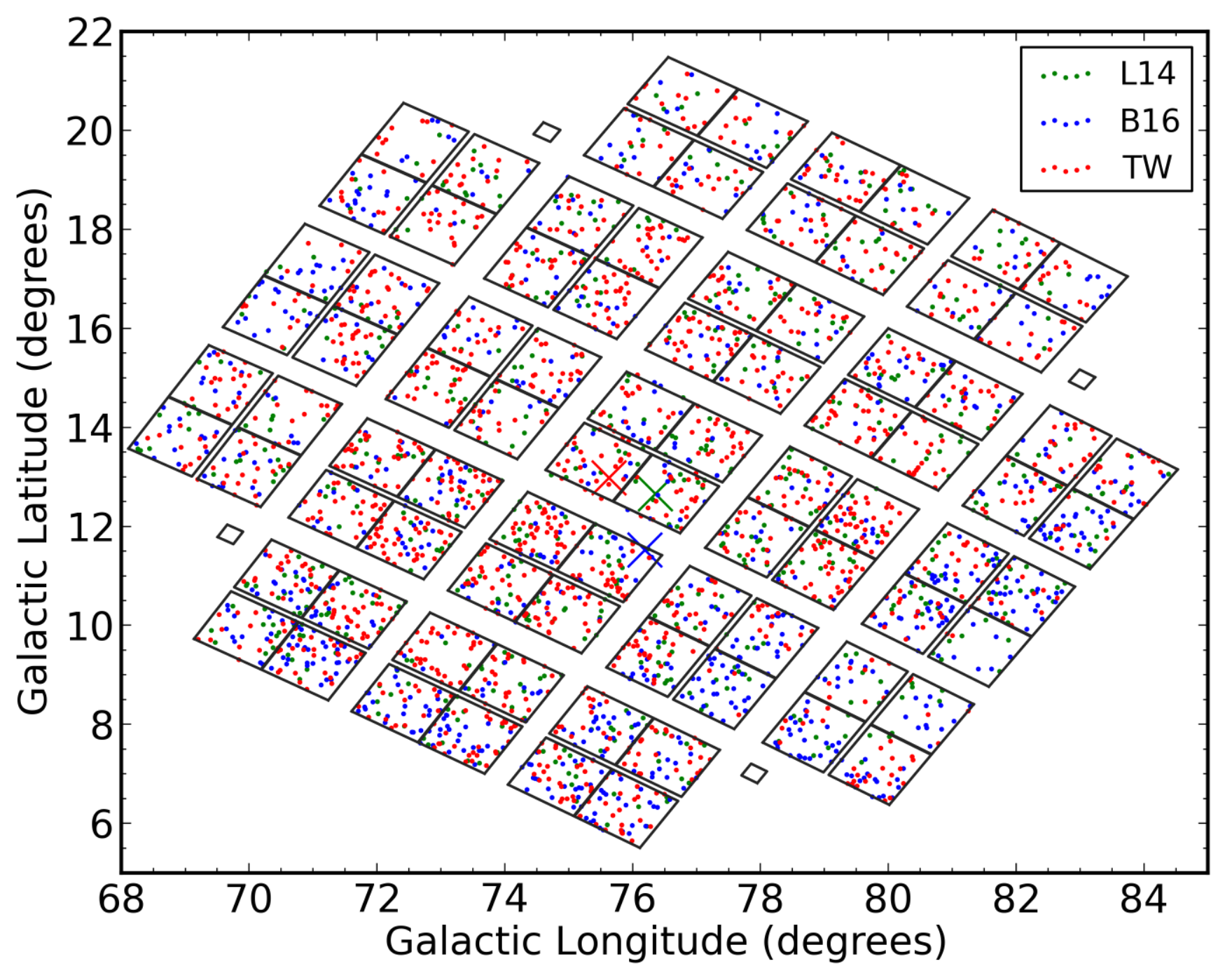}
\caption{Location on sky of targeted KOIs from Paper I (L14), Paper II (B16), and this work (TW).  The median coordinates of the targeted KOIs is designated by an `$\times$'.  A projection of the \textit{Kepler} field of view is provided for reference.}
\label{fig:fovplot}
\end{figure}

\section{Data Reduction}
\label{sec:datareduction}
With the largest adaptive optics dataset yet assembled by Robo-AO, the data reduction process was automated as much as possible for efficiency and consistency.  As in Paper I and \citet{ziegler15}, after initial pipeline reductions described in Section \ref{sec:pipeline}, the target stars were identified (Section \ref{sec:targetverification}), companions automatically identified (Section \ref{sec:compsearch}), PSF subtraction performed and companions again auto-identified (Section \ref{sec:psfsubtraction}), and constraints of the companion sensitivity of the survey measured (Section \ref{sec:imageperf}).  Finally, the properties of the detected companions are measured in Section \ref{sec:characterization}.

\subsection{Imaging Pipeline}
\label{sec:pipeline}

The Robo-AO imaging pipeline \citep{law09, law14} reduced the images: the raw EMCCD output frames are dark-subtracted and flat-fielded and then stacked and aligned using the Drizzle algorithm \citep{fruchter02}, which also up-samples the images by a factor of two.  To avoid tip/tilt anisoplanatasism effects, the image motion was corrected by using the KOI itself as the guide star in each observation.

\subsection{Target Verification}
\label{sec:targetverification}

To verify that the star viewed in the image is the desired KOI target, we created Digital Sky Survey cutouts of similar angular size around the target coordinates.  Each image was manually checked to assure no ambiguity in the target star with images with either poor performance or incorrect fields removed.  These bad images made up approximately 2$\%$ of all our images, and for all but 2 of the targets additional images were available. 

\begin{figure}
\centering
\includegraphics[width=180pt]{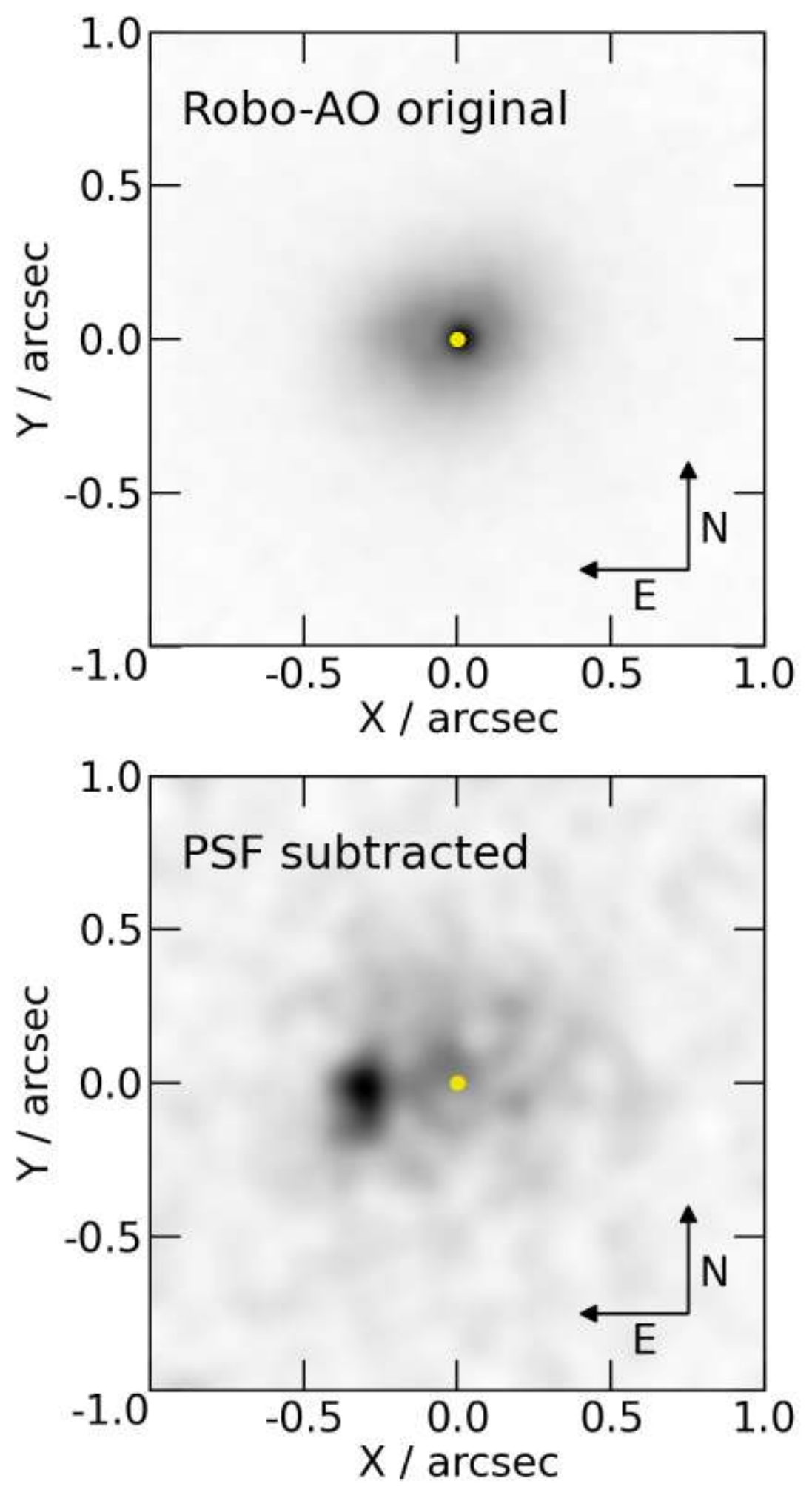}
\caption{Example of PSF subtraction on KOI-5762 with companion separation of 0$\farcs$34.  The yellow circle marks the position of the primary star's PSF peak. Both images have been scaled and smoothed for clarity.  Successful removal of the PSF leaves residuals consistent with photon noise.  The 2\arcsec square field shown here is approximately equal to half the \textit{Kepler} pixel size.  The close companion to KOI-5762 was confirmed with NIRC2/Keck images, shown in Figure$~\ref{fig:keckgrid}$.}
\label{fig:psf}
\end{figure}

\subsection{Image Preparation}
To facilitate the automation of the data reduction, centered 8$\farcs$5 square cutouts were created around the 1629 verified target KOIs.  We select a 4\arcsec separation cutoff for our companion search to detect all nearby stars that would blend with the target KOI in a \textit{Kepler} pixel.

\subsection{PSF Subtraction}
\label{sec:psfsubtraction}

To identify close companions, a custom locally optimized point spread function (PSF) subtraction routine based on the Locally Optimized Combination of Images algorithm \citep{lafreniere07} was applied to centered cutouts of all stars.  Detailed in Paper I, the code uses a set of twenty KOI observations, selected from the observations within the same filter closest to the target observation in time, as reference PSFs, as it is improbable that a companion would appear at the same position in two different images.  A locally optimized PSF is generated and subtracted from the original image, leaving residuals consistent with photon noise.

This procedure was performed on all KOI images out to 2$\arcsec$, and the results visually checked for companions.  Figure$~\ref{fig:psf}$ shows an example of the PSF subtraction performance.  The PSF subtracted images were subsequently run through the automated companion finding routine, as described in Section \ref{sec:compsearch}.

\subsection{Companion Detection}
\label{sec:compsearch}

An initial visual companion search was performed redundantly by three of the authors.  This search yielded a preliminary companion list, and filtered out bad images.

Continuing the companion search, we ran all images through a custom automated search algorithm, based on the code described in Paper I. The algorithm slides a 5-pixel diameter aperture within concentric annuli centered on the target star. Any aperture with $>$+5$\sigma$ outlier to the local noise is considered a potential astrophysical source. These are subsequently checked manually, eliminating spurious detections with dissimilar PSFs to the target star and those having characteristics of a cosmic ray hit, such as a single bright pixel or bright streak.  The detection significance of `secure' companions are listed in Tables$~\ref{tab:secure25}$ and $~\ref{tab:secure40}$.

Many possible companions were visually identified but fell beneath the formal 5$\sigma$ required for a discovery. Despite not reaching our formal significance level required for a discovery, previous results suggest that all but a small fraction are likely real: Keck/NIRC2 observations have confirmed all 15 `likely' detections in Paper I and all 38 re-observed `likely' companions in Paper II.  The detection significance of these `likely' companions are listed in Tables$~\ref{tab:likely25}$ and $~\ref{tab:likely40}$.

\begin{figure*}
\centering
\includegraphics[width=0.75\paperwidth]{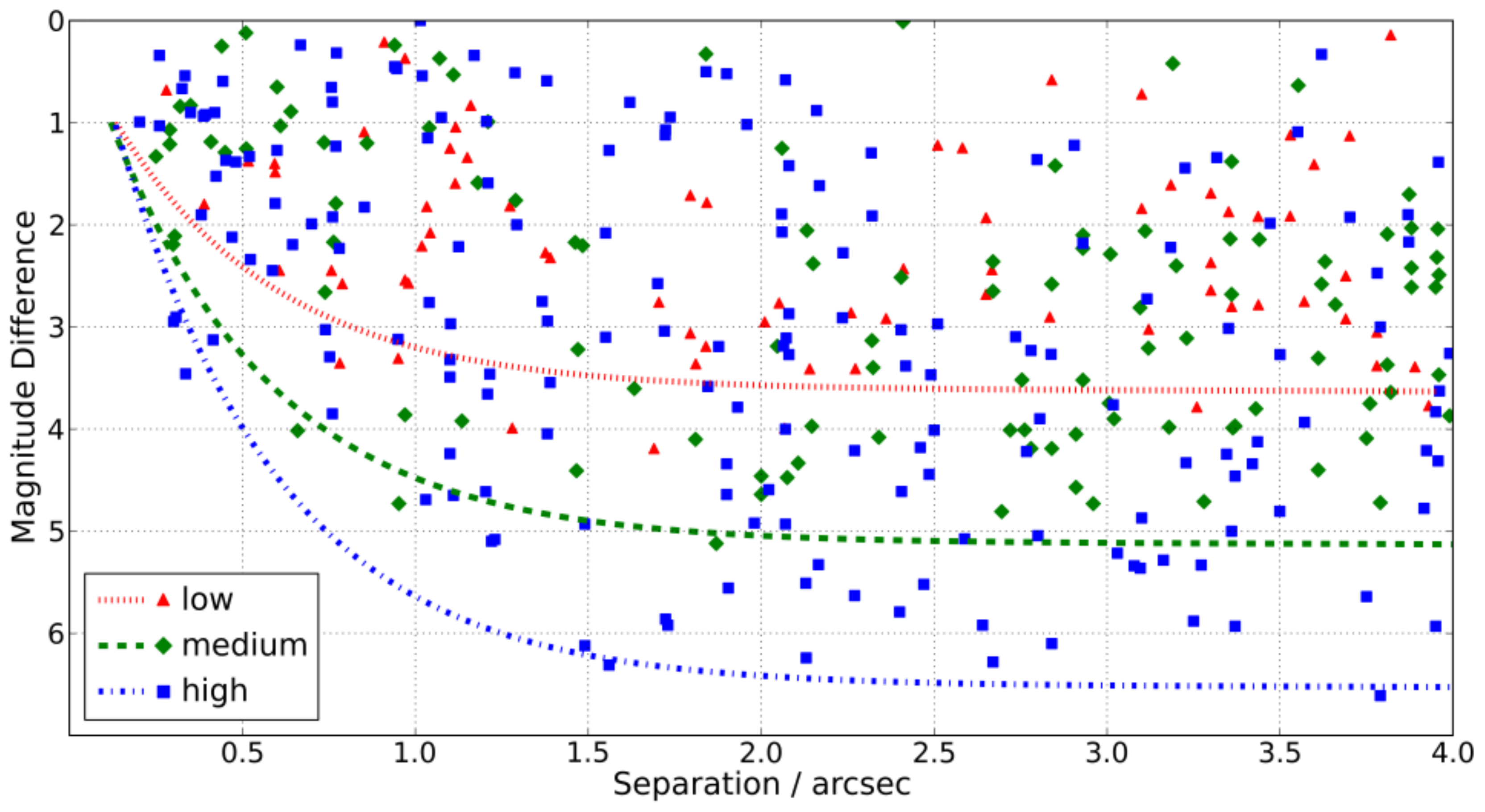}
\caption{Separations and magnitude differences of the detected companions in Paper II and this work, with the color and shape of each star denoting the associated typical low-, medium- and high-performance 5$\sigma$ contrast curve during the observation (as described in Section \ref{sec:imageperf}).}
\label{fig:contrastcurves}
\end{figure*}

\subsection{Imaging Performance Metrics}
\label{sec:imageperf}

The two dominant factors that affect the image performance of the Robo-AO system are seeing and target brightness.  An automated routine was used to classify the image performance for each target.  Described in detail in Paper I, the code uses PSF core size as a proxy for image performance.  Observations were binned into three performance groups, with 31\% fall in the low performance group, 41\% in the medium performance group, and 28\% in the high performance group.

We determine the angular separation and contrast consistent with a 5$\sigma$ detection by injecting artificial companions, a clone of the primary PSF.  For concentric annuli of 0$\farcs$1 width, the detection limit is calculated by steadily dimming the artificial companion until the auto-companion detection algorithm (Section \ref{sec:compsearch}) fails to detect it.  This process is subsequently performed at multiple random azimuths within each annulus and the limiting 5$\sigma$ magnitudes are averaged. For clarity, these average magnitudes for all radii measurements are fitted with functions of the form $a*sinh(b*r+c)+d$ (where \textit{r} is the radius from the target star and \textit{a, b, c} and \textit{d} are fitting variables).  Contrast curves for the three performance groups are shown in Section$~\ref{sec:Discoveries}$ in Figure$~\ref{fig:contrastcurves}$.

\subsection{Companion Characterization}
\label{sec:characterization}

\subsubsection{Contrast Ratios}
\label{sec:contrastratios}

For wide, resolved companions with little PSF overlap, the companion to primary star contrast ratio was determined using aperture photometry on the original images. The aperture radius was cycled in one-pixel increments from 1-5 FWHM for each system, with background measured opposite the primary from the companion (except in the few cases where another object falls near or within this region in the image).  Photometric uncertainties are estimated from the standard deviation of the contrast ratios measured for the various aperture sizes.

For close companions, the estimated PSF was used to remove the blended contributions of each star before aperture photometry was performed.  The locally optimized PSF subtraction algorithm can attempt to remove the flux from companions using other reference PSFs with excess brightness in those areas.  For detection purposes, we use many PSF core sizes for optimization, and the algorithm's ability to remove the companion light is reduced. However, the companion is artificially faint as some flux has still been subtracted. To avoid this, the PSF fit was redone excluding a six-pixel-diameter region around the detected companion.  The large PSF regions allow the excess light from the primary star to be removed, while not reducing the brightness of the companion.

\subsubsection{Separation and Position Angles}
\label{sec:separationposangles}

Separation and position angles were determined from the raw pixel positions.  Uncertainties were found using estimated systematic errors due to blending between components.  Typical uncertainty in the position for each star was 1-2 pixels.  Position angles and the plate scale were calculated using a distortion solution produced using Robo-AO measurements for the globular cluster M15.\footnote{S. Hildebrandt (2013, private communication)}

\subsubsection{Companion Spectral Types}
\label{sec:spectraltypes}

The approximate spectral type of the detected companions, assuming that they are bound to the primary and all stars are main sequence dwarfs, were estimated using an SED model \citep{kraus07} and the estimated stellar effective temperatures reported on the NASA Exoplanet Archive.  With the LP600 band closely matching the \textit{Kepler} bandpass, the magnitude differences between the primary star and nearby stars were converted to \textit{i}\textsuperscript{$\prime$}-band when necessary using the linear correlation found by \citet{lillo14}:
\begin{equation}
\Delta m_{i'}=0.947\cdot\Delta m_{LP600}
\end{equation}
In addition, we estimate the latest spectral type companion consistent with a 5$\sigma$ detection for each observed target based on the typical contrast curve for the three image performance groups (see Section$~\ref{sec:imageperf}$).  These estimates are listed in Table$~\ref{tab:whitelist}$ in the Appendix.

We caution that these spectral types are approximate and do not account for factors such as giant contamination, estimated at $\sim$12$\%$ by \citet{ciardi11}.  In addition, the use of a linear correlation in converting between passbands will result in an error with varying spectral types.  We estimate this error by calculating the flux of F0V to M5V stars \citep{pickles98} in the LP600 and \textit{i}\textsuperscript{$\prime$}-band inlcuding quantum efficiencies and instrumental effects (see Figure 2 in Paper I).  The maximum difference between the flux of spectral types in the two passbands results in an error of $\sim$0.15 mags, equivalent to approximately one subspectral type.

\begin{figure*}
\centering
\includegraphics[width=480pt]{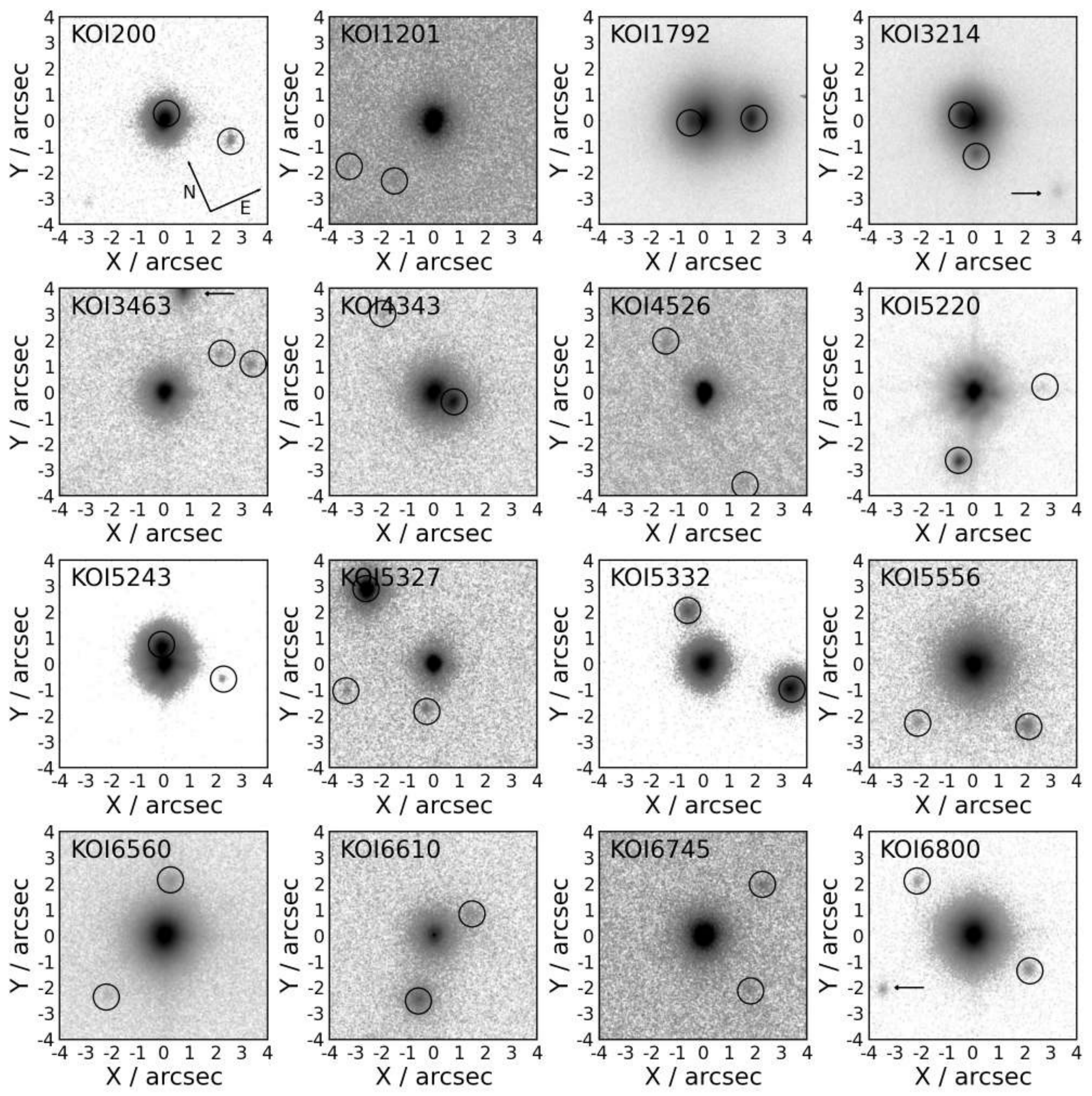}
\caption{Normalized log-scale cutouts of 16 KOIs with multiple companions with separations $<$4$\arcsec$ resolved with Robo-AO.  The angular scale and orientation (displayed in the first frame) is similar for each cutout, and circles are centered on the detected nearby stars.  Three targets (KOIs 3214, 3463, and 6800) have a possible third companion, marked with arrows, outside our 4$\arcsec$ separation cutoff, as described in Section$~\ref{sec:quadsystems}$.}
\label{fig:multiplesgrid}
\end{figure*}

\begin{figure*}
\centering
\includegraphics[width=480pt]{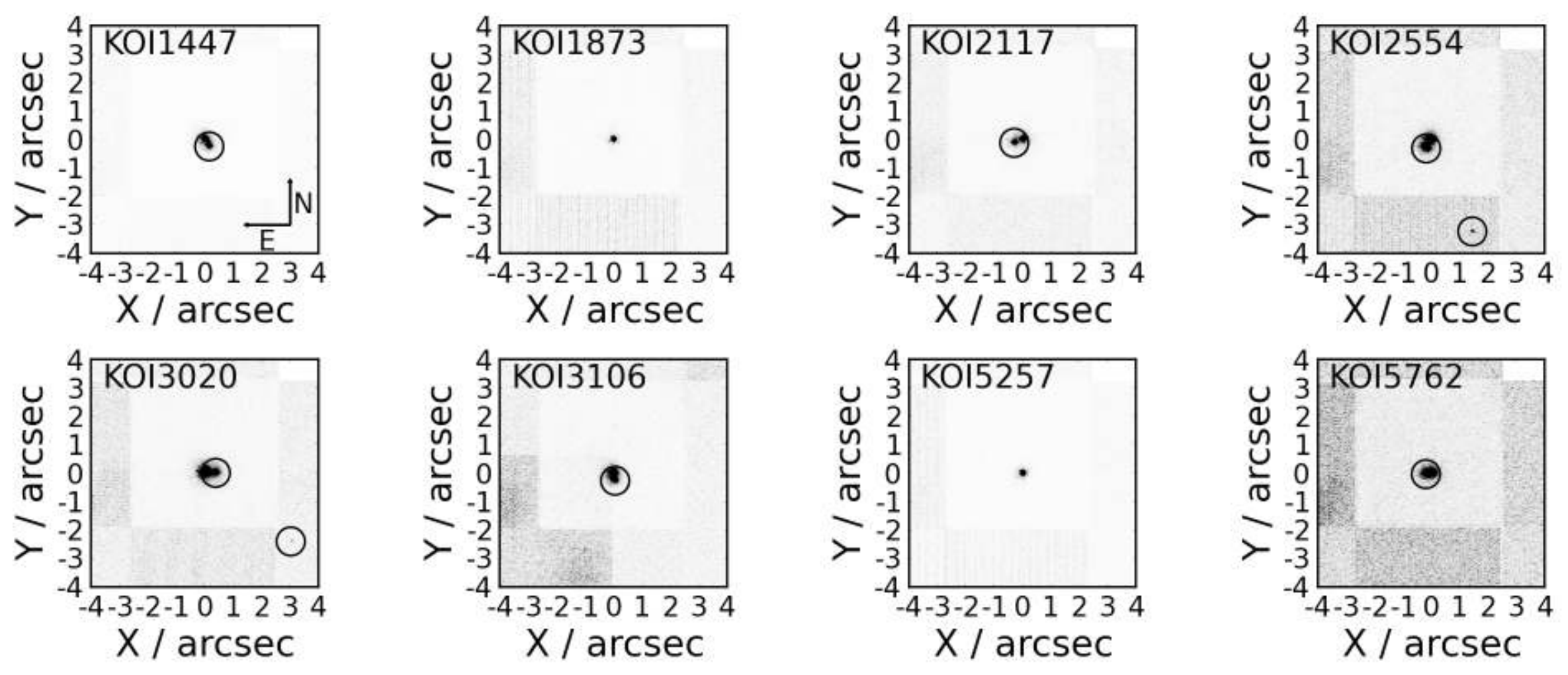}
\caption{Normalized log-scale cutouts of 8 KOIs observed with the NIRC2 instrument on Keck-II, as described in Section \ref{sec:keckao}.  The angular scale and orientation (displayed in the first frame) is similar for each cutout, and circles are centered on the detected nearby stars.}
\label{fig:keckgrid}
\end{figure*}

\section{Discoveries} 
\label{sec:Discoveries}

Of the 1629 KOI targets observed, 206 are apparent in multiple star systems for a nearby star fraction within 4\arcsec of 12.6\%$\pm$0.9$\%$\footnote{Error based on Poissonian statistics \citep{burgasser03}}. We also found 15 triple systems for a triplet fraction of $0.92^{+0.30}_{-0.18}\%$\footnote{\label{binomial}Error based on binomial statistics \citep{burgasser03}}, and 1 quadruple system for a quadruplet fraction of $0.06^{+0.14}_{-0.02}\%^{11}$.  Cutouts of the triple and quadruple systems are shown in Figure$~\ref{fig:multiplesgrid}$.  One quarter (25.8$\%$) of the companions would only be observable in a high-resolution survey ($<$1$\farcs$0 separation), and one half (49.8$\%$) of the companions are too close ($<$2$\farcs$0) for many seeing limited surveys to accurately measure binary properties (e.g. contrast ratios).  The detected companion separations and contrast ratios are plotted in Figure$~\ref{fig:contrastcurves}$, along with the calculated 5$\sigma$ detection limits as detailed in Section$~\ref{sec:imageperf}$.  Cutouts of all multiple star systems are shown in Figures$~\ref{fig:cutouts1}$,$~\ref{fig:cutouts2}$,$~\ref{fig:cutouts3}$, and$~\ref{fig:cutouts4}$.   For companions within 2$\farcs$5, measured properties are detailed in Tables$~\ref{tab:secure25}$ and$~\ref{tab:likely25}$.  For companions outside 2$\farcs$5 but within  4$\farcs$0, measured properties are detailed in Tables$~\ref{tab:secure40}$ and$~\ref{tab:likely40}$.

We confirmed six companions to eight Robo-AO detections with NIRC2 and AO on Keck-II \citep{wizinowich00}.  In addition, two new companions were found around KOIs 2554 and 3020.  These targets were selected for followup because of their faintness and/or closely separated detected companion.  Low-sigma, visually detected companions to KOIs 1873 and 5257 were not detected.  These non-detected companions are possibly a result of non-common path aberrations, as described in Section 5.1 of Paper II.  These spurious detections all have similar separations and position angles with respect to the target star, facilitating their identification and manual removal.  The PSF subtraction routine usually does not remove these false companions, as another star exhibiting the NCP error is unlikely to be within the set of twenty reference images.  The Keck-II observations are listed in Table$~\ref{tab:keck}$, with the measured separations and position angles of the confirmed companions using Keck-II images listed in Tables$~\ref{tab:secure25}$ and $~\ref{tab:likely25}$, and the follow-up images are shown in Figure$~\ref{fig:keckgrid}$.

\begin{figure*}
\centering
\includegraphics[width=480pt]{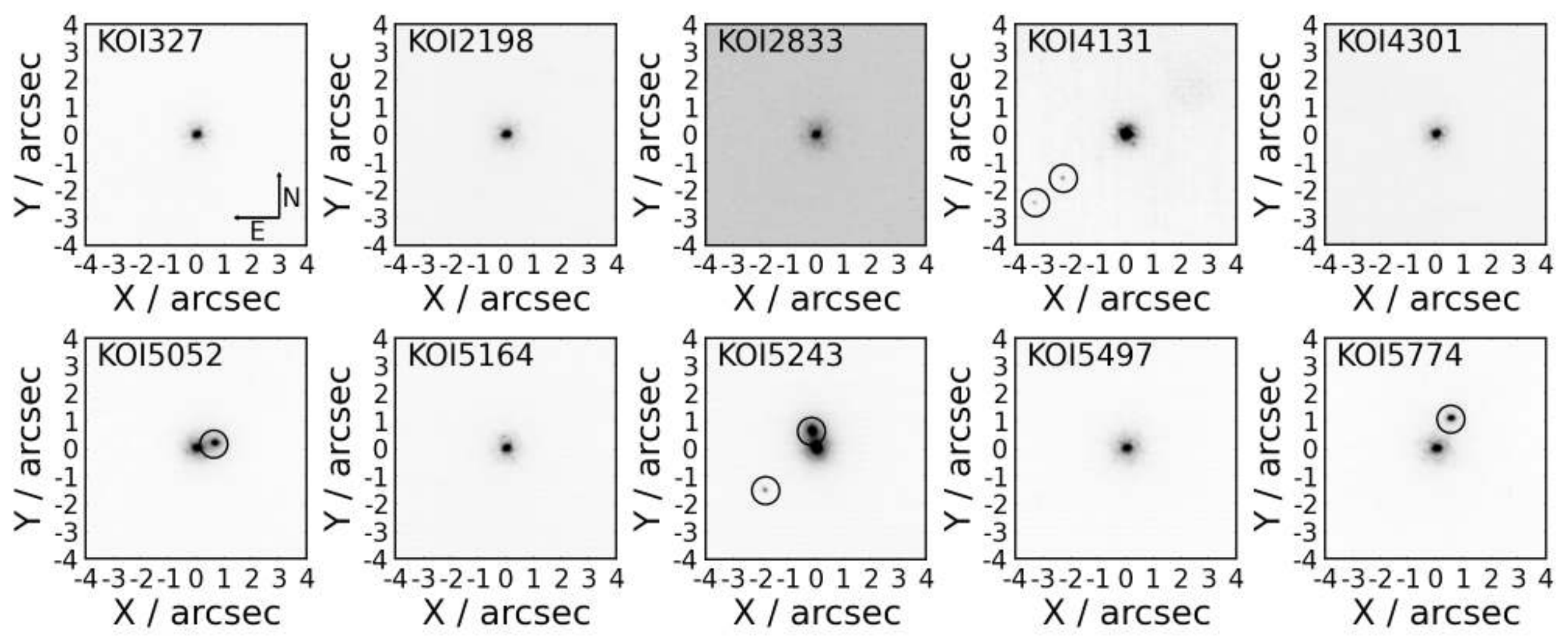}
\caption{Normalized log-scale cutouts of 10 KOIs observed with the NIRI instrument on Gemini North, as described in Section \ref{sec:gemini}.  The angular scale and orientation (displayed in the first frame) is similar for each cutout, and circles are centered on the detected nearby stars.}
\label{fig:geminigrid}
\end{figure*}

We confirmed five companions to seven KOIs observed in this paper with NIRI and AO on Gemini North.  We did not detect a possible companion to KOI-2198 that was visually detected, manifesting as an elongated PSF in the Robo-AO image. We observed three KOIs targeted in Paper I (KOI-327) and Paper II (KOIs 2833 and 4301) which displayed non-common path error aberrations.  No companions were observed to these three targets in the follow-up observations.  A new companion outside our separation cutoff ($\rho$=4$\farcs$24) was observed nearby KOI-4131.  The Gemini observations are listed in Table$~\ref{tab:gemini}$, and the follow-up images are shown in Figure$~\ref{fig:geminigrid}$.

\begin{table}
\renewcommand{\arraystretch}{1.3}
\begin{center}
\caption{Full Keck-AO Observation List}
\begin{tabular}{ccccc}
\hline
\hline
\noalign{\vskip 1pt}  
\text{KOI} & \text{m$_v$} & \text{ObsID} & \text{Companion?} & \text{$\Delta$\textit{K$_{p}$}}\\ [0.2ex]
\hline\
1447 & 13.2 & 2015 Jul 25 & yes & 0.63$\pm$0.06\\
1873 & 15.8 & 2015 Jul 25 & & \\
2117 & 16.2 & 2015 Jul 25 & yes & 0.53$\pm$0.06\\
2554 & 15.9 & 2015 Jul 25 & yes & 0.27$\pm$0.05\footnote{Companion at $\rho$=0$\farcs$37}\\
 & & & yes & 2.96$\pm$0.10\footnote{New companion at $\rho$=3$\farcs$55}\\
3020 & 13.8 & 2015 Jul 25 & yes & 1.27$\pm$0.06\footnote{Companion at $\rho$=0$\farcs$38}\\
 & & & yes & 5.01$\pm$0.07\footnote{New companion at $\rho$=3$\farcs$86}\\
3106 & 15.7 & 2015 Jul 25 & yes & 1.22$\pm$0.13\\
5257 & 15.5 & 2015 Jul 25 & & \\
5762 & 15.9 & 2015 Jul 25 & yes & 0.83$\pm$0.08\\
\hline
\end{tabular}
\label{tab:keck}
\end{center}
\end{table}

\begin{table}
\renewcommand{\arraystretch}{1.3}
\begin{center}
\caption{Full Gemini Observation List}
\begin{tabular}{ccccc}
\hline
\hline
\noalign{\vskip 1pt}  
\text{KOI} & \text{m$_v$} & \text{ObsID} & \text{Companion?} & \text{$\Delta$\textit{K$_{p}$}}\\ [0.2ex]
\hline\
327 & 13.1 & 2015 Aug 27 & & \\
2198 & 12.8 & 2015 Aug 27 & & \\
2833 & 12.8 & 2015 Aug 27 & & \\
4131 & 13.2 & 2015 Jul 31 & yes & 4.41$\pm$0.09\footnote{Companion at $\rho$=2$\farcs$85} \\
 & & & yes & 4.96$\pm$0.11\footnote{New companion at $\rho$=4$\farcs$24}\\
4301 & 13.3 & 2015 Aug 27 & & \\
5052 & 12.8 & 2015 Jul 31 & yes & 0.75$\pm$0.04 \\
5164 & 12.6 & 2015 Aug 27 & & \\
5243 & 12.5 & 2015 Jul 31 & yes & 0.53$\pm$0.05\footnote{Companion at $\rho$=0$\farcs$77}\\
 & & & yes & 4.11$\pm$0.09\footnote{Companion at $\rho$=2$\farcs$41}\\
5497 & 11.0 & 2015 Aug 27 & & \\
5774 & 11.1 & 2015 Aug 27 & yes & 1.54$\pm$0.04\\
\hline
\end{tabular}
\label{tab:gemini}
\end{center}
\end{table}

\begin{center}
\begin{table*}
\renewcommand{\arraystretch}{.8}
\setlength{\tabcolsep}{2pt}
\caption{Secure Detections of Objects within 2\farcs5 of \textit{Kepler} Planet Candidates}
\footnotesize
\centering
\begin{tabular}{l c c c c c c c c c c c c}
\hline
\hline
\noalign{\vskip 3pt}
KOI & $\rm m_{\textit{i}\textsuperscript{$\prime$}}$ & ObsID & Filter & Det. Significance & Separation & P.A. & Mag. Diff. & Approx. Comp. & Prev. High Res. & Prev. Low Res. & False Positive?\footnote{Probability that planetary transit signal is a false positive based on \textit{Kepler} data.} & $N_{KOI}$\footnote{Number of planet candidates detected orbiting KOI. } \\
& (mag) & & & $\sigma$ & ($\arcsec$) & (deg.) & (mag) & Spectral Type\footnote{Estimation method described in Section$~\ref{sec:spectraltypes}$.} &  Detection? & Detection? & & \\
\hline
\noalign{\vskip 3pt}
163 & 13.3 & 2012 Jul 18 & LP600 & 17 & 1.22$\pm$0.06 & 214$\pm$2 & --0.36$\pm$0.03 & K1 & & &  & 1 \\
454 & 14.5 & 2014 Jul 16 & LP600 & 6.6 & 1.49$\pm$0.06 & 204$\pm$2 & 2.08$\pm$0.04 & M0 & & &  & 1 \\
510 & 14.3 & 2014 Jul 14 & LP600 & 5.6 & 2.45$\pm$0.06 & 348$\pm$2 & 2.53$\pm$0.05 & M1 & & UKIRT &  & 4 \\
771 & 15.1 & 2014 Aug 27 & LP600 & 18 & 1.77$\pm$0.06 & 281$\pm$2 & 0.94$\pm$0.05 & K5 & W15 & &  & 1 \\
1137 & 13.8 & 2014 Jun 13 & LP600 & 6.1 & 0.75$\pm$0.06 & 197$\pm$2 & 0.81$\pm$0.08 & K5 & & & 0.01 & 1 \\
1409 & 15.0 & 2014 Jul 17 & LP600 & 27 & 2.17$\pm$0.06 & 312$\pm$2 & 2.58$\pm$0.02 & M0 & & UKIRT &  & 1 \\
1447 & 13.0 & 2012 Sep 04 & LP600 & 5.1 & 0.28$\pm$0.03\footnote{From Keck follow-up, described in Section$~\ref{sec:Discoveries}$.} & 212$\pm$2\textsuperscript{d} & 0.27$\pm$0.08 & F6 & & & 0.01,0.02 & 2 \\
1630 & 14.9 & 2014 Jul 16 & LP600 & 12 & 1.77$\pm$0.06 & 188$\pm$2 & 0.91$\pm$0.02 & K3 & & UKIRT &  & 1 \\
1687 & 14.9 & 2014 Jul 17 & LP600 & 6.3 & 2.11$\pm$0.06 & 209$\pm$2 & 4.10$\pm$0.16 & M5 & & &  & 1 \\
1792 & 11.9 & 2014 Sep 02 & LP600 & 9.6 & 1.99$\pm$0.06 & 111$\pm$2 & 0.98$\pm$0.05 & K4 & CFOP & UKIRT & 0.02 & 3 \\
2091 & 15.5 & 2014 Aug 27 & LP600 & 7.2 & 1.30$\pm$0.06 & 215$\pm$2 & 1.72$\pm$0.04 & M0 & & &  & 1 \\
2093 & 15.2 & 2014 Aug 27 & LP600 & 8.7 & 2.08$\pm$0.06 & 352$\pm$2 & 3.10$\pm$0.02 & M0 & & &  & 3 \\
2163 & 14.4 & 2014 Aug 31 & LP600 & 6.9 & 0.77$\pm$0.06 & 248$\pm$2 & 0.04$\pm$0.03 & G2 & & &  & 3 \\
2535 & 14.6 & 2014 Aug 23 & LP600 & 5.9 & 1.73$\pm$0.06 & 21$\pm$2 & 2.47$\pm$0.02 & M2 & & UKIRT &  & 1 \\
2554 & 15.0 & 2014 Sep 01 & LP600 & 5.4 & 0.37$\pm$0.03\textsuperscript{d} & 149$\pm$2\textsuperscript{d} & 0.37$\pm$0.08 & M0 &  & &  & 2 \\
2813 & 13.3 & 2013 Aug 15 & LP600 & 27 & 1.10$\pm$0.06 & 258$\pm$2 & 0.84$\pm$0.02 & K7 & D14, K16  & & 0.01 & 1 \\
2896 & 11.9 & 2015 Jun 05 & LP600 & 13 & 0.96$\pm$0.06 & 272$\pm$2 & 0.38$\pm$0.02 & F8 & CFOP & & 0.01,0.02 & 2 \\
2900 & 15.0 & 2014 Sep 03 & LP600 & 10 & 2.36$\pm$0.06 & 85$\pm$2 & 1.30$\pm$0.04 & M0 & & UKIRT &  & 1 \\
2976 & 15.6 & 2014 Aug 28 & LP600 & 66 & 2.02$\pm$0.06 & 198$\pm$2 & 2.66$\pm$0.06 & M2 & & UKIRT &  & 1 \\
3020 & 13.5 & 2013 Aug 13 & LP600 & 5.2 & 0.38$\pm$0.03\textsuperscript{d} & 272$\pm$2\textsuperscript{d} & 0.93$\pm$0.22 & G9 & & &  & 1 \\
3042 & 15.8 & 2014 Aug 31 & LP600 & 6.1 & 1.87$\pm$0.06 & 147$\pm$2 & 1.62$\pm$0.02 & K3 & & UKIRT &  & 1 \\
3112 & 15.6 & 2014 Sep 01 & LP600 & 14 & 1.87$\pm$0.06 & 151$\pm$2 & 0.49$\pm$0.03 & K3 & & UKIRT &  & 1 \\
3120 & 14.6 & 2014 Aug 29 & LP600 & 8.1 & 1.14$\pm$0.06 & 278$\pm$2 & 0.87$\pm$0.03 & G8 &  & &  & 1 \\
3214 & 11.8 & 2014 Aug 29 & LP600 & 10 & 1.41$\pm$0.06 & 198$\pm$2 & 2.50$\pm$0.04 & M0 & & &  & 2 \\
3413 & 15.0 & 2014 Aug 26 & LP600 & 54 & 2.18$\pm$0.06 & 12$\pm$2 & 3.79$\pm$0.03 & M2 & & UKIRT & 0.01 & 1 \\
3415 & 13.1 & 2013 Jul 27 & LP600 & 10 & 0.74$\pm$0.06 & 89$\pm$2 & 0.03$\pm$0.05 & G9 & & &  & 1 \\
3483 & 14.7 & 2014 Nov 09 & LP600 & 11 & 1.51$\pm$0.06 & 23$\pm$2 & 2.15$\pm$0.03 & K5 & & &  & 1 \\
3649 & 15.2 & 2014 Aug 23 & LP600 & 8.1 & 0.79$\pm$0.06 & 216$\pm$2 & 0.26$\pm$0.03 & F9 & LB14  & & 0.01 & 1 \\
3660 & 15.3 & 2014 Aug 24 & LP600 & 6.5 & 0.60$\pm$0.06 & 160$\pm$2 & 1.05$\pm$0.12 & K4 & & &  & 1 \\
3770 & 13.9 & 2014 Jun 19 & LP600 & 11 & 1.20$\pm$0.06 & 34$\pm$2 & 1.44$\pm$0.04 & K1 & & &  & 1 \\
3886 & 9.5 & 2014 Aug 20 & \textit{i}\textsuperscript{$\prime$} & 13 & 0.50$\pm$0.06 & 116$\pm$2 & 1.13$\pm$0.09 & M0 & LB14  & & 0.01 & 1 \\
4343 & 13.5 & 2014 Jun 19 & LP600 & 9.2 & 0.89$\pm$0.06 & 138$\pm$2 & 1.13$\pm$0.05 & M4 & & &  & 1 \\
4418 & 15.7 & 2014 Sep 03 & LP600 & 5.3 & 1.41$\pm$0.06 & 172$\pm$2 & 2.23$\pm$0.02 & M0 & & &  & 1 \\
4550 & 15.0 & 2014 Aug 29 & LP600 & 15 & 1.03$\pm$0.06 & 325$\pm$2 & 0.04$\pm$0.02 & K4 & & UKIRT &  & 1 \\
4713 & 13.4 & 2014 Jul 16 & LP600 & 39 & 1.72$\pm$0.06 & 251$\pm$2 & 0.27$\pm$0.04 & G7 & & UKIRT &  & 1 \\
4750 & 15.7 & 2014 Aug 29 & LP600 & 124 & 2.09$\pm$0.06 & 322$\pm$2 & 1.95$\pm$0.02 & M0 & & UKIRT &  & 1 \\
4895 & 14.5 & 2014 Aug 31 & LP600 & 5.8 & 2.27$\pm$0.06 & 75$\pm$2 & 2.28$\pm$0.02 & M0 & & UKIRT &  & 2 \\
5004 & 14.3 & 2014 Jul 16 & LP600 & 6.4 & 1.05$\pm$0.06 & 109$\pm$2 & 1.05$\pm$0.02 & K3 & & &  & 1 \\
5052 & 12.5 & 2014 Jun 17 & LP600 & 6.0 & 0.75$\pm$0.06 & 285$\pm$2 & 0.68$\pm$0.06 & F6 & & &  & 1 \\
5243 & 12.2 & 2014 Sep 03 & LP600 & 11 & 0.77$\pm$0.06 & 17$\pm$2 & 0.55$\pm$0.03 & G4 & & & 0.01 & 1 \\
5243 & 12.2 & 2014 Sep 03 & LP600 & 7.2 & 2.41$\pm$0.06 & 128$\pm$2 & 5.53$\pm$0.08 & M4 & & UKIRT & 0.01 & 1 \\
5570 & 14.5 & 2014 Aug 21 & LP600 & 7.7 & 2.06$\pm$0.06 & 236$\pm$2 & 4.64$\pm$0.06 & M4 & & &  & 1 \\
5578 & 10.9 & 2014 Nov 09 & LP600 & 7.4 & 0.33$\pm$0.06 & 89$\pm$2 & 1.78$\pm$0.22 & K7 & CFOP & &  & 1 \\
5665 & 11.3 & 2014 Jul 17 & LP600 & 106 & 2.11$\pm$0.06 & 91$\pm$2 & 3.24$\pm$0.03 & M1 & CFOP & &  & 1 \\
5671 & 13.4 & 2014 Jun 16 & LP600 & 61 & 2.17$\pm$0.06 & 225$\pm$2 & 1.79$\pm$0.05 & K4 & & UKIRT &  & 1 \\
5774 & 10.7 & 2014 Sep 01 & LP600 & 19 & 1.32$\pm$0.06 & 336$\pm$2 & 1.90$\pm$0.05 & G8 & & & 0.01 & 1 \\
5889 & 15.2 & 2014 Sep 01 & LP600 & 5.6 & 0.77$\pm$0.06 & 246$\pm$2 & 1.42$\pm$0.11 & K1 & & & 0.01 & 1 \\
6111 & 12.9 & 2015 Jun 04 & LP600 & 8.7 & 2.14$\pm$0.06 & 48$\pm$2 & 4.40$\pm$0.05 & M1 & & &  & 1 \\
6132 & 14.6 & 2015 Jun 12 & LP600 & 6.7 & 1.23$\pm$0.06 & 91$\pm$2 & 0.90$\pm$0.03 & G8 & & &  & 3 \\
6258 & 11.2 & 2015 Jun 04 & LP600 & 9.5 & 2.17$\pm$0.06 & 241$\pm$2 & 4.14$\pm$0.14 & M2 & CFOP & & 0.01 & 1 \\
6329 & 14.0 & 2015 Jun 04 & LP600 & 6.2 & 1.22$\pm$0.06 & 279$\pm$2 & 1.43$\pm$0.06 & K2 & & &  & 1 \\
6415 & 14.0 & 2015 Jun 03 & LP600 & 5.9 & 1.75$\pm$0.06 & 48$\pm$2 & 1.17$\pm$0.04 & K5 & & UKIRT &  & 1 \\
6475 & 13.7 & 2015 Jun 07 & LP600 & 14 & 1.31$\pm$0.06 & 57$\pm$2 & 0.50$\pm$0.02 & M2 & & &  & 1 \\
6482 & 13.6 & 2015 Jun 04 & LP600 & 5.8 & 0.52$\pm$0.06 & 271$\pm$2 & 0.58$\pm$0.07 & G7 & & & 0.01 & 1 \\
6527 & 12.3 & 2015 Jun 07 & LP600 & 224 & 2.21$\pm$0.06 & 353$\pm$2 & 1.60$\pm$0.02 & G7 & CFOP & & 0.01 & 1 \\
6560 & 12.9 & 2015 Jun 06 & LP600 & 28 & 2.20$\pm$0.06 & 30$\pm$2 & 5.38$\pm$0.07 & M5 & & & 0.01 & 1 \\
7205 & 14.1 & 2015 Jun 04 & LP600 & 6.0 & 1.04$\pm$0.06 & 42$\pm$2 & 0.44$\pm$0.03 & G8 & & &  & 1 \\
7448 & 11.3 & 2015 Jun 12 & LP600 & 12 & 0.87$\pm$0.06 & 260$\pm$2 & 1.40$\pm$0.09 & G0 & & & 0.01 & 1 \\
\hline
\end{tabular}
\small
\label{tab:secure25}
\begin{flushleft}
Notes. --- References for previous detections are denoted using the following codes: \citealt{dressing14} (D14), \citealt{lillo14} (LB14), \citealt{kraus16} (K16), \citealt{wang15a} (W15), visible in United Kingdom InfraRed Telescope images (UKIRT), high angular resolution images available on \textit{Kepler} Community FollowUp Observing Program (CFOP).
\end{flushleft}
\end{table*}
\end{center}
\unskip
\begin{center}
\begin{table*}
\renewcommand{\arraystretch}{.6}
\setlength{\tabcolsep}{2pt}
\caption{Secure Detections of Objects outside 2\farcs5 and within 4\farcs0 of \textit{Kepler} Planet Candidates}
\footnotesize
\centering
\begin{tabular}{l c c c c c c c c c c c c}
\hline
\hline
\noalign{\vskip 3pt}
KOI & $\rm m_{\textit{i}\textsuperscript{$\prime$}}$ & ObsID & Filter & Det. Significance & Separation & P.A. & Mag. Diff. & Approx. Comp. & Prev. High Res. & Prev. Low Res. & False Positive?\footnote{Probability that planetary transit signal is a false positive based on \textit{Kepler} data.} & $N_{KOI}$\footnote{Number of planet candidates detected orbiting KOI. } \\
& (mag) & & & $\sigma$ & ($\arcsec$) & (deg.) & (mag) & Spectral Type\footnote{Estimation method described in Section$~\ref{sec:spectraltypes}$.} &  Detection? & Detection? & & \\
\hline
\noalign{\vskip 3pt}
255 & 14.5 & 2014 Jul 17 & LP600 & 5.8 & 3.41$\pm$0.06 & 357$\pm$2 & 2.14$\pm$0.04 & M4 & K16 & UKIRT &  & 2 \\
734 & 15.1 & 2014 Sep 02 & LP600 & 6.3 & 3.51$\pm$0.06 & 175$\pm$2 & 2.05$\pm$0.04 & K7 & & UKIRT &  & 2 \\
1558 & 15.0 & 2014 Jul 11 & LP600 & 6.2 & 3.61$\pm$0.06 & 308$\pm$2 & 1.09$\pm$0.04 & K0 & & J19401085+4658310 & 0.01 & 1 \\
1593 & 15.6 & 2014 Aug 24 & LP600 & 7.6 & 3.24$\pm$0.06 & 80$\pm$2 & 1.60$\pm$0.03 & K4 & & UKIRT &  & 2 \\
1846 & 15.5 & 2014 Sep 02 & LP600 & 5.2 & 3.77$\pm$0.06 & 136$\pm$2 & 1.07$\pm$0.03 & K7 & & J19192894+4643440 &  & 1 \\
2213 & 15.1 & 2014 Aug 24 & LP600 & 7.1 & 3.94$\pm$0.06 & 91$\pm$2 & 1.67$\pm$0.02 & M0 & & J19411432+4302399&  & 1 \\
2744 & 14.9 & 2014 Jul 17 & LP600 & 5.7 & 3.50$\pm$0.06 & 257$\pm$2 & 2.12$\pm$0.03 & M0 & & UKIRT &  & 2 \\
3791 & 13.6 & 2014 Aug 22 & \textit{i}\textsuperscript{$\prime$} & 7.9 & 3.50$\pm$0.06 & 258$\pm$2 & 1.89$\pm$0.04 & K3 & & UKIRT &  & 2 \\
3928 & 13.1 & 2014 Jul 14 & LP600 & 21 & 2.96$\pm$0.06 & 265$\pm$2 & 1.21$\pm$0.03 & G6 & & UKIRT &  & 1 \\
4343 & 13.5 & 2014 Jun 19 & LP600 & 6.1 & 3.68$\pm$0.06 & 350$\pm$2 & 4.81$\pm$0.15 & M3 & & UKIRT &  & 1 \\
4630 & 14.7 & 2014 Jul 17 & LP600 & 6.6 & 3.94$\pm$0.06 & 53$\pm$2 & 2.17$\pm$0.05 & M0 & & J19422364+4335492&  & 1 \\
4743 & 14.7 & 2014 Sep 03 & LP600 & 7.9 & 3.06$\pm$0.06 & 98$\pm$2 & 2.29$\pm$0.04 & M0 & & UKIRT & 0.01 & 1 \\
4993 & 12.5 & 2014 Sep 01 & LP600 & 8.8 & 3.49$\pm$0.06 & 148$\pm$2 & 4.13$\pm$0.02 & M2 & & UKIRT &  & 1 \\
5220 & 11.8 & 2014 Sep 03 & LP600 & 29 & 2.89$\pm$0.06 & 216$\pm$2 & 3.27$\pm$0.05 & M3 & & UKIRT &  & 1 \\
5327 & 15.0 & 2014 Sep 01 & LP600 & 31 & 3.96$\pm$0.06 & 342$\pm$2 & --0.12$\pm$0.03 & M1 & & J19261347+4212546&  & 1 \\
5332 & 14.3 & 2015 Jun 12 & LP600 & 15 & 3.61$\pm$0.06 & 129$\pm$2 & 0.63$\pm$0.03 & G7 & & J19405741+4219181&  & 1 \\
5465 & 13.7 & 2014 Jun 19 & LP600 & 19 & 2.85$\pm$0.06 & 158$\pm$2 & 1.36$\pm$0.05 & K3 & & UKIRT &  & 1 \\
7020 & 13.5 & 2015 Jun 12 & LP600 & 14 & 3.28$\pm$0.06 & 23$\pm$2 & 1.43$\pm$0.04 & G9 & & UKIRT & 0.01 & 1 \\
7395 & 11.7 & 2015 Jun 12 & LP600 & 9.0 & 3.41$\pm$0.06 & 212$\pm$2 & 3.00$\pm$0.04 & G8 & CFOP & UKIRT & 0.01 & 1 \\
\hline
\end{tabular}
\small
\label{tab:secure40}
\begin{flushleft}
Notes. --- References for previous detections are denoted using the following codes: \citealt{kraus16} (K16),  visible in United Kingdom InfraRed Telescope images (UKIRT), high angular resolution images available on \textit{Kepler} Community FollowUp Observing Program (CFOP), companions visible in UKIRT and with 2MASS designations (J*).
\end{flushleft}
\end{table*}
\end{center}
\unskip
\begin{center}
\begin{table*}
\renewcommand{\arraystretch}{.8}
\setlength{\tabcolsep}{2pt}
\caption{Likely Detections of Objects within 2\farcs5 of \textit{Kepler} Planet Candidates}
\footnotesize
\centering
\begin{tabular}{l c c c c c c c c c c c c}
\hline
\hline
\noalign{\vskip 3pt}
KOI & $\rm m_{\textit{i}\textsuperscript{$\prime$}}$ & ObsID & Filter & Det. Significance & Separation & P.A. & Mag. Diff. & Approx. Comp. & Prev. High Res. & Prev. Low Res. & False Positive?\footnote{Probability that planetary transit signal is a false positive based on \textit{Kepler} data.} & $N_{KOI}$\footnote{Number of planet candidates detected orbiting KOI.} \\
& (mag) & & & $\sigma$ & ($\arcsec$) & (deg.) & (mag) & Spectral Type\footnote{Estimation method described in Section$~\ref{sec:spectraltypes}$.} &  Detection? & Detection? & & \\
\hline
\noalign{\vskip 3pt}
126 & 13.1 & 2015 Jun 08 & LP600 & 3.3 & 0.34$\pm$0.06 & 36$\pm$2 & 0.97$\pm$0.15 & K1 & & & 0.01,0.02 & 2 \\
200 & 14.2 & 2014 Sep 01 & LP600 & 3.0 & 0.30$\pm$0.06 & 44$\pm$2 & 0.52$\pm$0.23 & G7 & & &  & 1 \\
225 & 14.6 & 2014 Jul 16 & LP600 & 3.5 & 0.53$\pm$0.06 & 338$\pm$2 & 0.93$\pm$0.15 & G8 & & & 0.01 & 1 \\
532 & 14.5 & 2014 Jul 17 & LP600 & 2.8 & 0.97$\pm$0.06 & 232$\pm$2 & 3.44$\pm$0.25 & M1 && &  & 1 \\
841 & 15.8 & 2012 Sep 02 & LP600 & 2.6 & 2.00$\pm$0.06 & 69$\pm$2 & 3.60$\pm$0.04 & M3 & LB12 & &  & 5 \\
1261 & 14.9 & 2014 Aug 22 & \textit{i}\textsuperscript{$\prime$} & 2.9 & 1.83$\pm$0.06 & 340$\pm$2 & 1.58$\pm$0.05 & K3 & & UKIRT &  & 2 \\
1503 & 14.6 & 2014 Aug 22 & \textit{i}\textsuperscript{$\prime$} & 3.1 & 0.77$\pm$0.06 & 107$\pm$2 & 1.52$\pm$0.16 & M0 & & &  & 1 \\
1506 & 14.8 & 2014 Sep 02 & LP600 & 2.8 & 1.15$\pm$0.06 & 14$\pm$2 & 3.14$\pm$0.16 & M1 & & &  & 1 \\
1656 & 14.8 & 2014 Jun 13 & LP600 & 4.5 & 1.06$\pm$0.06 & 189$\pm$2 & 1.65$\pm$0.09 & K4 & & &  & 1 \\
1660 & 15.4 & 2014 Aug 28 & LP600 & 2.7 & 1.40$\pm$0.06 & 23$\pm$2 & 2.00$\pm$0.08 & K7 & & &  & 1 \\
1695 & 13.6 & 2014 Aug 31 & LP600 & 2.8 & 0.31$\pm$0.06 & 215$\pm$2 & 0.61$\pm$0.26 & G7 & & &  & 1 \\
1792 & 11.9 & 2014 Sep 02 & LP600 & 3.7 & 0.53$\pm$0.06 & 284$\pm$2 & 1.06$\pm$0.16 & K4 & CFOP & & 0.02 & 3 \\
1908 & 14.2 & 2014 Aug 22 & \textit{i}\textsuperscript{$\prime$} & 4.2 & 1.29$\pm$0.06 & 260$\pm$2 & 4.11$\pm$0.13 & M5 & K16 &&  & 2 \\
1973 & 15.3 & 2014 Aug 28 & LP600 & 3.7 & 0.79$\pm$0.06 & 31$\pm$2 & 1.69$\pm$0.19 & M2 & & &  & 1 \\
2048 & 15.5 & 2014 Aug 28 & LP600 & 2.9 & 1.84$\pm$0.06 & 353$\pm$2 & 3.33$\pm$0.17 & M4 & & UKIRT & 0.02 & 2 \\
2117 & 15.2 & 2014 Nov 09 & LP600 & 4.9 & 0.33$\pm$0.03\footnote{From Keck follow-up, described in Section$~\ref{sec:Discoveries}$.} & 111$\pm$2\textsuperscript{d} & 0.71$\pm$0.17 & M0 & & &  & 1 \\
2283 & 14.7 & 2014 Sep 01 & LP600 & 3.5 & 1.05$\pm$0.06 & 21$\pm$2 & 1.46$\pm$0.10 & M2 & & & 0.01 & 1 \\
2376 & 15.0 & 2014 Aug 21 & LP600 & 3.1 & 0.40$\pm$0.06 & 213$\pm$2 & 0.46$\pm$0.12 & K4 & & &  & 1 \\
2445 & 15.6 & 2014 Aug 28 & LP600 & 4.9 & 2.10$\pm$0.06 & 25$\pm$2 & 3.21$\pm$0.03 & M2 & & UKIRT &  & 1 \\
2460 & 14.6 & 2014 Aug 29 & LP600 & 3.3 & 2.36$\pm$0.06 & 192$\pm$2 & 3.41$\pm$0.02 & M4 & & UKIRT &  & 1 \\
2482 & 14.8 & 2014 Aug 24 & LP600 & 3.2 & 0.31$\pm$0.06 & 212$\pm$2 & 0.59$\pm$0.28 & G9 & & &  & 1 \\
2580 & 15.5 & 2014 Aug 31 & LP600 & 4.2 & 0.60$\pm$0.06 & 154$\pm$2 & 0.86$\pm$0.13 & K4 & & &  & 1 \\
2688 & 16.1 & 2014 Aug 31 & LP600 & 3.6 & 1.09$\pm$0.06 & 205$\pm$2 & 0.86$\pm$0.04 & M0 & & &  & 1 \\
2760 & 14.5 & 2014 Aug 23 & LP600 & 3.7 & 0.45$\pm$0.06 & 142$\pm$2 & 0.84$\pm$0.16 & M0 & & &  & 1 \\
2797 & 15.6 & 2014 Aug 28 & LP600 & 2.6 & 0.35$\pm$0.06 & 222$\pm$2 & 0.72$\pm$0.25 & G7 & & &  & 1 \\
2851 & 15.2 & 2014 Aug 26 & LP600 & 3.1 & 0.39$\pm$0.06 & 223$\pm$2 & 0.45$\pm$0.08 & K2 & & &  & 2 \\
2856 & 15.1 & 2014 Aug 26 & LP600 & 3.6 & 2.31$\pm$0.06 & 287$\pm$2 & 3.44$\pm$0.03 & M1 & & UKIRT &  & 1 \\
2862 & 15.3 & 2014 Aug 27 & LP600 & 3.0 & 0.68$\pm$0.06 & 20$\pm$2 & 0.17$\pm$0.05 & M2 & & &  & 1 \\
2926 & 15.7 & 2014 Aug 28 & LP600 & 3.4 & 0.33$\pm$0.06 & 16$\pm$2 & 0.27$\pm$0.09 & M1 & & &  & 4 \\
2927 & 15.7 & 2014 Aug 28 & LP600 & 3.5 & 1.39$\pm$0.06 & 36$\pm$2 & 2.65$\pm$0.04 & M0 & & &  & 1 \\
2958 & 14.6 & 2014 Sep 02 & LP600 & 2.3 & 1.15$\pm$0.06 & 302$\pm$2 & 2.47$\pm$0.14 & K7 & & &  & 1 \\
3043 & 14.6 & 2014 Jul 12 & LP600 & 3.2 & 1.14$\pm$0.06 & 68$\pm$2 & 1.94$\pm$0.07 & K5 & & &  & 2 \\
3106 & 15.2 & 2014 Aug 26 & LP600 & 3.0 & 0.30$\pm$0.03\textsuperscript{d} & 189$\pm$2\textsuperscript{d} & 0.76$\pm$0.16 & G9 & & &  & 1 \\
3136 & 15.4 & 2014 Aug 28 & LP600 & 4.5 & 1.83$\pm$0.06 & 238$\pm$2 & 2.91$\pm$0.04 & M3 & & &  & 1 \\
3214 & 11.8 & 2014 Aug 29 & LP600 & 3.2 & 0.49$\pm$0.06 & 320$\pm$2 & 0.73$\pm$0.13 & G8 & CFOP & &  & 2 \\
3263 & 15.3 & 2014 Aug 23 & LP600 & 3.0 & 0.80$\pm$0.06 & 276$\pm$2 & 2.01$\pm$0.16 & M5 & LB14 & & 0.01 & 1 \\
3335 & 15.6 & 2014 Sep 01 & LP600 & 3.3 & 2.40$\pm$0.06 & 61$\pm$2 & 2.89$\pm$0.04 & M0 & & UKIRT &  & 1 \\
3372 & 15.2 & 2014 Aug 23 & LP600 & 4.3 & 2.36$\pm$0.06 & 127$\pm$2 & 1.95$\pm$0.02 & K5 & & UKIRT &  & 1 \\
3418 & 15.2 & 2014 Aug 23 & LP600 & 3.9 & 1.13$\pm$0.06 & 43$\pm$2 & 1.29$\pm$0.10 & K2 & & &  & 1 \\
3432 & 14.8 & 2014 Jul 16 & LP600 & 2.8 & 0.66$\pm$0.06 & 113$\pm$2 & 1.37$\pm$0.17 & M0 & & &  & 1 \\
3471 & 13.0 & 2014 Jul 11 & LP600 & 2.9 & 0.63$\pm$0.06 & 224$\pm$2 & 3.05$\pm$0.12 & M3 & CFOP & & 0.01 & 1 \\
3480 & 15.7 & 2014 Sep 03 & LP600 & 3.6 & 0.40$\pm$0.06 & 210$\pm$2 & 0.75$\pm$0.20 & K4 & & &  & 1 \\
3611 & 16.3 & 2014 Aug 26 & LP600 & 3.1 & 2.30$\pm$0.06 & 267$\pm$2 & 2.77$\pm$0.05 & M0 & & &  & 1 \\
3626 & 16.2 & 2014 Sep 03 & LP600 & 4.3 & 1.96$\pm$0.06 & 310$\pm$2 & 3.82$\pm$0.14 & M2 & & UKIRT & 0.01 & 1 \\
3783 & 12.8 & 2014 Aug 21 & LP600 & 4.9 & 1.13$\pm$0.06 & 272$\pm$2 & 3.53$\pm$0.15 & K5 & CFOP && 0.01 & 1 \\
4062 & 13.9 & 2014 Aug 29 & LP600 & 3.4 & 1.49$\pm$0.06 & 28$\pm$2 & 3.66$\pm$0.13 & M0 & & & 0.01 & 1 \\
4267 & 15.0 & 2014 Jun 19 & LP600 & 3.9 & 1.66$\pm$0.06 & 194$\pm$2 & 3.29$\pm$0.08 & M0 & & &  & 1 \\
4323 & 13.4 & 2014 Jun 13 & LP600 & 3.8 & 1.12$\pm$0.06 & 96$\pm$2 & 2.22$\pm$0.10 & K5 & & & 0.02 & 2 \\
4366 & 15.3 & 2014 Aug 28 & LP600 & 2.7 & 2.46$\pm$0.06 & 303$\pm$2 & 3.38$\pm$0.02 & M3 & & &  & 1 \\
4421 & 12.6 & 2014 Jul 12 & LP600 & 3.4 & 2.45$\pm$0.06 & 322$\pm$2 & 4.62$\pm$0.02 & M3 & & UKIRT &  & 2 \\
4549 & 15.7 & 2014 Aug 27 & LP600 & 3.0 & 0.75$\pm$0.06 & 149$\pm$2 & 1.99$\pm$0.14 & K7 & & &  & 1 \\
4590 & 15.5 & 2014 Sep 02 & LP600 & 4.8 & 0.87$\pm$0.06 & 340$\pm$2 & 0.38$\pm$0.03 & K3 & & &  & 1 \\
4653 & 13.4 & 2014 Jul 19 & LP600 & 2.7 & 0.77$\pm$0.06 & 324$\pm$2 & 2.02$\pm$0.24 & K3 & & &  & 1 \\
4759 & 14.8 & 2014 Jul 19 & LP600 & 2.6 & 0.67$\pm$0.06 & 4$\pm$2 & 2.12$\pm$0.28 & K7 & & &  & 1 \\
4810 & 15.0 & 2014 Aug 24 & LP600 & 3.0 & 2.36$\pm$0.06 & 146$\pm$2 & 3.16$\pm$0.03 & M1 & & UKIRT &  & 1 \\
4923 & 13.0 & 2014 Jul 14 & LP600 & 4.7 & 0.78$\pm$0.06 & 123$\pm$2 & 1.46$\pm$0.10 & K2 & & &  & 1 \\
4974 & 15.5 & 2014 Aug 26 & LP600 & 2.9 & 1.23$\pm$0.06 & 242$\pm$2 & 3.33$\pm$0.13 & M2 & & &  & 1 \\
5101 & 12.9 & 2014 Jul 17 & LP600 & 3.1 & 1.24$\pm$0.06 & 99$\pm$2 & 3.33$\pm$0.19 & M0 & & &  & 1 \\
5143 & 15.7 & 2014 Nov 09 & LP600 & 3.8 & 1.22$\pm$0.06 & 222$\pm$2 & 3.83$\pm$0.18 & M3 & & &  & 1 \\
5232 & 13.5 & 2014 Aug 31 & LP600 & 4.0 & 1.75$\pm$0.06 & 200$\pm$2 & 4.67$\pm$0.19 & M2 & & &  & 1 \\
5327 & 15.0 & 2014 Sep 01 & LP600 & 4.8 & 1.88$\pm$0.06 & 211$\pm$2 & 3.43$\pm$0.05 & M5 & & &  & 1 \\
5332 & 14.3 & 2015 Jun 12 & LP600 & 4.1 & 2.19$\pm$0.06 & 7$\pm$2 & 2.37$\pm$0.04 & K7 & & UKIRT &  & 1 \\
5340 & 15.0 & 2014 Jun 19 & LP600 & 3.0 & 1.24$\pm$0.06 & 217$\pm$2 & 2.66$\pm$0.17 & M0 & & &  & 1 \\
5373 & 11.5 & 2015 Jun 05 & LP600 & 3.1 & 0.21$\pm$0.06 & 81$\pm$2 & 0.12$\pm$0.03 & K3 & & &  & 1 \\
5440 & 15.1 & 2014 Aug 28 & LP600 & 3.2 & 2.45$\pm$0.06 & 345$\pm$2 & 3.04$\pm$0.02 & M0 & & UKIRT &  & 1 \\
5482 & 15.0 & 2014 Aug 31 & LP600 & 3.6 & 0.62$\pm$0.06 & 270$\pm$2 & 1.44$\pm$0.18 & K2 & & & 0.01 & 1 \\
5486 & 12.6 & 2015 Jun 12 & LP600 & 2.9 & 0.34$\pm$0.06 & 333$\pm$2 & 0.73$\pm$0.28 & F6 && &  & 1 \\
5553 & 15.3 & 2014 Aug 23 & LP600 & 3.3 & 0.97$\pm$0.06 & 346$\pm$2 & 2.52$\pm$0.18 & M1 & & &  & 1 \\
5695 & 14.9 & 2015 Jun 12 & LP600 & 3.0 & 0.60$\pm$0.06 & 163$\pm$2 & 1.47$\pm$0.20 & K2 & & & 0.01 & 1 \\
5762 & 15.4 & 2014 Sep 03 & LP600 & 3.7 & 0.23$\pm$0.03\textsuperscript{d} & 95$\pm$2\textsuperscript{d} & 0.65$\pm$0.25 & K3 & & &  & 1 \\
6109 & 11.9 & 2015 Jun 07 & LP600 & 4.6 & 0.60$\pm$0.06 & 322$\pm$2 & 1.30$\pm$0.17 & G6 & CFOP  & & 0.01,0.02 & 2 \\
6202 & 11.4 & 2014 Aug 23 & \textit{i}\textsuperscript{$\prime$} & 2.9 & 0.77$\pm$0.06 & 322$\pm$2 & 2.49$\pm$0.27 & M2 & & &  & 1 \\
6311 & 9.0 & 2015 Jun 04 & LP600 & 3.1 & 1.75$\pm$0.06 & 290$\pm$2 & 0.83$\pm$0.10 & F3 & & & 0.01 & 1 \\
6464 & 13.7 & 2015 Jun 04 & LP600 & 3.9 & 0.75$\pm$0.06 & 122$\pm$2 & 1.72$\pm$0.15 & K0 & && 0.01,0.02,0.03 & 3 \\
6483 & 12.5 & 2015 Jun 05 & LP600 & 3.2 & 1.41$\pm$0.06 & 272$\pm$2 & 2.78$\pm$0.14 & M0 & && 0.01 & 1 \\
6539 & 12.5 & 2015 Jun 12 & LP600 & 3.1 & 1.58$\pm$0.06 & 175$\pm$2 & 3.89$\pm$0.17 & K7 & && 0.01 & 1 \\
6602 & 10.2 & 2015 Jun 03 & LP600 & 4.8 & 0.77$\pm$0.06 & 322$\pm$2 & 0.54$\pm$0.05 & K4 & & & 0.01 & 1 \\
6610 & 15.3 & 2015 Jun 12 & LP600 & 4.0 & 1.73$\pm$0.06 & 84$\pm$2 & 2.68$\pm$0.04 & K3 & & & & 1 \\
6654 & 13.5 & 2015 Jun 12 & LP600 & 3.9 & 1.41$\pm$0.06 & 195$\pm$2 & 2.88$\pm$0.08 & K7 & & & 0.01 & 1 \\
6706 & 13.8 & 2015 Jun 04 & LP600 & 3.0 & 1.04$\pm$0.06 & 339$\pm$2 & 1.44$\pm$0.10 & G6 & & & 0.01 & 1 \\
6728 & 13.9 & 2015 Jun 12 & LP600 & 3.3 & 1.94$\pm$0.06 & 134$\pm$2 & 5.04$\pm$0.11 & M4 & & &  & 1 \\
7426 & 15.4 & 2015 Jun 05 & LP600 & 3.2 & 2.45$\pm$0.06 & 212$\pm$2 & 2.37$\pm$0.02 & M2 & & UKIRT &  & 1 \\
\hline
\end{tabular}
\small
\label{tab:likely25}
\begin{flushleft}
Notes. --- References for previous detections are denoted using the following codes: \citealt{lillo12} (LB12), \citealt{lillo14} (LB14), \citealt{kraus16} (K16), visible in United Kingdom InfraRed Telescope images (UKIRT).
\end{flushleft}
\end{table*}
\end{center}

\begin{center}
\begin{table*}
\renewcommand{\arraystretch}{.6}
\setlength{\tabcolsep}{2pt}
\caption{Likely Detections of Objects outside 2\farcs5 and within 4\farcs0 of \textit{Kepler} Planet Candidates}
\footnotesize
\centering
\begin{tabular}{l c c c c c c c c c c c c}
\hline
\hline
\noalign{\vskip 3pt}
KOI & $\rm m_{\textit{i}\textsuperscript{$\prime$}}$ & ObsID & Filter & Det. Significance & Separation & P.A. & Mag. Diff. & Approx. Comp. & Prev. High Res. & Prev. Low Res. & False Positive?\footnote{Probability that planetary transit signal is a false positive based on \textit{Kepler} data.} & $N_{KOI}$\footnote{Number of planet candidates detected orbiting KOI. } \\
& (mag) & & & $\sigma$ & ($\arcsec$) & (deg.) & (mag) & Spectral Type\footnote{Estimation method described in Section$~\ref{sec:spectraltypes}$.} &  Detection? & Detection? & & \\
\hline
\noalign{\vskip 3pt}
51 & 13.4 & 2013 Jul 25 & LP600 & 2.6 & 3.51$\pm$0.06 & 161$\pm$2 & 2.63$\pm$0.07 & M0 & & & & 1 \\
193 & 14.7 & 2014 Aug 21 & LP600 & 4.8 & 2.78$\pm$0.06 & 137$\pm$2 & 3.07$\pm$0.02 & M0 & & UKIRT & 0.01 & 1 \\
200 & 14.2 & 2014 Sep 01 & LP600 & 3.8 & 2.81$\pm$0.06 & 130$\pm$2 & 4.00$\pm$0.02 & M2 & & UKIRT &  & 1 \\
240 & 14.8 & 2014 Aug 21 & LP600 & 4.2 & 2.71$\pm$0.06 & 272$\pm$2 & 3.46$\pm$0.03 & M1 & & UKIRT &  & 1 \\
326 & 13.0 & 2013 Aug 15 & LP600 & 4.9 & 3.53$\pm$0.06 & 267$\pm$2 & 2.01$\pm$0.02 & M2 & LB12 & &  & 2 \\
541 & 14.5 & 2014 Aug 29 & LP600 & 4.5 & 2.80$\pm$0.06 & 246$\pm$2 & 3.50$\pm$0.02 & M3 & & UKIRT &  & 1 \\
598 & 14.5 & 2014 Jul 11 & LP600 & 4.3 & 3.17$\pm$0.06 & 357$\pm$2 & 2.73$\pm$0.04 & M2 & & UKIRT & 0.02 & 2 \\
757 & 15.5 & 2014 Sep 03 & LP600 & 3.7 & 2.94$\pm$0.06 & 243$\pm$2 & 3.37$\pm$0.04 & M3 & & J19244737+4718244 & 0.02 & 3 \\
814 & 15.3 & 2014 Aug 24 & LP600 & 2.7 & 3.40$\pm$0.06 & 346$\pm$2 & 4.16$\pm$0.07 & M4 & & UKIRT &  & 1 \\
816 & 15.4 & 2014 Aug 24 & LP600 & 3.6 & 3.50$\pm$0.06 & 120$\pm$2 & 2.66$\pm$0.03 & M0 & & UKIRT &  & 1 \\
1193 & 15.0 & 2014 Aug 26 & LP600 & 3.5 & 3.08$\pm$0.06 & 7$\pm$2 & 2.81$\pm$0.02 & M1 & & UKIRT & 0.01 & 1 \\
1201 & 14.9 & 2012 Aug 04 & LP600 & 2.9 & 2.81$\pm$0.06 & 236$\pm$2 & 4.26$\pm$0.08 & M6 & K16 & UKIRT &  & 1 \\
1201 & 14.9 & 2012 Aug 04 & LP600 & 2.6 & 3.76$\pm$0.06 & 265$\pm$2 & 5.17$\pm$0.12 & M7 & K16 & UKIRT &  & 1 \\
1441 & 14.9 & 2014 Aug 31 & LP600 & 4.0 & 3.06$\pm$0.06 & 333$\pm$2 & 3.73$\pm$0.03 & M2 & & UKIRT &  & 1 \\
1804 & 15.3 & 2014 Sep 02 & LP600 & 3.2 & 2.88$\pm$0.06 & 168$\pm$2 & 2.84$\pm$0.03 & M3 & & UKIRT & 0.01 & 1 \\
1995 & 15.0 & 2014 Aug 24 & LP600 & 3.1 & 2.96$\pm$0.06 & 355$\pm$2 & 5.34$\pm$0.04 & M4 & & UKIRT & 0.01 & 1 \\
2050 & 12.2 & 2015 Jun 07 & LP600 & 4.6 & 3.33$\pm$0.06 & 215$\pm$2 & 5.33$\pm$0.04 & M5 & CFOP & & 0.01,0.02 & 2 \\
2206 & 15.0 & 2014 Jul 19 & LP600 & 4.6 & 3.28$\pm$0.06 & 87$\pm$2 & 1.28$\pm$0.05 & K4 & & UKIRT &  & 1 \\
2379 & 14.9 & 2014 Aug 29 & LP600 & 4.8 & 3.59$\pm$0.06 & 139$\pm$2 & 1.89$\pm$0.03 & K3 & & UKIRT &  0.01 & 1 \\
2579 & 15.0 & 2014 Jul 12 & LP600 & 2.7 & 3.48$\pm$0.06 & 355$\pm$2 & 3.69$\pm$0.03 & M2 & & UKIRT &  & 3 \\
3066 & 15.6 & 2014 Aug 24 & LP600 & 4.3 & 3.41$\pm$0.06 & 335$\pm$2 & 1.86$\pm$0.02 & M0 & & UKIRT &  & 1 \\
3111 & 12.7 & 2014 Aug 20 & \textit{i}\textsuperscript{$\prime$} & 3.8 & 3.36$\pm$0.06 & 234$\pm$2 & 5.87$\pm$0.13 & M5 & D14 & &  & 2 \\
3161 & 9.6 & 2015 Jun 03 & LP600 & 2.7 & 2.68$\pm$0.06 & 67$\pm$2 & 3.04$\pm$0.14 & K5 & CFOP & & 0.01 & 1 \\
3264 & 15.6 & 2014 Aug 28 & LP600 & 3.1 & 3.66$\pm$0.06 & 217$\pm$2 & 1.37$\pm$0.02 & M0 & & UKIRT &  & 1 \\
3341 & 14.7 & 2014 Jul 17 & LP600 & 3.2 & 3.23$\pm$0.06 & 107$\pm$2 & 4.27$\pm$0.08 & M3 & & UKIRT &  & 2 \\
3347 & 15.2 & 2014 Aug 28 & LP600 & 4.2 & 3.24$\pm$0.06 & 295$\pm$2 & 2.20$\pm$0.02 & M0 & & UKIRT &  & 1 \\
3354 & 14.9 & 2014 Jul 16 & LP600 & 4.2 & 3.71$\pm$0.06 & 227$\pm$2 & 2.55$\pm$0.06 & M0 & & UKIRT &  & 1 \\
3463 & 14.6 & 2015 Jun 07 & LP600 & 4.3 & 3.67$\pm$0.06 & 96$\pm$2 & 4.41$\pm$0.04 & M3 & & UKIRT &  & 1 \\
3463 & 14.6 & 2015 Jun 07 & LP600 & 3.2 & 2.74$\pm$0.06 & 79$\pm$2 & 4.79$\pm$0.02 & M4 & & &  & 1 \\
3533 & 14.4 & 2014 Nov 09 & LP600 & 4.1 & 3.08$\pm$0.06 & 10$\pm$2 & 5.21$\pm$0.03 & M3 & & UKIRT & 0.01 & 1 \\
3678 & 12.6 & 2014 Jun 17 & LP600 & 2.8 & 2.63$\pm$0.06 & 170$\pm$2 & 5.08$\pm$0.04 & M5 & W15 & UKIRT &  & 1 \\
4131 & 13.2 & 2014 Jun 19 & LP600 & 4.7 & 2.85$\pm$0.06 & 124$\pm$2 & 5.04$\pm$0.02 & K7 & & UKIRT & 0.01 & 2 \\
4268 & 14.8 & 2014 Aug 31 & LP600 & 3.0 & 3.56$\pm$0.06 & 263$\pm$2 & 4.77$\pm$0.04 & M6 & & UKIRT &  & 1 \\
4334 & 15.5 & 2014 Sep 01 & LP600 & 2.6 & 3.32$\pm$0.06 & 15$\pm$2 & 3.79$\pm$0.06 & M4 & & UKIRT &  & 1 \\
4345 & 13.2 & 2014 Jul 13 & LP600 & 3.9 & 3.17$\pm$0.06 & 242$\pm$2 & 3.22$\pm$0.02 & M2 & & UKIRT & 0.01 & 1 \\
4353 & 15.4 & 2014 Aug 24 & LP600 & 2.8 & 3.50$\pm$0.06 & 36$\pm$2 & 2.75$\pm$0.04 & M0 & & UKIRT & 0.01 & 1 \\
4405 & 14.5 & 2014 Jul 17 & LP600 & 3.4 & 2.95$\pm$0.06 & 249$\pm$2 & 3.19$\pm$0.02 & M0 & & UKIRT & 0.01 & 1 \\
4467 & 15.6 & 2014 Aug 26 & LP600 & 4.0 & 3.99$\pm$0.06 & 131$\pm$2 & 4.21$\pm$0.04 & M4 & & UKIRT &  & 1 \\
4526 & 15.1 & 2014 Aug 24 & LP600 & 4.4 & 2.53$\pm$0.06 & 346$\pm$2 & 4.44$\pm$0.02 & M3 & & UKIRT &  & 2 \\
4526 & 15.1 & 2014 Aug 24 & LP600 & 3.1 & 3.98$\pm$0.06 & 179$\pm$2 & 4.80$\pm$0.09 & M4 & & UKIRT &  & 2 \\
4655 & 15.2 & 2014 Aug 23 & LP600 & 2.9 & 3.17$\pm$0.06 & 116$\pm$2 & 3.02$\pm$0.05 & M0 & & UKIRT &  & 1 \\
4661 & 14.5 & 2014 Jul 18 & LP600 & 4.1 & 3.93$\pm$0.06 & 198$\pm$2 & 2.32$\pm$0.05 & M2 & & J19295122+4117529&  & 1 \\
4700 & 15.7 & 2014 Aug 31 & LP600 & 3.8 & 3.77$\pm$0.06 & 49$\pm$2 & 1.89$\pm$0.05 & M0 & & UKIRT &  & 1 \\
4710 & 15.4 & 2014 Sep 01 & LP600 & 3.5 & 2.70$\pm$0.06 & 168$\pm$2 & 3.50$\pm$0.05 & M2 & & UKIRT &  & 1 \\
4881 & 12.7 & 2014 Aug 21 & LP600 & 4.9 & 3.42$\pm$0.06 & 30$\pm$2 & 3.30$\pm$0.03 & M0 & & UKIRT & 0.01,0.02 & 2 \\
5210 & 14.9 & 2014 Jul 14 & LP600 & 3.8 & 2.71$\pm$0.06 & 267$\pm$2 & 2.22$\pm$0.05 & M1 & & UKIRT &  & 1 \\
5216 & 15.3 & 2014 Aug 31 & LP600 & 3.2 & 3.67$\pm$0.06 & 96$\pm$2 & 3.31$\pm$0.03 & M1 & & UKIRT &  & 1 \\
5220 & 11.8 & 2014 Sep 03 & LP600 & 3.0 & 2.83$\pm$0.06 & 109$\pm$2 & 7.22$\pm$0.08 & M7 & & UKIRT &  & 1 \\
5327 & 15.0 & 2014 Sep 01 & LP600 & 3.0 & 3.63$\pm$0.06 & 277$\pm$2 & 3.92$\pm$0.02 & M5 & & UKIRT &  & 1 \\
5331 & 14.9 & 2014 Aug 31 & LP600 & 3.0 & 3.67$\pm$0.06 & 351$\pm$2 & 3.72$\pm$0.03 & M4 & & UKIRT &  & 1 \\
5480 & 16.3 & 2014 Aug 29 & LP600 & 3.6 & 3.52$\pm$0.06 & 174$\pm$2 & 1.24$\pm$0.05 & K1 & & UKIRT &  & 1 \\
5556 & 13.2 & 2014 Jun 13 & LP600 & 3.6 & 3.28$\pm$0.06 & 162$\pm$2 & 4.31$\pm$0.03 & M4 & CFOP &&  & 1 \\
5556 & 13.2 & 2014 Jun 13 & LP600 & 3.4 & 3.22$\pm$0.06 & 247$\pm$2 & 5.29$\pm$0.02 & M5 & CFOP  & &  & 1 \\
5707 & 15.0 & 2014 Aug 23 & LP600 & 3.3 & 2.71$\pm$0.06 & 239$\pm$2 & 2.43$\pm$0.02 & K7 & & UKIRT &  & 1 \\
5885 & 14.7 & 2014 Aug 21 & LP600 & 2.8 & 3.42$\pm$0.06 & 127$\pm$2 & 4.03$\pm$0.04 & M3 & & UKIRT &  & 1 \\
6120 & 15.4 & 2015 Jun 08 & LP600 & 3.3 & 3.85$\pm$0.06 & 128$\pm$2 & 2.48$\pm$0.02 & M0 & & J19214830+3951405&  & 2 \\
6560 & 12.9 & 2015 Jun 06 & LP600 & 4.8 & 3.28$\pm$0.06 & 246$\pm$2 & 6.04$\pm$0.12 & M6 & & UKIRT & 0.01 & 1 \\
6605 & 11.3 & 2015 Jun 08 & LP600 & 4.2 & 2.53$\pm$0.06 & 320$\pm$2 & 3.46$\pm$0.04 & M0 & CFOP  & &  & 1 \\
6610 & 15.3 & 2015 Jun 12 & LP600 & 3.5 & 2.63$\pm$0.06 & 216$\pm$2 & 1.22$\pm$0.02 & G3 & & UKIRT &  & 1 \\
6745 & 15.2 & 2015 Jun 12 & LP600 & 2.7 & 3.07$\pm$0.06 & 72$\pm$2 & 3.78$\pm$0.02 & M0 & & UKIRT &  & 1 \\
6745 & 15.2 & 2015 Jun 12 & LP600 & 2.5 & 2.85$\pm$0.06 & 163$\pm$2 & 3.92$\pm$0.03 & M0 & & UKIRT &  & 1 \\
6800 & 12.8 & 2015 Jun 12 & LP600 & 2.7 & 2.62$\pm$0.06 & 145$\pm$2 & 5.10$\pm$0.04 & M4 & & UKIRT & 0.01 & 1 \\
6800 & 12.8 & 2015 Jun 12 & LP600 & 2.9 & 3.11$\pm$0.06 & 337$\pm$2 & 5.41$\pm$0.04 & M5 & & UKIRT & 0.01 & 1 \\
6925 & 15.7 & 2015 Jun 03 & LP600 & 3.1 & 2.66$\pm$0.06 & 125$\pm$2 & 1.71$\pm$0.12 & M4 & & UKIRT &  & 1 \\

\hline
\end{tabular}
\small
\label{tab:likely40}
\begin{flushleft}
Notes. --- References for previous detections are denoted using the following codes: \citealt{lillo12} (LB12), \citealt{dressing14} (D14), \citealt{kraus16} (K16), visible in United Kingdom InfraRed Telescope images (UKIRT), companions visible in UKIRT and with 2MASS designations (J*).
\end{flushleft}
\end{table*}
\end{center}

\subsection{Comparison to Other Surveys}
\label{sec:othersurveys}

Two detected companions (KOI-326 and KOI-841) in our survey were previously found in \citet{lillo12}, who observed 98 KOIs using the AstraLux Lucky Imaging system on the 2.2m telescope at the Calar Alto Observatory.  \citet{lillo14} also previously detected companions to KOI-3263, 3649, and 3886 in a survey of 174 KOIs.  \citet{adams12} and \citet{adams13} observed 87 and 13 KOIs, respectively, with the instruments ARIES and PHARO on the MMT and Palomar telescopes, respectively.  They detect companions to KOI-126 and 266 that are fainter than our survey sensitivity. Observing 87 KOIs with ARIES at the MMT, \citet{dressing14} previously detected companions to KOI-2813 and KOI-3111 and also detected a companion to KOI-266 ($\Delta$m$_{K_s}$=6.32) that is outside our detection sensitivity.  \citet{gilliland15} found two companions to KOI-829 using the \textit{Hubble Space Telescope} (HST) with $\Delta$m$_{K_p}$ of 2.4 and 6.0 and separations of 0$\farcs$11 and 3$\farcs$31, respectively, which were outside the detection limits of our Robo-AO image. \citet{wang15a} observed 84 KOIs using the PHARO and NIRC2 instruments at Palomar and Keck, respectively, with one discovered companion (KOI-3678) appearing in our survey.  Two of our targets (KOI-1411 and KOI-3823) have companions detected by \citeauthor{wang15a}, both with $\Delta$m$_{K}>$5, which fall outside our detection sensitivity.  \citet{wang15b} observed 73 multiple transiting planet KOI systems at Palomar and Keck, with the only overlapping system being a companion observed near KOI-1806 which we did not detect.  The companion to KOI-1806, measured by \citet{wang15b} as $\Delta$K=1.45 at 3$\farcs$43 separation, is well within out survey sensitivity, and the reason for the non-detection is unclear.  The reported companion is also not visible in UKIRT images, although it would be detectable.  We detected companions to KOI-126 and 200 not detected by \citet{howell11}; both companions are within the stated sensitivity limits for their respective targets, so the reason for the earlier non-detection is unclear.  None of our nearby-star detections overlap with the discoveries of \citet{everett15}.

\citet{kraus16} observed 382 KOIs with AO on the Keck-II telescope.  They detected single companions to KOI-255, 1908, 2705, and 2813, and both companions to KOI-1201 that were detected in our survey.  They also detected single companions to KOI-1298, 1681, 2179 2453, and 2862, and double companions to KOI-1361 and 2813, that all fall outside of our reported sensitivity.

\citet{kolbl15} searched for the blended spectra of KOIs with secondary stars within $\sim$0$\farcs$8 using Keck-HIRES optical echelle spectra of 1160 California \textit{Kepler} Survey KOIs. Of the 63 KOIs the authors found with evidence of a secondary star, we found companions to seven (KOIs 1137, 2813, 3161, 3415, 3471, 4345, 4713) and did not detect companions to eight (KOIs 1121, 1326, 1645, 3515, 3527, 3605, 3606, 3853). The companions we did not detect likely lie at small separations inside the limits of our survey sensitivity. Two of our companions (KOIs 1137 and 3415) fall within their calculated flux ratio uncertainty and within their $\sim$0$\farcs$8 separation limit.  Without known separations and position angles, however, it is not clear that these are the same companion stars.

Nine of the widest nearby stars we detected have 2MASS \citep{skrutskie06} designations.  102 of our wide ($\rho>$2$\arcsec$) nearby star detections are noted on the \textit{Kepler} Community Follow-up Observing Program (CFOP) using J-band, $\sim$1$\arcsec$ seeing-limited imaging from United Kingdom InfraRed Telescope (UKIRT) \citep{lawrence07}.  However, with high-acuity imaging to resolve blended companions, providing greater precision photometry, and a filter that better simulates the \textit{Kepler} bandpass, the Robo-AO survey can better evaluate the effect of the companion on the observed transit signal.

\subsection{Multiplicity and Other Surveys}

There have been multiple past high-resolution surveys of KOIs performed, allowing our results to be put into context with the overall community follow-up program.  A comparison of the observed nearby-star rates from various surveys with differing methodologies may also provide convergence on the intrinsic multiplicity rate of planet hosting stars. With varying sensitivities between surveys, we use a lower separation cutoff than in this paper for a uniform comparison between surveys. This also has the added benefit of using only the nearest stars that have the highest probability of association.  An exact comparison between surveys is still hindered, however, by the use of dissimilar instruments, passbands, and target selection criteria;  in comparing results in this section we attempt to highlight major differences when comparing multiplicity rates, however we caution that in each case there are inherent biases in the coverage of the different surveys which requires detailed analysis not covered in this work.

We find that 6.8$\%$ of KOIs have nearby stars within 2$\arcsec$, in agreement with other visible light surveys: 6.4$\%$ in Paper I, 8.2$\%$ in Paper II, and 6.4$\%$ \citep{howell11}. \citet{horch14} found 7.0$\%$ of KOI targets had nearby stars within 1$\arcsec$ separation, a range where we showed a 3.4$\%$ nearby star rate.  \citet{horch14} do not report their target list, so it is not possible to identify the source of this discrepancy. It is possible that this is a result of our target selection of every KOI, resulting in a dimmer overall sample than surveys which prioritize brighter targets.  The targets in this paper have a median \textit{K$_{p,med}$} = 14.9, significantly fainter overall than the targets in \citet[\textit{K$_{p,med}$} = 12.2]{adams12}, \citet[\textit{K$_{p,med}$} = 13.3]{dressing14}, \citet[\textit{K$_{p}$} $<$ 14]{wang15a}, Paper I (K$_{p,med}$ = 13.7), and in Paper II (K$_{p,med}$ = 14.2).  \citet{horch14} note that their \textit{Kepler} targets mainly are between 11th and 14th magnitude.  There are several reasons a brighter overall target list will inflate binarity rates: the target stars are intrinsically more luminous, which results in more physically associated companion stars as binarity correlates with luminosity \citep{duchene13}; the target stars are less distant, so physically associated companion of a given spectral type is brighter, thus easier to detect; brighter stars tend to have deeper detectable contrast ratios.

The disparity in multiplicity between papers in the Robo-AO survey was explored in Section 6 of Paper II as a possible result of the bias in the KOI selection process between data releases, with the median observed KOI in Paper II located nearer the Galactic plane than in Paper I.  KOIs near the Galactic plane lie in denser stellar fields, increasing the likelihood of unassociated nearby stars with the separation cutoff.  Plotting the \textit{Kepler} field of view with our targeted KOIs in Figure$~\ref{fig:fovplot}$, the median position of KOIs in this work is closer to the center of the field than in Paper II, and further from the Galactic plane than Paper I or Paper II.

Surveys in the NIR find higher multiplicity rates within 2$\arcsec$: 13$\%$ \citep{dressing14}, 17$\%$ \citep{adams12}, 20$\%$ \citep{adams13}. This is likely caused by many companions being cool, red dwarf stars that are faint in the optical \citep{ngo15}, and deeper, higher angular resolution imaging.

\section{Discussion}
\label{sec:Discussion}

In this section, we delve further into the combined datasets of Paper II and this work to further explore the implications of stellar multiplicity on the planetary candidates (Section \ref{sec:implications}), expand on the planetary candidates found in higher order multiple systems or orbiting within the habitable zone (Section \ref{sec:interestingsystems}), and search for insight into the role that multiple stellar bodies play on planetary formation and evolution (Section \ref{sec:multiplicitystudy}). 

\subsection{Implications for \textit{Kepler} Planet Candidates}
\label{sec:implications}

When a close companion is detected near a KOI host star, there are several potential implications.  If the planet does indeed transit the purported target star, the consequences may be relatively mild: the planet's radius will be slightly larger than had previously been thought---at most by a factor of $\sqrt{2}$ in the case of an equal-brightness companion \citep{ciardi15}.  If the eclipsed star is a faint companion, however, the radius of the eclipsing object may be many times larger, potentially turning a small planet into a giant planet or a planet into a false-positive eclipsing binary star.  Additionally, as the properties of most of the host stars in the \textit{Kepler} stellar catalog are based on broad-band photometry assuming that they are single, the derived stellar radii may well be incorrect if the system actually contains multiple stars.  Re-fitting the stellar properties of all the companion stars, as well as for the \textit{Kepler} target stars accounting for the presence of the companions, is beyond the scope of this work, but will be addressed in a future paper in this series.

Finally, if a KOI system has multiple transiting planets detected, it might be the case that the planets are distributed around multiple stars in the system.  KOI-284/Kepler-132 is a good example of such a case \citep{fabrycky12,lissauer14}: its multiple planets would be unstable if they all orbited a single star, but it turns out to be a close visual binary, with the only sensible interpretation being that some of the planets transit one star and some transit the other.  While such ``split multiple'' systems are predicted to be relatively rare among the population of \textit{Kepler} multiple stellar systems \citep{fabrycky12}, any multi-planet system with a close companion has a higher chance of being split, and thus deserves close consideration.  \citet{barclay15} presents a model of how such systems might be analyzed, investigating the KOI-1422/Kepler-296.

\subsection{Particularly Interesting Systems}
\label{sec:interestingsystems}

Several KOIs with detected companions are of particular interest, either for displaying unusual system characteristics, rare false-positive scenarios, or planetary attributes that satisfy habitability requirements.

\subsubsection{Possible Quadruple Systems}
\label{sec:quadsystems}

\textit{KOI-5327} hosts a 2.24 $R_{\oplus}$ planetary candidate on a 5.4 day orbit.  We detect two nearby stellar companions, with angular separations of 1$\farcs$88 and 3$\farcs$63 and magnitude differences of 3.43 and 3.92, respectively.  A possible fourth component of the system lies at 3$\farcs$96 and is 0.12 mags brighter than the KOI target. Further multiple passband observations and radial velocity measurements are needed to understand the hierarchy of this system. In the full 44$\arcsec$ square image, a total of 8 stars appear, including the target and possible companions.  With nearly equal brightness, the third companion has a high probability of being associated.  With few stars found in the full field, it is unlikely that any unassociated stars would be found within 4\arcsec of the KOI. The likelihood that the other two stars are in fact bound is 97$\%$.  If physical association is confirmed for all four components, KOI-5327 would be the third known planet residing in a quadruple star system \citep{schwamb13, roberts15}.  Following the analysis of Section 5.1 in Paper I, with the planet assumed to orbit the bright target star, the updated planetary radius estimate for the planetary candidate with all three stars in the aperture is 3.3 $R_{\oplus}$.  Without the 0.12 mags brighter star in the photometric aperture, the updated radius estimate is 2.3 $R_{\oplus}$.  The second scenario detailed in Paper I, with the planet orbiting one of the fainter companions, will be further explored in future papers in this survey for all detected companions.

\textit{KOI-4495}, first detected in Paper II, has three nearby stars, with angular separations of 3$\farcs$04, 3$\farcs$06, and 3$\farcs$41 and magnitude differences of 4.73, 3.90, and 2.68, respectively.  The system hosts a planetary candidate with period of 5.92 days and estimated radius of 1.49 $R_{\oplus}$.  The system lies in a relatively dense stellar field, thus it is probable that at least one of the stars is an unassociated asterism.  The Robo-AO discovery image is available in Figure 5 of Paper II.  With all three stars in the photometric aperture diluting the transit signal, and assuming the planet does indeed orbit the bright star, the updated planetary radius estimate is 1.6 $R_{\oplus}$.

\textit{KOI-3214} hosts planetary candidates with radii of 2.59 $R_{\oplus}$ and 2.02 $R_{\oplus}$ on 11.5 and 25.1 day orbits, respectively.  We detect two nearby stellar companions, with angular separations of 0$\farcs$49 and 1$\farcs$41 and magnitude differences of 0.73 and 2.50, respectively.  Outside our 4$\arcsec$ separation cutoff, another 5.33 mag dimmer star appears at and 4$\farcs$34.  With multiple stars in the same \textit{Kepler} pixel, the probability of an eclipsing binary resulting in a false planetary transit signal is increased.  KOI-3214 lies in a relatively sparse stellar field, with only 6 additional stars in the full 44$\arcsec$ square image, including the target and possible companions.  The probability based on the background star density that all three stars are bound is approximately 98$\%$.  Assuming the planets orbit the brightest star, the two close stars likely dilute the observed transit signal, leading to updated  planetary radii estimates of 3.3 $R_{\oplus}$ and 2.6 $R_{\oplus}$ for the planet candidates on orbits with periods of 11.5 and 25.1 days, respectively.

\textit{KOI-3463} hosts a 1.3 $R_{\oplus}$ planetary candidate on a 32.5 day orbit.  We detect two nearby stellar companions, with angular separations of 2$\farcs$74 and 3$\farcs$67 and magnitude differences of 4.79 and 4.41, respectively.  Just outside our 4$\arcsec$ separation cutoff, another 2.44 mag dimmer star appears at 4$\farcs$11. KOI-3463 lies in a relatively dense stellar field, with at least 16 stars in the full 44$\arcsec$ square image, including the target and possible companions.  The probability based on the background star density that all three stars are bound is approximately 86$\%$.  If the planet candidate orbits the bright star, the additional two nearby stars in the photometric aperture only marginally dilute the transit signal, leading to an updated planetary radius estimate of 1.3 $R_{\oplus}$.

\textit{KOI-6800} hosts a 27.5 $R_{\oplus}$ planetary candidate on a 2.5 day orbit.  We detect two nearby stellar companions, with angular separations of 2$\farcs$62 and 3$\farcs$11 and magnitude differences of 5.10 and 5.41, respectively.  Outside our 4$\arcsec$ separation cutoff, another 5.27 mag dimmer object appears at 4$\farcs$13, although UKIRT photometry suggests that this is highly likely ($>$99$\%$) a background galaxy.  In the full 44$\arcsec$ square image of KOI-6800, 9 stars are visible, including the target and possible companions.  The probability based on the background star density that all three stars are bound is approximately 97$\%$. The two dim nearby stars within the photometric aperture only slightly increases the estimated planetary radius to 27.7 $R_{\oplus}$, assuming the planet orbits the brightest star.  If the planet candidate orbits one of the fainter stars, the corrected planetary radius would be large enough to make it highly probable the transiting event is in fact a background eclipsing binary.

The Robo-AO images of the four possible quadruple systems from this work are displayed in Figure$~\ref{fig:multiplesgrid}$. 

\subsubsection{Habitable Zone Candidates}

The discovery of habitable exoplanets is a major goal of the \textit{Kepler} mission, and an accurate knowledge of the host star's properties is required to establish unambiguously whether an exoplanet possesses the two habitability conditions -- rocky and in a location where water can be found in a liquid state on the surface (habitable zone; HZ). The exact requirements for habitability are still debated \citep{kasting93, selis07, seager13, zsom13}, however it has been shown that the transition between ``rocky'' and ``non-rocky'' occurs rather sharply at R$_{P}$=1.6$R_{\oplus}$ \citep{rogers15}. For this analysis, we will use a large cutoff of 4 $R_{\oplus}$, as the presence of a stellar companion may dramatically alter the estimated radius, and even a gaseous planet in the HZ may host a rocky exomoon \citep{heller12}.  Overall, the existence of an unknown stellar companion within the same photometric aperture as the KOI will increase the calculated radius of the planet, as the observed transit signal will be diluted by the constant light of the nearby star.

In Paper II and this work, we detected companions to 26 KOIs which host planetary candidates with equilibrium temperatures, from the NEA, in the HZ range (273 K$\le$T$_{eq}\le$373 K) and R$<$4 $R_{\oplus}$, displayed in Table$~\ref{tab:hz_table}$.  All are newly detected in this survey.  Corrected planetary radii estimates are included, as described in Section 5.1 of Paper I, with the assumption that the planet orbits the bright star.

KOI-2926 hosts two planetary candidates within the HZ, and KOI-6745 is a possible triple system hosting a planet in the HZ.  The equilibrium temperature calculation is based on an estimate of the stellar effective temperature of the host star, thus if the planet orbits the dimmer companion it is unlikely to be in the HZ.

KOI-1503 hosts a planetary candidate with initial diluted radius estimate of 3.79 $R_{\oplus}$ in a 0.51 AU orbit.  If the planet orbits the primary star, the corrected radius of the planet is 4.23$\pm$0.07 $R_{\oplus}$, decreasing the probability that it is rocky. 

KOI-5101, a sun-like star, hosts a near-Earth analog, with a calculated radius 64\% larger than Earth and an orbit of 1.12 AU.  With a 3.33 magnitude dimmer companion at 1$\farcs$22, the KOI is likely a 1.68 $R_{\oplus}$ rocky planet if it orbits the primary.

\begin{table}
\renewcommand{\arraystretch}{1.3}
\begin{center}
\caption{Habitable Zone Candidates with Robo-AO Detected Companions}
\begin{tabular}{ccccccc}
\hline
\hline
\noalign{\vskip 1pt}  
\text{Planet} & \text{Period} & \text{R$_{p,i}$}\footnote{Initial planetary radius estimate} & \text{R$_{p,c}$}\footnote{Corrected planetary radius estimate} & \text{Equil. Temp.} & \text{Sep} & \text{$\Delta$m}\\ [0.2ex]
\text{candidate} & \text{(d)} & \text{(R$_\oplus$)} & \text{(R$_\oplus$)} & \text{(K)} & \text{($\arcsec$)} & \text{(mag)}\\ [0.2ex]
\hline
227.01\footnote{Detected in Paper II of this survey.} & 17.7 & 2.45 & 2.96 & 350 & 0.33 & 0.84\\ 
255.01 & 27.5 & 2.51 & 2.67 & 313 & 3.36 & 2.14\\
438.02\textsuperscript{c} & 52.7 & 1.76 & 1.81 & 271 & 3.28 & 3.11\\
1503.01 & 150.2 & 3.79 & 4.23 & 291 & 0.76 & 1.52\\ 
1846.01 & 106.0 & 3.81 & 4.46 & 322 & 3.7 & 1.07\\ 
1989.01\textsuperscript{c} & 201.1 & 1.84 & 1.88 & 297 & 1.12 & 3.49\\
2174.02\textsuperscript{c} & 33.1 & 1.88 & 2.53 & 343 & 0.92 & 0.21\\
2744.01 & 109.6 & 2.46 & 2.63 & 340 & 3.44 & 2.12\\ 
2760.01 & 56.6 & 2.19 & 2.64 & 317 & 0.44 & 0.84\\ 
2862.01 & 24.6 & 1.79 & 2.44 & 321 & 0.67 & 0.17\\ 
2926.03 & 21.0 & 2.43 & 3.24 & 357 & 0.33 & 0.27\\ 
2926.04 & 37.6 & 2.09 & 2.79 & 294 & 0.33 & 0.27\\ 
3255.01\textsuperscript{c} & 66.7 & 1.37 & 1.38 & 294 & 3.15 & 4.87\\
3284.01\textsuperscript{c} & 35.2 & 0.98 & 1.03 & 272 & 3.94 & 2.42\\ 
3401.02\textsuperscript{c} & 326.7 & 2.20 & 2.64 & 283 & 0.65 & 0.89\\ 
3946.01\textsuperscript{c} & 308.5 & 2.36 & 2.37 & 298 & 4.27 & 5.26\\ 
4550.01 & 140.3 & 1.73 & 2.42 & 257 & 1.03 & 0.04\\ 
4810.01 & 115.2 & 2.07 & 2.13 & 353 & 2.32 & 3.16\\ 
5101.01 & 436.2 & 1.64 & 1.68 & 331 & 1.22 & 3.33\\ 
5553.01 & 120.9 & 2.59 & 2.71 & 333 & 0.95 & 2.52\\ 
5671.01 & 190.9 & 1.73 & 1.89 & 356 & 2.13 & 1.79\\ 
5707.01 & 208.8 & 2.88 & 3.03 & 347 & 2.67 & 2.43\\ 
5885.01 & 111.1 & 1.87 & 1.89 & 364 & 3.36 & 4.03\\ 
6120.02 & 205.4 & 1.67 & 1.75 & 323 & 3.78 & 2.48\\ 
6745.01 & 383.9 & 2.78 & 2.82 & 314 & 3.02 & 3.78\\ 
6745.01 & 383.9 & 2.78 & 2.82 & 314 & 2.81 & 3.92\\ 
\hline
\end{tabular}
\label{tab:hz_table}
\end{center}
\end{table}

\begin{figure}
\centering
\includegraphics[width=245pt]{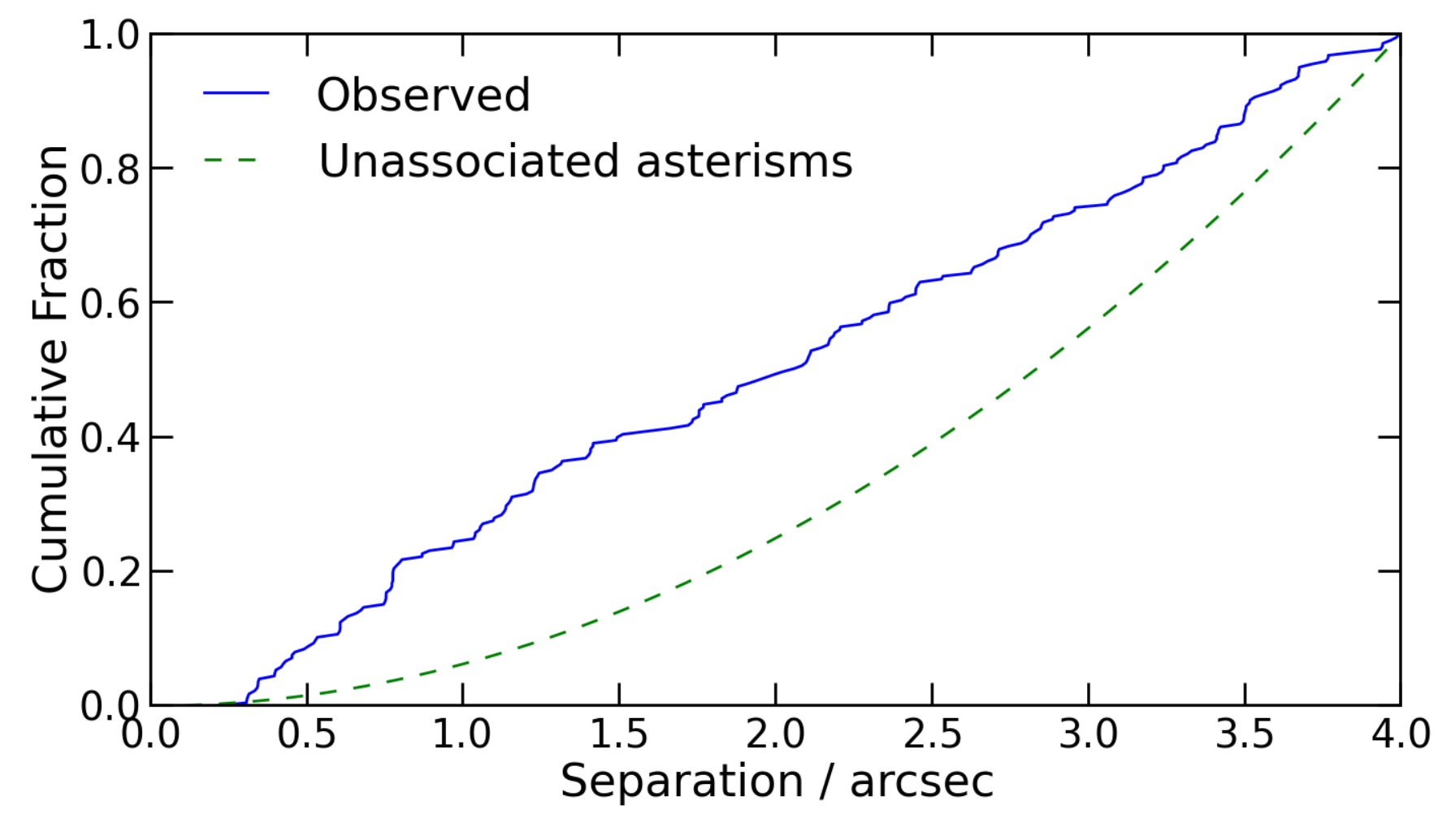}
\caption{The cumulative distribution of nearby stars within a given separation from our observations in Paper II and this work, and the expected distribution from a set of the same number of unassociated stars.  For all separations, the observed number of companions we detected is above the expected number if all stars were unassociated.}
\label{fig:sep_dist}
\end{figure}

\subsection{Stellar Multiplicity and \textit{Kepler} Planet Candidates}
\label{sec:multiplicitystudy}

We detect 206 planetary candidate hosts with nearby stars from 1629 targets, for an overall multiplicity fraction of 12.6\%$\pm$0.9\% within the detectability range of our survey ($\sim$0$\farcs$15--4$\farcs$0, $\Delta$m$\le$6). For this analysis, we will combined the results in this work with the results from Paper II.  With this large dataset we continue our search that began in Paper I for broad-scale correlations between the observed stellar multiplicity and planetary candidate properties. Such correlations provide an avenue to constrain and test planet formation and evolution models.  

Any individual companion found may not be physically bound, however we expect a small number of unassociated asterisms within our complete set of observed targets.  An argument for the majority of nearby stars being associated is derived from the observed distribution of companion separations: if all companions were unassociated background or foreground stars, we would expect a quadratic distribution of companions (i.e., $\sim$4$\times$ the number of objects at 4$\arcsec$ as at 2$\arcsec$).  Instead we find a near linear distribution. The dissimilarity between the observed distribution and the distribution of all unassociated objects is shown in Figure$~\ref{fig:sep_dist}$. In addition, a recent follow-up study with the NIRC2 instrument on the Keck-II telescope \citep{atkinson16} observed 84 KOI systems, finding that at least $14.5^{+3.8}_{-3.4}\%$ of companions within $\sim$4$\arcsec$ are inconsistent with being physically associated based on multi-band photometric parallax.  We therefore expect the overall multiplicity trends to remain relatively unchanged when the unassociated objects are removed.

A summary and analysis paper in the Robo-AO survey will investigate the multiplicity properties of \textit{Kepler} candidates in more detail, including quantifying the effects of association probability and incompleteness.

All stellar and planetary properties for the KOIs in this section were obtained from the cumulative planet candidate list at the NASA Exoplanet Archive\footnote{\url{http://exoplanetarchive.ipac.caltech.edu/}} and have not been corrected for possible dilution due to the presence of nearby stars.

\subsubsection{Stellar Multiplicity and KOI Number}

\begin{figure}
\centering
\includegraphics[width=245pt]{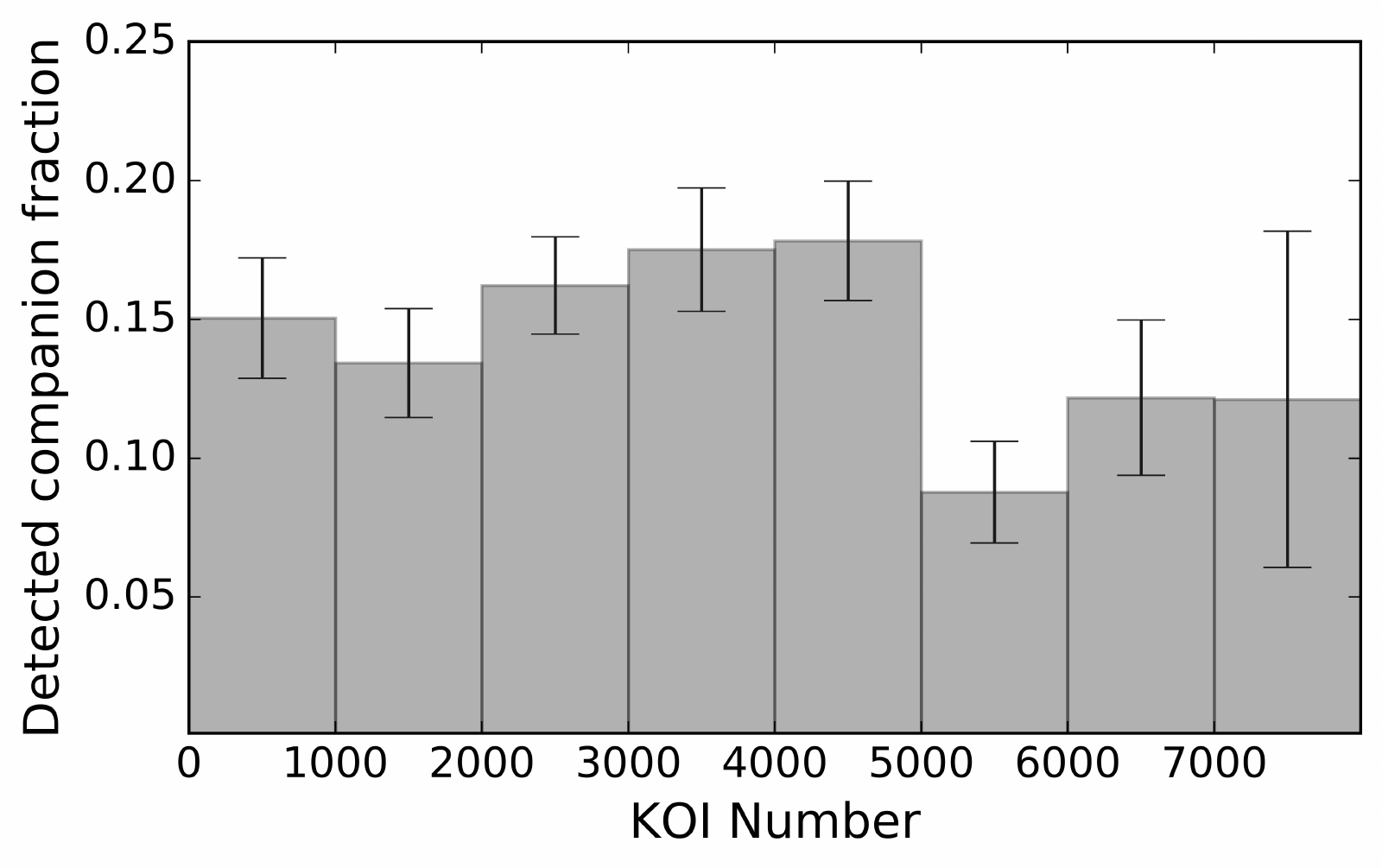}
\caption{Multiplicity fraction within 4\arcsec of KOIs as a function of KOI number. A 2.9$\sigma$ decrease in the fraction of nearby stars between KOIs numbered less than 5000 and greater than 5000 is apparent.}
\label{fig:koinumberbinaryfrac}
\end{figure}

The early and late public releases of KOIs \citep{borucki11, batalha13, burke14, coughlin15} could conceivably have a built-in bias, either astrophysical in origin or as a result of the initial vetting process by the \textit{Kepler} team. This bias might appear as a variation in multiplicity with respect to KOI number.  With a target list of KOIs in Paper II and this work widely dispersed in the full KOI dataset, we can search for such a trend.  The fraction of KOIs with companions as a function of KOI number, as displayed in Figure$~\ref{fig:koinumberbinaryfrac}$, shows a sharp decrease at approximately KOI-5000.  We find KOI numbers less than 5000 have a nearby star fraction of 16.1\%$\pm$0.9\% and KOI numbers greater than 5000 have a nearby star fraction of 10.2\%$\pm$1.5\%, a 2.9$\sigma$ disparity.  The exact mechanism for this is unclear, however this may be a result of better false positive detection in the later data releases due to automation of the vetting process \citep{mullally15}.  There is no significant corresponding variation in the separations or contrasts of stellar companions between the two populations.

\subsubsection{Stellar Multiplicity Rates and Host-star Temperature Revisited}

\begin{figure}
\centering
\includegraphics[width=245pt]{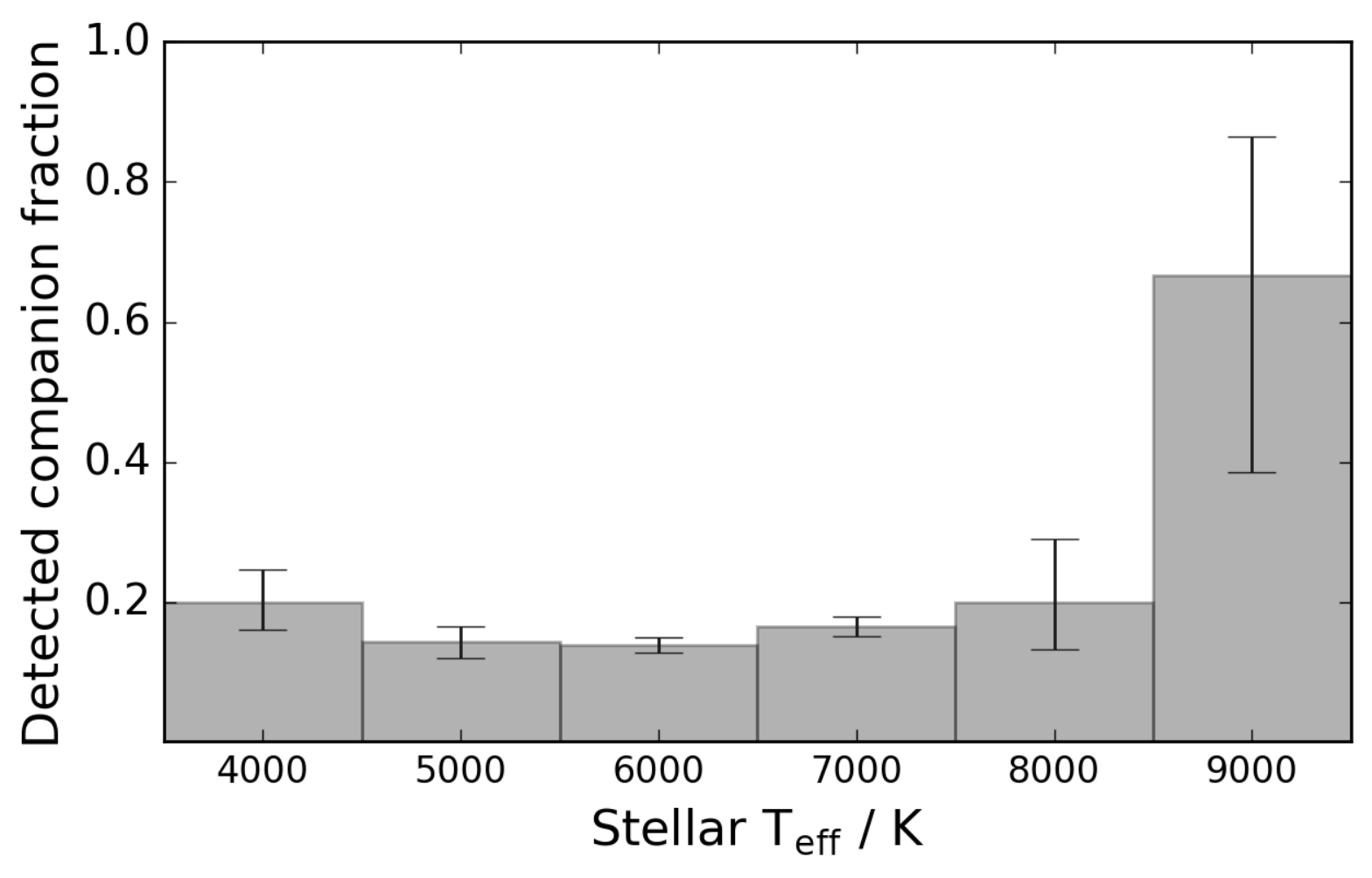}
\caption{Fraction of KOIs with detected nearby ($\le$4\arcsec) stars as a function of stellar effective temperature.}
\label{fig:teffbinaryfrac}
\end{figure}

It has been well established that stellar multiplicity correlates with stellar mass and temperature \citep{duchene13}.  In Paper I, it was found at low significance that this trend appears to also be true for the observed KOIs.  \citet{ngo15} found in a sample of stars hosting close-in giant planets that, with 2.9$\sigma$ significance, stars hotter than 6200 K have a companion rate two times larger than their cool counterparts.  We find in the combined target sample of Paper II and this work that 14.7\%$\pm$0.9\% of KOIs below 6200 K have a companion, compared to 17.2\%$\pm$2.0\% above 6200 K. A Fisher exact test gives an 83$\%$ probability that the two samples are indeed from two distinct populations.  The trend towards higher multiplicity with higher stellar temperatures is still visually evident, as seen in Figure$~\ref{fig:teffbinaryfrac}$.  With an emphasis on solar analogs in the input catalog, the majority of KOIs are FGK-type stars \citep{batalha13}, thus the small number of early type stars in our sample prevents any high significance conclusions.

\subsubsection{Stellar Multiplicity and Multiple-planet Systems Revisited}

Multiple star systems are thought to more commonly host single transiting planets than multiple planet systems.  Perturbations from the companion star will change the mutual inclinations of planets in the same system \citep{wang14}, therefore a lower number of multiple transiting planet systems are expected to have stellar companions.  Multiple planet systems are also subject to planet-planet effects \citep{rasio96, wang15a}.

\begin{figure}
\centering
\includegraphics[width=240pt]{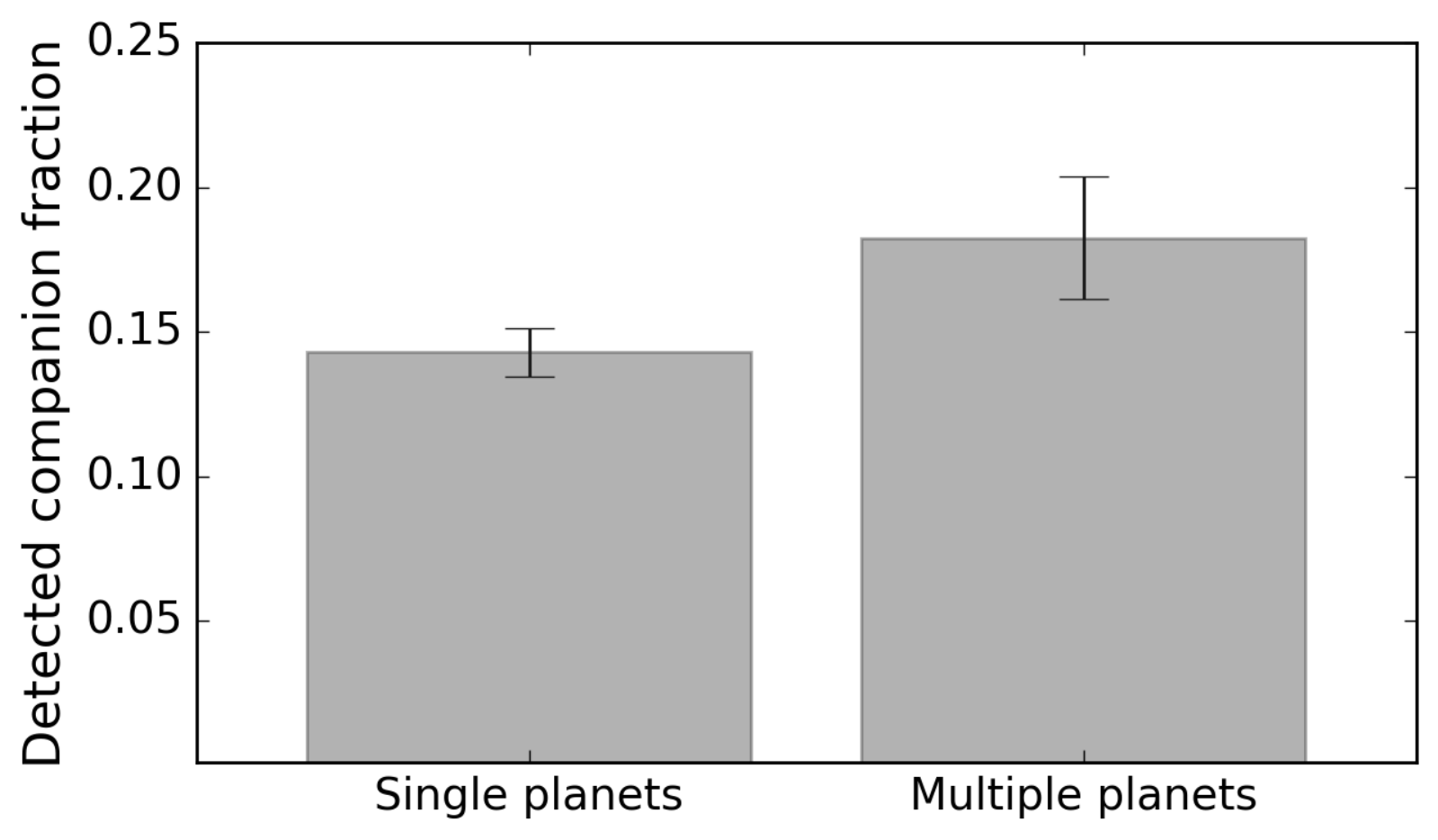}
\caption{The multiplicity fraction within 4\arcsec of KOIs hosting detected single- and multiple-planetary systems.}
\label{fig:numplanetsbinaryfrac}
\end{figure}

In Paper I, we found a low-sigma disparity in multiplicity rates between single- and multiple-planet systems, with single-planet systems exhibiting a slightly higher nearby star fraction.  With our combined sample from Paper II and this work, we revisit this result with over three times more targets.  We find a slightly higher nearby star fraction for multiple planetary systems, displayed in Figure$~\ref{fig:numplanetsbinaryfrac}$.  A Fisher exact test gives an 8.7\% probability of this being a chance difference.  With the expectation, given the effects of stellar perturbations and the higher false positive rate for single star systems, of a higher nearby-star fraction for single-planet candidate hosting stars, even this low-significance result is surprising.  A possible explanation is that the additional stellar body in the system is causing orbital migration of outer planets, moving them to shorter period orbits where \textit{Kepler} has higher sensitivity to transit events.  Also, multiple star systems have at least twice as many stars that could host transiting planets, resulting in a higher probability of observing multiple planetary transits.  Lastly, with relatively low-significance, this result could also be a consequence of the ``look-elsewhere'' effect inherent to any multi-comparison study \citep{gross10};  with the parameter space explored in this section, a result of this significance is expected to arise approximately 50\% of the time out of per chance.

\citet{wang15b} studied the influence of stellar companions on multiple-planet systems, finding a 3.2$\sigma$ deficit in multiplicity rate in multi-planet systems compared to a control sample of field stars.  However, they also found no significant disparity in multiplicity rates between single- and multi-planet systems.

\subsubsection{Stellar Multiplicity and Close-in Planets Revisited}

The presence of stellar companions is hypothesized to shape the formation and evolution of planetary systems.  Overall, there is evidence that planetary formation is disrupted in close binary systems \citep{fragner11, roell12}.  The third body in the system can lead to Kozai oscillations causing orbital migration of the planets \citep{fabrycky07, katz11, naox12} or tilt the circumstellar disk \citep{batygin12}.  Smaller planets are also more prone to the influence of a stellar companion because of weaker planet-planet dynamical coupling \citep{wang15a}.  These dynamical interactions between small and large planets in the same system tend to differentially eject small planets more frequently than large planets \citep{xie14}. The presence of a stellar companion increases the frequency of these interactions, leading to higher loss of small planets.  Consequently, we would expect a correlation between binarity and planetary period for different sized planets.

We previously reported a low-significance result of stellar third bodies increasing the rate of close-in giant planets, possible evidence of orbital migration of the planet caused by the stellar companion.  We revisit the discussion and analysis from Paper I in search of this correlation using the results of Paper II and this work. This analysis splits the ``small'' and ``giant'' planets at the arbitrary value of Neptune's radius (3.9 R$_\oplus$).  The exact value does not significantly affect the results as just 11 of the detected systems have planetary radii within 20$\%$ of the cutoff value, with 1635 small and 395 giant planets in total.

\begin{figure}
\centering
\includegraphics[width=245pt]{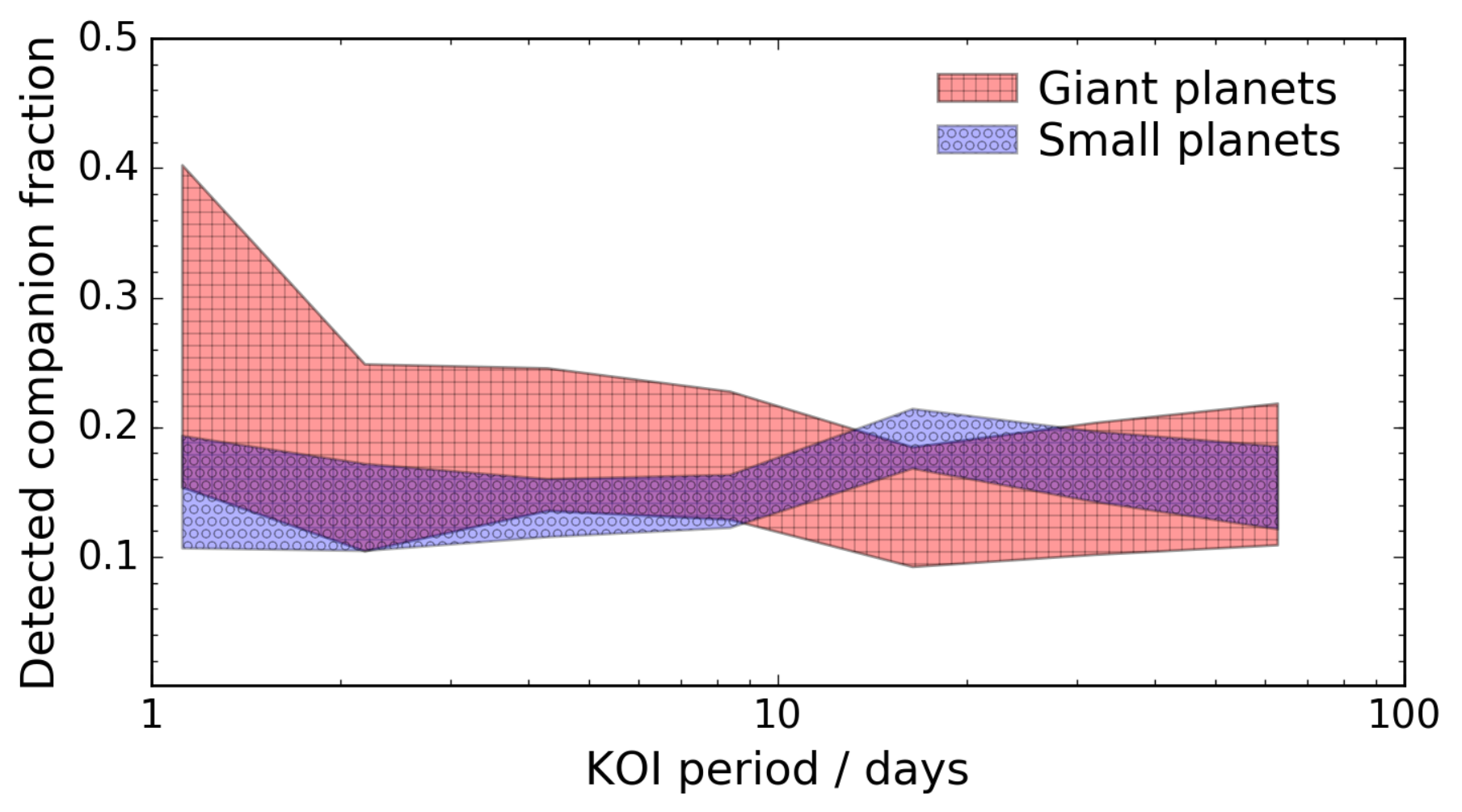}
\caption{1$\sigma$ uncertainty regions for the binarity fraction as a function of KOI period for two different planetary populations.}
\label{fig:periodfrac}
\end{figure}

\begin{figure}
\centering
\includegraphics[width=245pt]{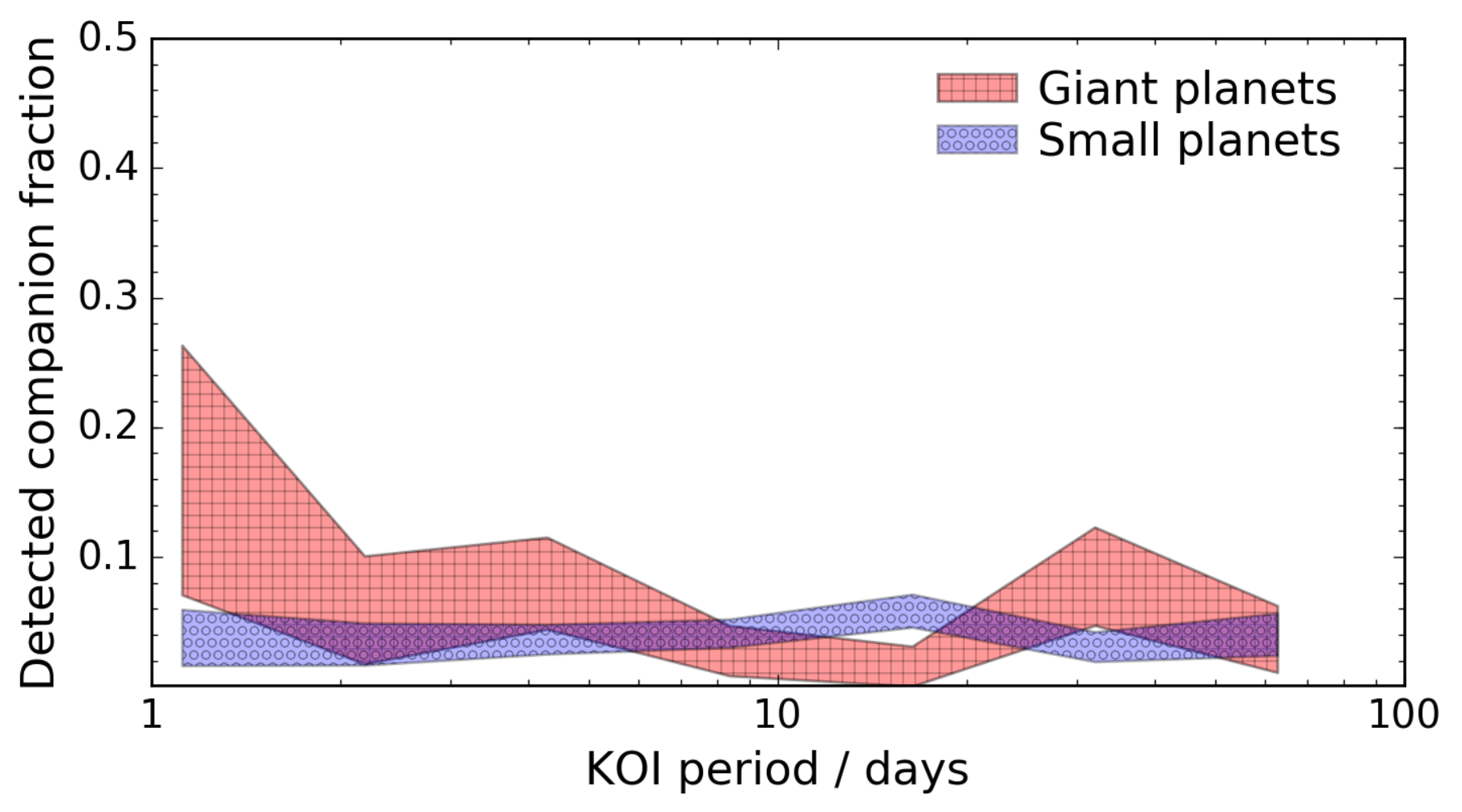}
\caption{1$\sigma$ uncertainty regions for the binarity fraction as a function of KOI period for two different planetary populations, with only companions with separations $<$2$\farcs$5 used to align with Paper I.}
\label{fig:25periodfrac}
\end{figure}

In Figure$~\ref{fig:periodfrac}$ the fraction of \textit{Kepler} planet candidates with nearby stars is shown, with planets grouped into two different size ranges. We again see a small increase in the nearby star fraction for giants with periods $<$15 days, however the $>$2$\sigma$ spike at periods of 2-4 days seen in Paper I is not present.  If our sample is reduced to correspond to the separation range of Paper I ($\rho$$<$2$\farcs$5) in Figure$~\ref{fig:25periodfrac}$, again no binarity spike at periods $<$10 days is apparent.

\begin{figure}
\centering
\includegraphics[width=245pt]{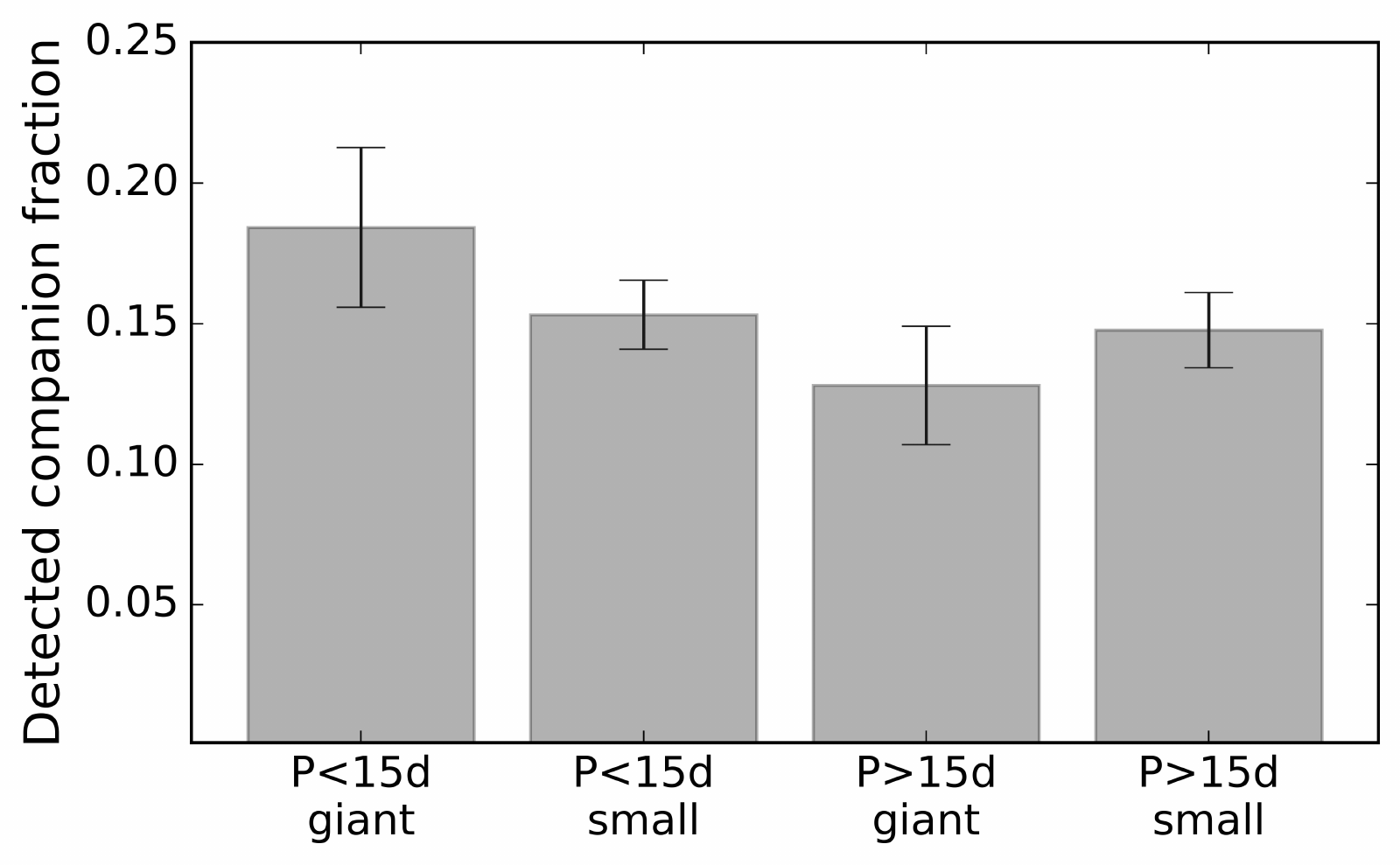}
\caption{Multiplicity fraction of KOIs with four planetary populations, with all contrast ratios and separations $\le$4\arcsec.  A planet is considered giant if its radius is equal to or larger to that of Neptune (3.9 R$_\oplus$).  Multi-planet systems can be assigned to multiple populations.}
\label{fig:periodradiusbinaryfrac}
\end{figure}

\begin{figure}
\centering
\includegraphics[width=245pt]{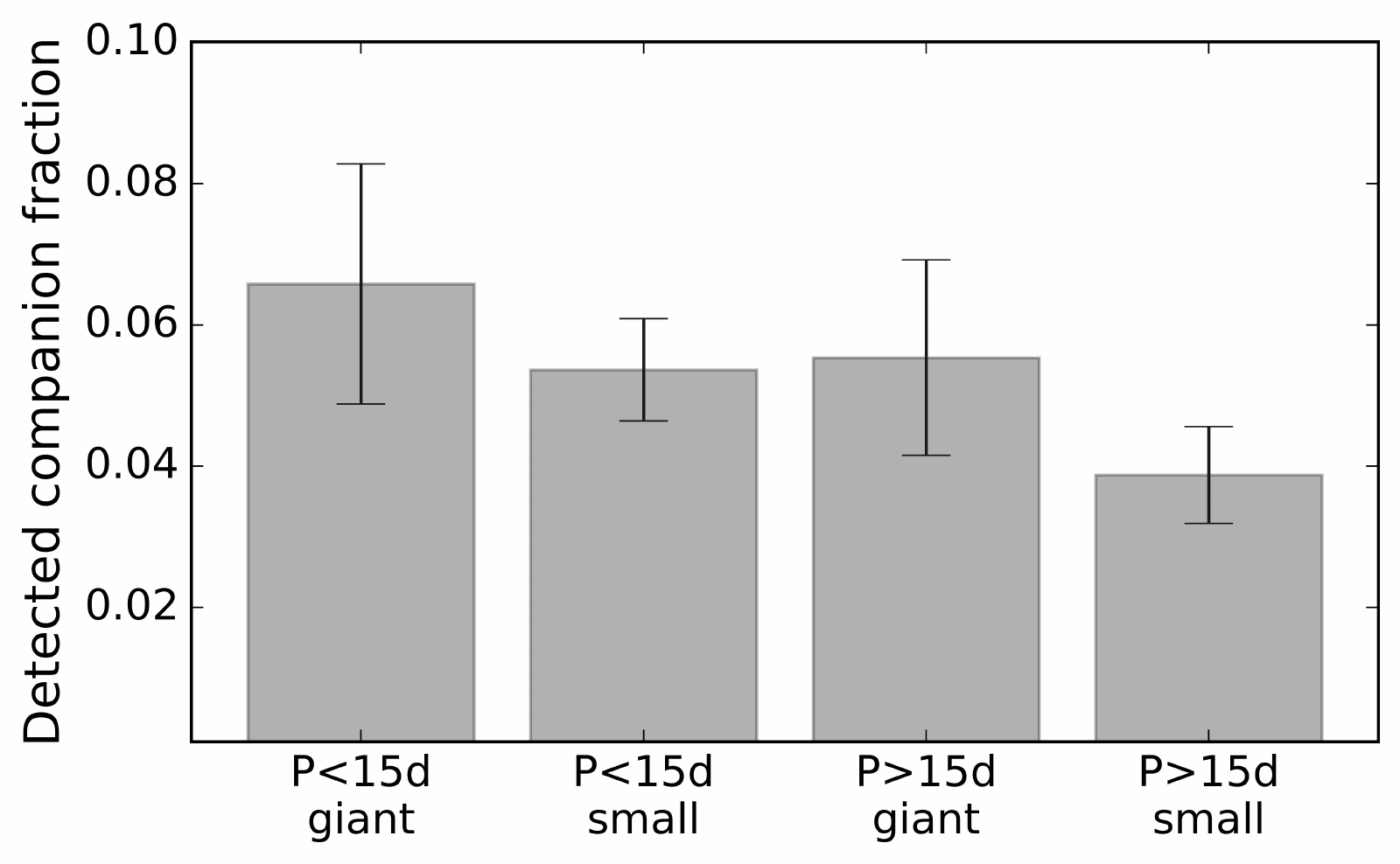}
\caption{Multiplicity fraction of KOIs with four planetary populations, with only companions with $\Delta$m$\le$2 and separations $\le$1\farcs5, removing the faint nearby stars that are less likely to be physically associated.}
\label{fig:closeperiodradiusbinaryfrac}
\end{figure}

Binning our targets into four population groups in Figure$~\ref{fig:periodradiusbinaryfrac}$ suggests no significant difference in the binarity rate of short period giants.  We also attempt to decrease the occurrence of unassociated asterisms by only using close, bright companions ($\Delta$m$\le$2, $\rho$$\le$1\farcs5).  As in Paper I, we detected an excess of close-separation bright companions (Figure$~\ref{fig:contrastcurves}$), which suggests a higher probability of association for these nearby stars.  We show the binarity fraction of the four populations in Figure$~\ref{fig:closeperiodradiusbinaryfrac}$.  As with the complete set of nearby stars, no significant differences between the four populations is evident.

\begin{figure}
\centering
\includegraphics[width=245pt]{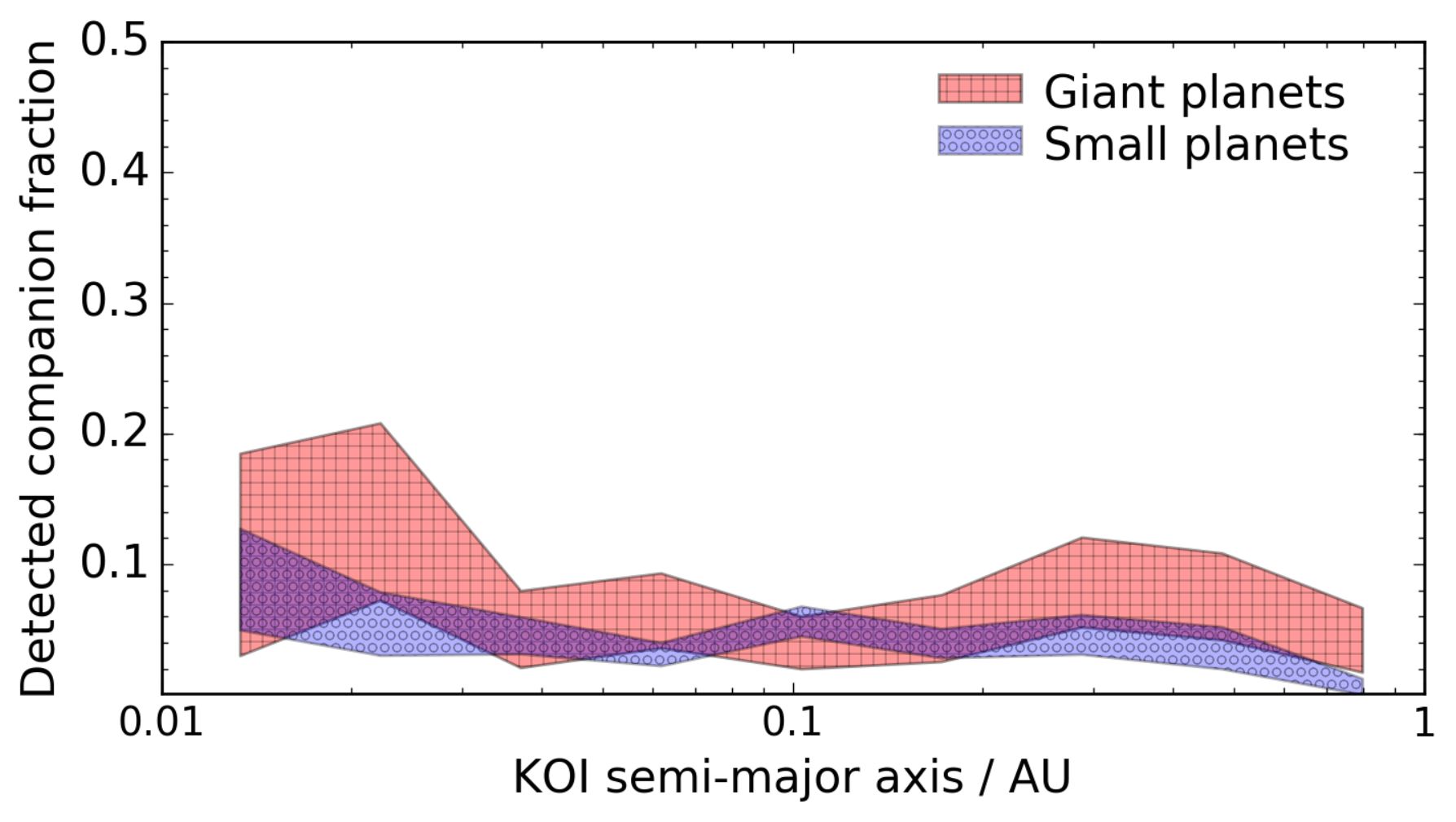}
\caption{1$\sigma$ uncertainty regions for the binarity fraction as a function of KOI semi-major axis between 0.01 and 1.0 AU for two different planetary populations.}
\label{fig:semimajoraxis}
\end{figure}

\begin{figure*}
\centering
\includegraphics[width=500pt]{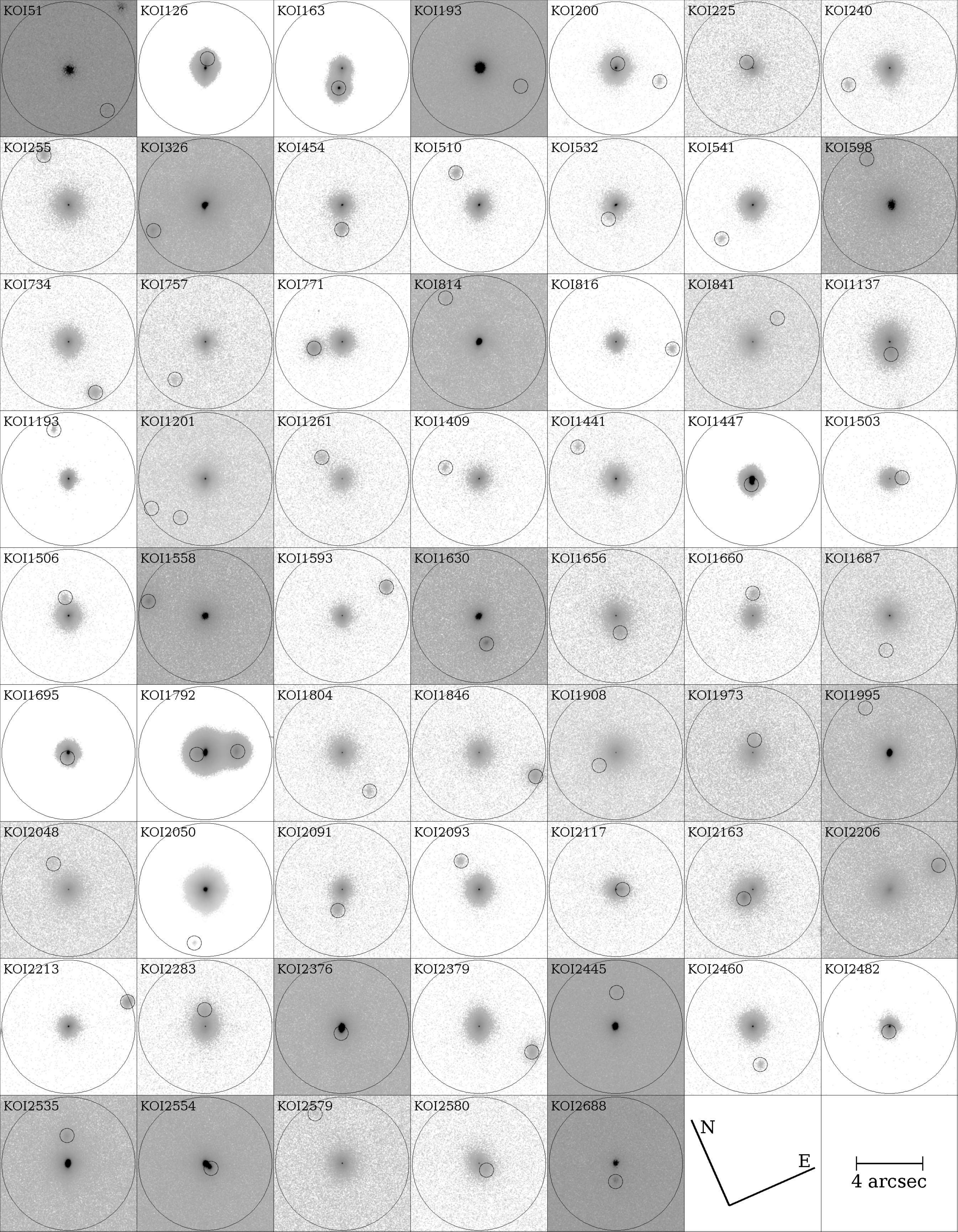}
\caption{Color inverted, normalized log-scale cutouts of 61 multiple KOI systems [KOI-51 to KOI-2688] with separations $<$4$\arcsec$ resolved with Robo-AO.  The angular scale and orientation is similar for each cutout.  The smaller circles are centered on the detected nearby star, and the larger circle is the limit of the survey's 4\arcsec separation range.}
\label{fig:cutouts1}
\end{figure*}

\begin{figure*}
\centering
\includegraphics[width=500pt]{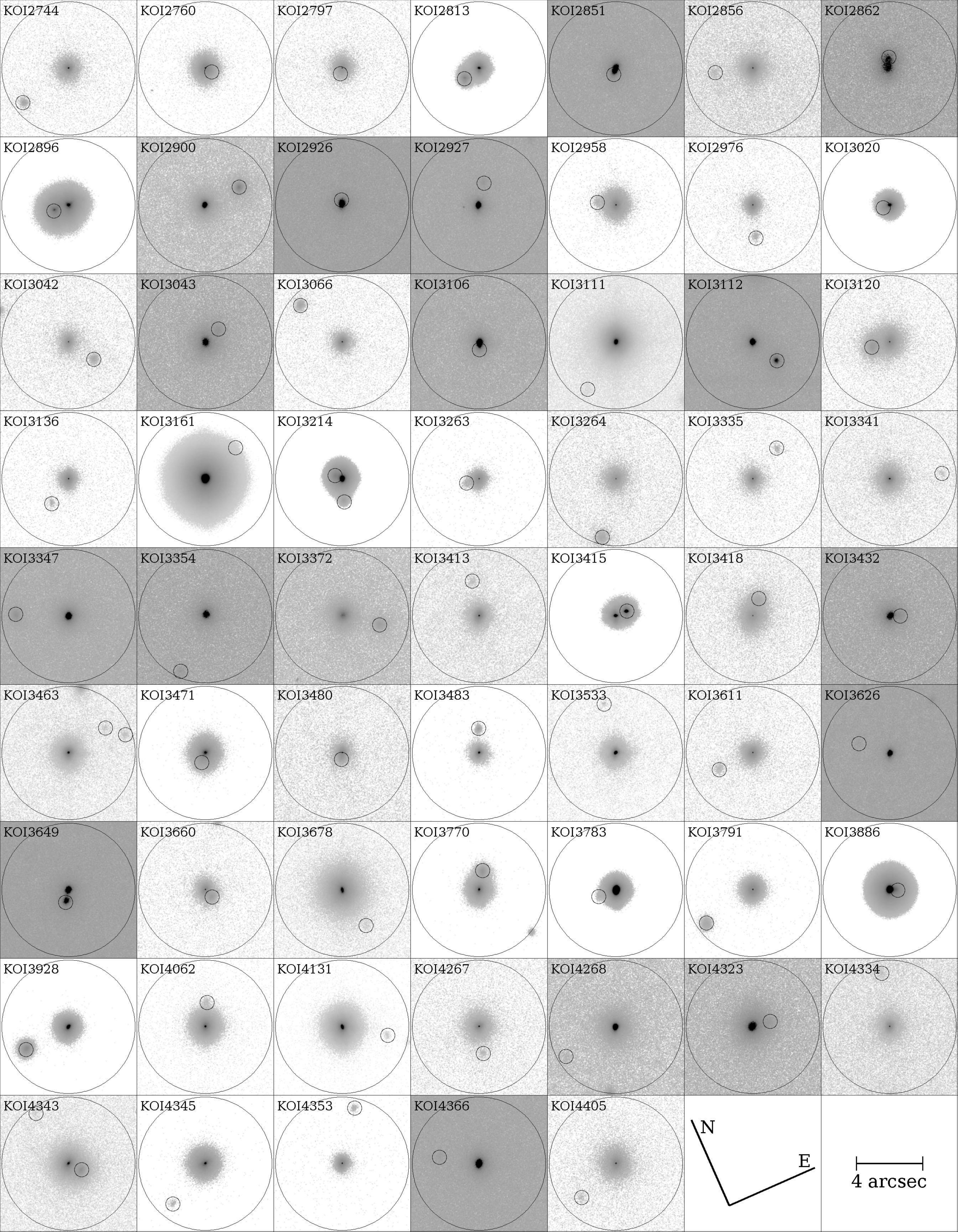}
\caption{Color inverted, normalized log-scale cutouts of 61 multiple KOI systems [KOI-2744 to KOI-4405] with separations $<$4$\arcsec$ resolved with Robo-AO.  The angular scale and orientation is similar for each cutout.  The smaller circles are centered on the detected nearby star, and the larger circle is the limit of the survey's 4\arcsec separation range.}
\label{fig:cutouts2}
\end{figure*}

\begin{figure*}
\centering
\includegraphics[width=500pt]{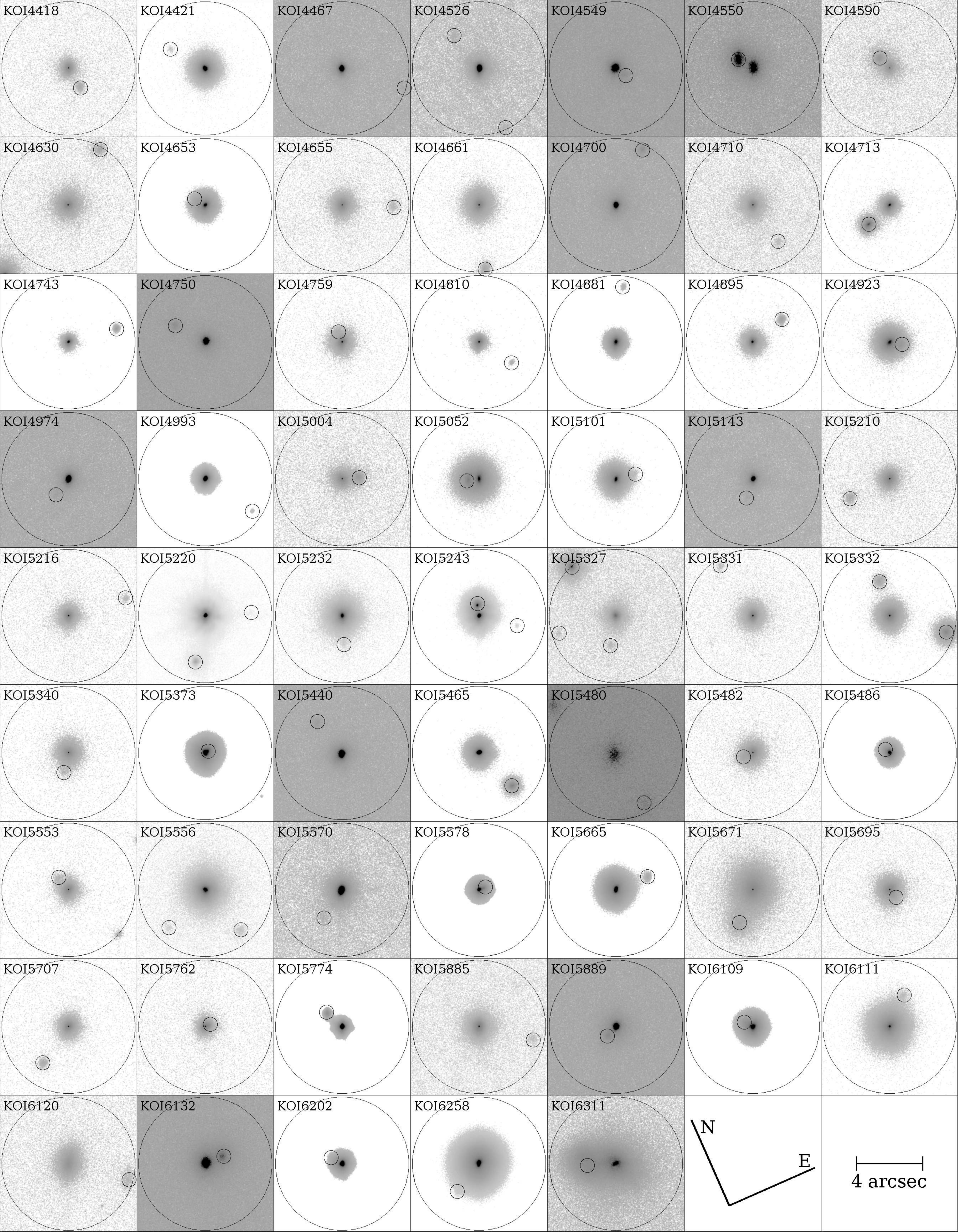}
\caption{Color inverted, normalized log-scale cutouts of 61 multiple KOI systems [KOI-4418 to KOI-6311] with separations $<$4$\arcsec$ resolved with Robo-AO.  The angular scale and orientation is similar for each cutout.  The smaller circles are centered on the detected nearby star, and the larger circle is the limit of the survey's 4\arcsec separation range.}
\label{fig:cutouts3}
\end{figure*}

\begin{figure*}
\centering
\includegraphics[width=500pt]{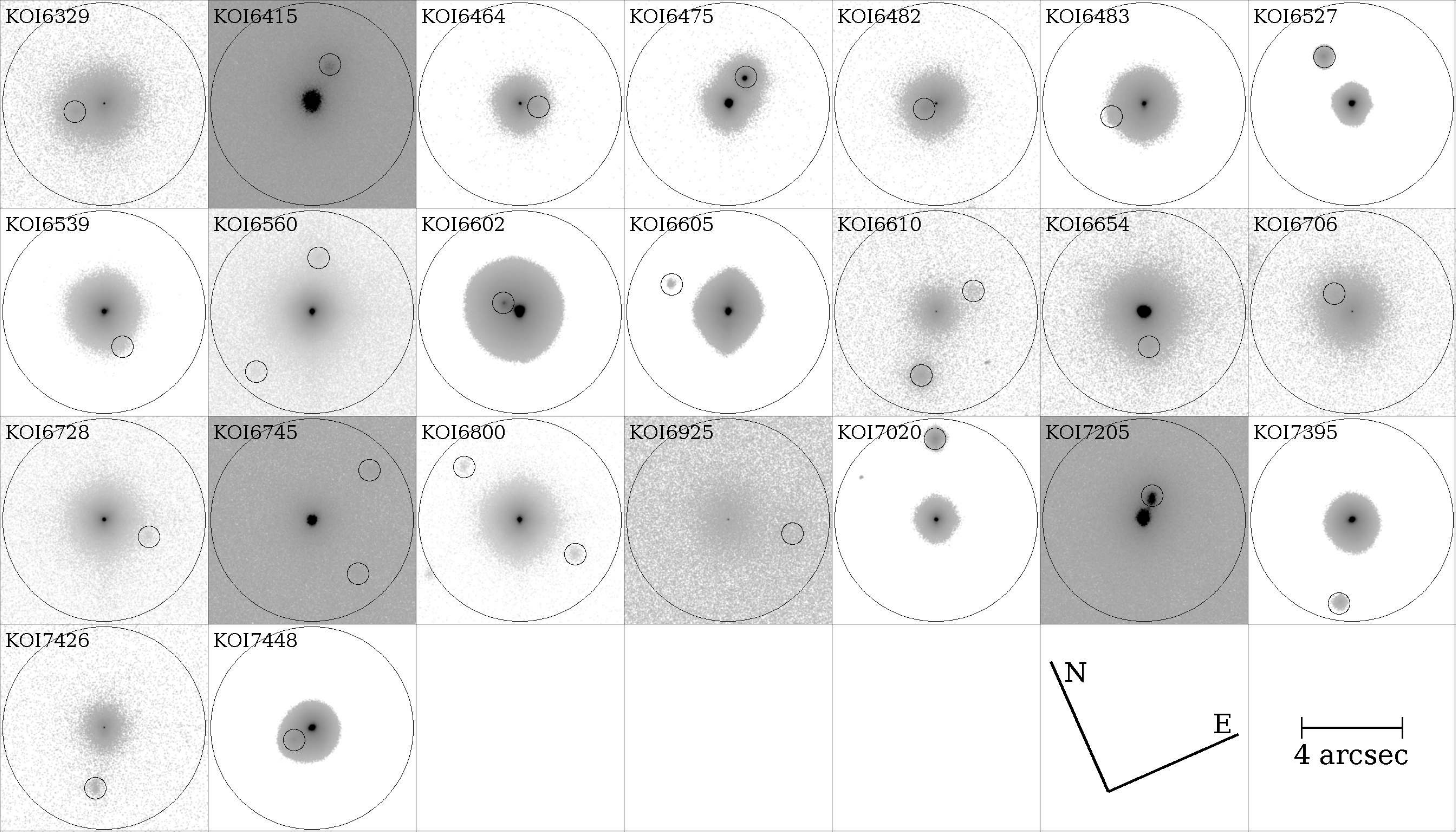}
\caption{Color inverted, normalized log-scale cutouts of 23 multiple KOI systems [KOI-6329 to KOI-7448] with separations $<$4$\arcsec$ resolved with Robo-AO.  The angular scale and orientation is similar for each cutout.  The smaller circles are centered on the detected nearby star, and the larger circle is the limit of the survey's 4\arcsec separation range.}
\label{fig:cutouts4}
\end{figure*}

Any real disparity between the populations would also manifest in the physical orbital semi-major axis, which is related to the observable periods by the stellar mass. In Figure$~\ref{fig:semimajoraxis}$ we plot the two population's binarity fraction as a function of the calculated semi-major axis of the planetary candidates between 0.01 and 1.0 AU.  No significant giant planet binarity spike is observed as in the periods plot.

Our updated study using the targets in Paper II and this work suggests that the presence of a second stellar body in planetary systems does not appreciably affect the number of close-in giant planets.  This agrees with the analysis of \citet{wang15a} who find a relatively uniform multiplicity rate for planets with short and long periods.  They note that our previous tentative result may have been due to short-period giants with brighter stellar companions in the visible biasing our detections.  Subject to the same potential biases, the larger survey in this analysis does not indicate a period-multiplicity correlation for the two planetary populations, suggesting that our previous low-sigma result may have instead been an artifact of small-number statistics.

\citet{kraus16} find a 6.6$\sigma$ deficit in binary stars with separation $\rho$$<$50 AU in KOIs compared to field stars, again suggesting that close-in stellar companions disrupt the formation and/or evolution of planets, as had been previously hypothesized \citep{wang14}.  Indeed, a quarter of all solar-type stars in the Milky Way are disallowed from hosting planetary systems due to the influence of binary companions.

Some evidence remains, however, that stellar binarity may encourage the presence of hot Jupiters.  A recent NIR survey \citep{ngo15} of exoplanetary systems with known close-in giants finds that hot Jupiter hosts are twice as likely as field stars to be found in a multiple star system, with a significance of 2.8$\sigma$.  However, the binarity rates of systems containing hot Jupiters remains unclear: 12$\%$ \citep{roell12}, 38$\%$ \citep{evans16}, 51$\%$ \citep{ngo15}. 

We will revisit this discussion in the last paper in this series where we will combine the full Robo-AO KOI survey dataset.

\section{Conclusion}
\label{sec:conclusion}

We observed 1629 \textit{Kepler} planetary candidates with the Robo-AO robotic laser adaptive optics system.  We detected 206 planetary candidates with nearby stars, implying an overall nearby-star probability of 12.6\%$\pm$0.9\% at separations between $\sim$0$\farcs$15 and 4$\farcs$0 and $\Delta$m$\le$6.  

Many of our newly found companions are of particular interest, including 26 habitable zone candidates found within possible multiple star systems.  In addition, we found 16 KOIs with multiple nearby stars, and 5 new candidate quadruple star systems hosting planet candidates, including KOI-4495 from Paper II.  We looked at broad correlations between the presence of nearby stars and planetary characteristics.  We find a higher detected companion rate of systems with multiple planets than in single planet systems.   Our previous tentative result of a deficit of close-in giant planets when a third stellar body appears in the system is not apparent in this dataset.

The Robo-AO system was installed on the 2.1-m telescope at Kitt Peak in November 2015, and a new low-noise infrared camera that will allow observations of redder companion stars will be added in the future.  In addition, a second generation Robo-AO instrument on the University of Hawai`i 2.2-m telescope on Maunakea \citep{Robo-AO2} is being built.  The two systems will together image up to $\sim$500 objects per night and have access to three-quarters of the sky over the course of a year.  A southern analog to Robo-AO, mounted on the Southern Astrophysical Research Telescope (SOAR) at CTIO and capable of twice HST resolution imaging, is also in development.  With unmatched efficiency, Robo-AO and its lineage of instruments are uniquely able to perform high-acuity imaging of the hundreds of K2 \citep{K2} planetary candidates, ground-based transit surveys such as MEarth \citep{mearth}, KELT \citep{kelt1, kelt2}, HATNet \citep{hatnet}, SuperWASP \citep{superwasp}, NGTS \citep{ngts}, XO \citep{xo}, and the Evryscope \citep{evryscope}, as well as the thousands of expected exoplanet hosts discovered by the forthcoming NASA Transiting Exoplanet Survey Satellite \citep[TESS,][]{TESS} and ESA PLAnetary Transits and Oscillations of stars 2.0 \citep[PLATO,][]{PLATO} missions.  

The Robo-AO survey has completed observations of over $90\%$ of the \textit{Kepler} planet candidates, with the remaining targets to be observed at the Kitt Peak telescope.  Future papers in this survey will present these final KOI targets, and perform a full probability of association analysis.  With the entire survey soon to be completed, providing us with an unprecedented dataset of thousands of high angular resolution imaged planetary candidates, we can continue our search for clues to planetary formation and evolution.

\section*{Acknowledgements}
We thank the anonymous referee for careful analysis and useful comments on the manuscript.

This research is supported by the NASA Exoplanets Research Program, grant $\#$NNX 15AC91G. C.B. acknowledges support from the Alfred P. Sloan Foundation. T.M is supported by NASA grant $\#$NNX 14AE11G under the Kepler Participating Scientist Program. D.A. is supported by a NASA Space Technology Research Fellowship, grant $\#$NNX 13AL75H.

The Robo-AO system is supported by collaborating partner institutions, the California Institute of Technology and the Inter-University Centre for Astronomy and Astrophysics, and by the National Science Foundation under Grant Nos. AST-0906060, AST-0960343, and AST-1207891, by the Mount Cuba Astronomical Foundation, and by a gift from Samuel Oschin. We are grateful to the Palomar Observatory staff for their ongoing support of Robo-AO on the 1.5-m telescope, particularly S. Kunsman, M. Doyle, J. Henning, R. Walters, G. Van Idsinga, B. Baker, K. Dunscombe and D. Roderick.

Some of the data presented herein were obtained at the W.M. Keck Observatory, which is operated as a scientific partnership among the California Institute of Technology, the University of California and the National Aeronautics and Space Administration. The Observatory was made possible by the generous financial support of the W.M. Keck Foundation. Some of the data presented herein  is based on observations obtained at the Gemini Observatory, operated by the Association of Universities for Research in Astronomy, Inc., under a cooperative agreement with the NSF on behalf of the Gemini partnership.   We recognize and acknowledge the very significant cultural role and reverence that the summit of Maunakea has always had within the indigenous Hawaiian community. We are most fortunate to have the opportunity to conduct observations from this mountain.

We thank Adam Kraus et al. for sharing a preprint of their paper.

This research has made use of the SIMBAD database, operated by Centre des Donn\'ees Stellaires (Strasbourg, France), and bibliographic references from the Astrophysics Data System maintained by SAO/NASA.  This research has made use of the \textit{Kepler} Community FollowUp Observing Program Web site (https://cfop.ipac.caltech.edu) and the NASA Exoplanet Archive, which is operated by the California Institute of Technology, under contract with the National Aeronautics and Space Administration under the Exoplanet Exploration Program.  This work used the K2fov \citep{k2fov} Python package.

{\it Facilities:} \facility{PO:1.5m (Robo-AO)}, \facility{Keck:II (NIRC2-LGS)}, \facility{Gemini:Gillett (NIRI)}

\bibliography{koi3.bbl}

\begin{thebibliography}{97}
\expandafter\ifx\csname natexlab\endcsname\relax\def\natexlab#1{#1}\fi

\bibitem[{{Adams} {et~al.}(2012){Adams}, {Ciardi}, {Dupree}, {Gautier},
  {Kulesa}, \& {McCarthy}}]{adams12}
{Adams}, E.~R., {Ciardi}, D.~R., {Dupree}, A.~K., {Gautier}, III, T.~N.,
  {Kulesa}, C., \& {McCarthy}, D. 2012, \aj, 144, 42

\bibitem[{{Adams} {et~al.}(2013){Adams}, {Dupree}, {Kulesa}, \&
  {McCarthy}}]{adams13}
{Adams}, E.~R., {Dupree}, A.~K., {Kulesa}, C., \& {McCarthy}, D. 2013, \aj,
  146, 9

\bibitem[{{Atkinson} {et~al.}(2016){Atkinson}, {Baranec}, {Ziegler}, {Law},
  {Riddle}, \& {Morton}}]{atkinson16}
{Atkinson}, D., {Baranec}, C., {Ziegler}, C., {Law}, N.~M., {Riddle}, R., \&
  {Morton}, T. 2016

\bibitem[{{Bakos} {et~al.}(2004){Bakos}, {Noyes}, {Kov{\'a}cs}, {Stanek},
  {Sasselov}, \& {Domsa}}]{hatnet}
{Bakos}, G., {Noyes}, R.~W., {Kov{\'a}cs}, G., {Stanek}, K.~Z., {Sasselov},
  D.~D., \& {Domsa}, I. 2004, \pasp, 116, 266

\bibitem[{{Baranec} {et~al.}(2014{\natexlab{a}}){Baranec}, {Riddle}, {Law},
  {Chun}, {Lu}, {Connelley}, {Hall}, {Atkinson}, \& {Jacobson}}]{Robo-AO2}
{Baranec}, C., {Riddle}, R., {Law}, N.~M., {Chun}, M.~R., {Lu}, J.~R.,
  {Connelley}, M.~S., {Hall}, D., {Atkinson}, D., \& {Jacobson}, S.
  2014{\natexlab{a}}, in \procspie, Vol. 9148, Adaptive Optics Systems IV,
  914812

\bibitem[{{Baranec} {et~al.}(2013){Baranec}, {Riddle}, {Law}, {Ramaprakash},
  {Tendulkar}, {Bui}, {Burse}, {Chordia}, {Das}, {Davis}, {Dekany}, {Kasliwal},
  {Kulkarni}, {Morton}, {Ofek}, \& {Punnadi}}]{baranec13}
{Baranec}, C., {Riddle}, R., {Law}, N.~M., {Ramaprakash}, A.~N., {Tendulkar},
  S.~P., {Bui}, K., {Burse}, M.~P., {Chordia}, P., {Das}, H.~K., {Davis},
  J.~T.~C., {Dekany}, R.~G., {Kasliwal}, M.~M., {Kulkarni}, S.~R., {Morton},
  T.~D., {Ofek}, E.~O., \& {Punnadi}, S. 2013, Journal of Visualized
  Experiments, 72, e50021

\bibitem[{{Baranec} {et~al.}(2014{\natexlab{b}}){Baranec}, {Riddle}, {Law},
  {Ramaprakash}, {Tendulkar}, {Hogstrom}, {Bui}, {Burse}, {Chordia}, {Das},
  {Dekany}, {Kulkarni}, \& {Punnadi}}]{baranec14}
{Baranec}, C., {Riddle}, R., {Law}, N.~M., {Ramaprakash}, A.~N., {Tendulkar},
  S.~P., {Hogstrom}, K., {Bui}, K., {Burse}, M., {Chordia}, P., {Das}, H.,
  {Dekany}, R.~G., {Kulkarni}, S., \& {Punnadi}, S. 2014{\natexlab{b}}, \apjl,
  790, L8

\bibitem[{{Baranec} {et~al.}(2016){Baranec}, {Ziegler}, {Law}, {Morton},
  {Riddle}, {Atkinson}, {Schonhut}, \& {Crepp}}]{baranec16}
{Baranec}, C., {Ziegler}, C., {Law}, N.~M., {Morton}, T., {Riddle}, R.,
  {Atkinson}, D., {Schonhut}, J., \& {Crepp}, J. 2016, ArXiv e-prints

\bibitem[{{Barclay} {et~al.}(2015){Barclay}, {Quintana}, {Adams}, {Ciardi},
  {Huber}, {Foreman-Mackey}, {Montet}, \& {Caldwell}}]{barclay15}
{Barclay}, T., {Quintana}, E.~V., {Adams}, F.~C., {Ciardi}, D.~R., {Huber}, D.,
  {Foreman-Mackey}, D., {Montet}, B.~T., \& {Caldwell}, D. 2015, \apj, 809, 7

\bibitem[{{Batalha} {et~al.}(2010){Batalha}, {Borucki}, {Koch}, {Bryson},
  {Haas}, {Brown}, {Caldwell}, {Hall}, {Gilliland}, {Latham}, {Meibom}, \&
  {Monet}}]{batalha10}
{Batalha}, N.~M., {Borucki}, W.~J., {Koch}, D.~G., {Bryson}, S.~T., {Haas},
  M.~R., {Brown}, T.~M., {Caldwell}, D.~A., {Hall}, J.~R., {Gilliland}, R.~L.,
  {Latham}, D.~W., {Meibom}, S., \& {Monet}, D.~G. 2010, \apjl, 713, L109

\bibitem[{{Batalha} {et~al.}(2013){Batalha}, {Rowe}, {Bryson}, {Barclay},
  {Burke}, {Caldwell}, {Christiansen}, {Mullally}, {Thompson}, {Brown},
  {Dupree}, {Fabrycky}, {Ford}, {Fortney}, {Gilliland}, {Isaacson}, {Latham},
  {Marcy}, {Quinn}, {Ragozzine}, {Shporer}, {Borucki}, {Ciardi}, {Gautier},
  {Haas}, {Jenkins}, {Koch}, {Lissauer}, {Rapin}, {Basri}, {Boss}, {Buchhave},
  {Carter}, {Charbonneau}, {Christensen-Dalsgaard}, {Clarke}, {Cochran},
  {Demory}, {Desert}, {Devore}, {Doyle}, {Esquerdo}, {Everett}, {Fressin},
  {Geary}, {Girouard}, {Gould}, {Hall}, {Holman}, {Howard}, {Howell},
  {Ibrahim}, {Kinemuchi}, {Kjeldsen}, {Klaus}, {Li}, {Lucas}, {Meibom},
  {Morris}, {Pr{\v s}a}, {Quintana}, {Sanderfer}, {Sasselov}, {Seader},
  {Smith}, {Steffen}, {Still}, {Stumpe}, {Tarter}, {Tenenbaum}, {Torres},
  {Twicken}, {Uddin}, {Van Cleve}, {Walkowicz}, \& {Welsh}}]{batalha13}
{Batalha}, N.~M., {Rowe}, J.~F., {Bryson}, S.~T., {Barclay}, T., {Burke},
  C.~J., {Caldwell}, D.~A., {Christiansen}, J.~L., {Mullally}, F., {Thompson},
  S.~E., {Brown}, T.~M., {Dupree}, A.~K., {Fabrycky}, D.~C., {Ford}, E.~B.,
  {Fortney}, J.~J., {Gilliland}, R.~L., {Isaacson}, H., {Latham}, D.~W.,
  {Marcy}, G.~W., {Quinn}, S.~N., {Ragozzine}, D., {Shporer}, A., {Borucki},
  W.~J., {Ciardi}, D.~R., {Gautier}, III, T.~N., {Haas}, M.~R., {Jenkins},
  J.~M., {Koch}, D.~G., {Lissauer}, J.~J., {Rapin}, W., {Basri}, G.~S., {Boss},
  A.~P., {Buchhave}, L.~A., {Carter}, J.~A., {Charbonneau}, D.,
  {Christensen-Dalsgaard}, J., {Clarke}, B.~D., {Cochran}, W.~D., {Demory},
  B.-O., {Desert}, J.-M., {Devore}, E., {Doyle}, L.~R., {Esquerdo}, G.~A.,
  {Everett}, M., {Fressin}, F., {Geary}, J.~C., {Girouard}, F.~R., {Gould}, A.,
  {Hall}, J.~R., {Holman}, M.~J., {Howard}, A.~W., {Howell}, S.~B., {Ibrahim},
  K.~A., {Kinemuchi}, K., {Kjeldsen}, H., {Klaus}, T.~C., {Li}, J., {Lucas},
  P.~W., {Meibom}, S., {Morris}, R.~L., {Pr{\v s}a}, A., {Quintana}, E.,
  {Sanderfer}, D.~T., {Sasselov}, D., {Seader}, S.~E., {Smith}, J.~C.,
  {Steffen}, J.~H., {Still}, M., {Stumpe}, M.~C., {Tarter}, J.~C., {Tenenbaum},
  P., {Torres}, G., {Twicken}, J.~D., {Uddin}, K., {Van Cleve}, J.,
  {Walkowicz}, L., \& {Welsh}, W.~F. 2013, \apjs, 204, 24

\bibitem[{{Batygin}(2012)}]{batygin12}
{Batygin}, K. 2012, \nat, 491, 418

\bibitem[{{Borucki} {et~al.}(2011{\natexlab{a}}){Borucki}, {Koch}, {Basri},
  {Batalha}, {Boss}, {Brown}, {Caldwell}, {Christensen-Dalsgaard}, {Cochran},
  {DeVore}, {Dunham}, {Dupree}, {Gautier}, {Geary}, {Gilliland}, {Gould},
  {Howell}, {Jenkins}, {Kjeldsen}, {Latham}, {Lissauer}, {Marcy}, {Monet},
  {Sasselov}, {Tarter}, {Charbonneau}, {Doyle}, {Ford}, {Fortney}, {Holman},
  {Seager}, {Steffen}, {Welsh}, {Allen}, {Bryson}, {Buchhave},
  {Chandrasekaran}, {Christiansen}, {Ciardi}, {Clarke}, {Dotson}, {Endl},
  {Fischer}, {Fressin}, {Haas}, {Horch}, {Howard}, {Isaacson}, {Kolodziejczak},
  {Li}, {MacQueen}, {Meibom}, {Prsa}, {Quintana}, {Rowe}, {Sherry},
  {Tenenbaum}, {Torres}, {Twicken}, {Van Cleve}, {Walkowicz}, \&
  {Wu}}]{borucki11a}
{Borucki}, W.~J., {Koch}, D.~G., {Basri}, G., {Batalha}, N., {Boss}, A.,
  {Brown}, T.~M., {Caldwell}, D., {Christensen-Dalsgaard}, J., {Cochran},
  W.~D., {DeVore}, E., {Dunham}, E.~W., {Dupree}, A.~K., {Gautier}, III, T.~N.,
  {Geary}, J.~C., {Gilliland}, R., {Gould}, A., {Howell}, S.~B., {Jenkins},
  J.~M., {Kjeldsen}, H., {Latham}, D.~W., {Lissauer}, J.~J., {Marcy}, G.~W.,
  {Monet}, D.~G., {Sasselov}, D., {Tarter}, J., {Charbonneau}, D., {Doyle}, L.,
  {Ford}, E.~B., {Fortney}, J., {Holman}, M.~J., {Seager}, S., {Steffen},
  J.~H., {Welsh}, W.~F., {Allen}, C., {Bryson}, S.~T., {Buchhave}, L.,
  {Chandrasekaran}, H., {Christiansen}, J.~L., {Ciardi}, D., {Clarke}, B.~D.,
  {Dotson}, J.~L., {Endl}, M., {Fischer}, D., {Fressin}, F., {Haas}, M.,
  {Horch}, E., {Howard}, A., {Isaacson}, H., {Kolodziejczak}, J., {Li}, J.,
  {MacQueen}, P., {Meibom}, S., {Prsa}, A., {Quintana}, E.~V., {Rowe}, J.,
  {Sherry}, W., {Tenenbaum}, P., {Torres}, G., {Twicken}, J.~D., {Van Cleve},
  J., {Walkowicz}, L., \& {Wu}, H. 2011{\natexlab{a}}, \apj, 728, 117

\bibitem[{{Borucki} {et~al.}(2011{\natexlab{b}}){Borucki}, {Koch}, {Basri},
  {Batalha}, {Brown}, {Bryson}, {Caldwell}, {Christensen-Dalsgaard}, {Cochran},
  {DeVore}, {Dunham}, {Gautier}, {Geary}, {Gilliland}, {Gould}, {Howell},
  {Jenkins}, {Latham}, {Lissauer}, {Marcy}, {Rowe}, {Sasselov}, {Boss},
  {Charbonneau}, {Ciardi}, {Doyle}, {Dupree}, {Ford}, {Fortney}, {Holman},
  {Seager}, {Steffen}, {Tarter}, {Welsh}, {Allen}, {Buchhave}, {Christiansen},
  {Clarke}, {Das}, {D{\'e}sert}, {Endl}, {Fabrycky}, {Fressin}, {Haas},
  {Horch}, {Howard}, {Isaacson}, {Kjeldsen}, {Kolodziejczak}, {Kulesa}, {Li},
  {Lucas}, {Machalek}, {McCarthy}, {MacQueen}, {Meibom}, {Miquel}, {Prsa},
  {Quinn}, {Quintana}, {Ragozzine}, {Sherry}, {Shporer}, {Tenenbaum}, {Torres},
  {Twicken}, {Van Cleve}, {Walkowicz}, {Witteborn}, \& {Still}}]{borucki11b}
{Borucki}, W.~J., {Koch}, D.~G., {Basri}, G., {Batalha}, N., {Brown}, T.~M.,
  {Bryson}, S.~T., {Caldwell}, D., {Christensen-Dalsgaard}, J., {Cochran},
  W.~D., {DeVore}, E., {Dunham}, E.~W., {Gautier}, III, T.~N., {Geary}, J.~C.,
  {Gilliland}, R., {Gould}, A., {Howell}, S.~B., {Jenkins}, J.~M., {Latham},
  D.~W., {Lissauer}, J.~J., {Marcy}, G.~W., {Rowe}, J., {Sasselov}, D., {Boss},
  A., {Charbonneau}, D., {Ciardi}, D., {Doyle}, L., {Dupree}, A.~K., {Ford},
  E.~B., {Fortney}, J., {Holman}, M.~J., {Seager}, S., {Steffen}, J.~H.,
  {Tarter}, J., {Welsh}, W.~F., {Allen}, C., {Buchhave}, L.~A., {Christiansen},
  J.~L., {Clarke}, B.~D., {Das}, S., {D{\'e}sert}, J.-M., {Endl}, M.,
  {Fabrycky}, D., {Fressin}, F., {Haas}, M., {Horch}, E., {Howard}, A.,
  {Isaacson}, H., {Kjeldsen}, H., {Kolodziejczak}, J., {Kulesa}, C., {Li}, J.,
  {Lucas}, P.~W., {Machalek}, P., {McCarthy}, D., {MacQueen}, P., {Meibom}, S.,
  {Miquel}, T., {Prsa}, A., {Quinn}, S.~N., {Quintana}, E.~V., {Ragozzine}, D.,
  {Sherry}, W., {Shporer}, A., {Tenenbaum}, P., {Torres}, G., {Twicken}, J.~D.,
  {Van Cleve}, J., {Walkowicz}, L., {Witteborn}, F.~C., \& {Still}, M.
  2011{\natexlab{b}}, \apj, 736, 19

\bibitem[{{Borucki} {et~al.}(2011{\natexlab{c}}){Borucki}, {Koch}, {Basri},
  {Batalha}, {Brown}, {Bryson}, {Caldwell}, {Christensen-Dalsgaard}, {Cochran},
  {DeVore}, {Dunham}, {Gautier}, {Geary}, {Gilliland}, {Gould}, {Howell},
  {Jenkins}, {Latham}, {Lissauer}, {Marcy}, {Rowe}, {Sasselov}, {Boss},
  {Charbonneau}, {Ciardi}, {Doyle}, {Dupree}, {Ford}, {Fortney}, {Holman},
  {Seager}, {Steffen}, {Tarter}, {Welsh}, {Allen}, {Buchhave}, {Christiansen},
  {Clarke}, {Das}, {D{\'e}sert}, {Endl}, {Fabrycky}, {Fressin}, {Haas},
  {Horch}, {Howard}, {Isaacson}, {Kjeldsen}, {Kolodziejczak}, {Kulesa}, {Li},
  {Lucas}, {Machalek}, {McCarthy}, {MacQueen}, {Meibom}, {Miquel}, {Prsa},
  {Quinn}, {Quintana}, {Ragozzine}, {Sherry}, {Shporer}, {Tenenbaum}, {Torres},
  {Twicken}, {Van Cleve}, {Walkowicz}, {Witteborn}, \& {Still}}]{borucki11}
---. 2011{\natexlab{c}}, \apj, 736, 19

\bibitem[{{Borucki} {et~al.}(2010){Borucki}, {Koch}, {Brown}, {Basri},
  {Batalha}, {Caldwell}, {Cochran}, {Dunham}, {Gautier}, {Geary}, {Gilliland},
  {Howell}, {Jenkins}, {Latham}, {Lissauer}, {Marcy}, {Monet}, {Rowe}, \&
  {Sasselov}}]{borucki10}
{Borucki}, W.~J., {Koch}, D.~G., {Brown}, T.~M., {Basri}, G., {Batalha}, N.~M.,
  {Caldwell}, D.~A., {Cochran}, W.~D., {Dunham}, E.~W., {Gautier}, III, T.~N.,
  {Geary}, J.~C., {Gilliland}, R.~L., {Howell}, S.~B., {Jenkins}, J.~M.,
  {Latham}, D.~W., {Lissauer}, J.~J., {Marcy}, G.~W., {Monet}, D., {Rowe},
  J.~F., \& {Sasselov}, D. 2010, \apjl, 713, L126

\bibitem[{{Brown} {et~al.}(2011){Brown}, {Latham}, {Everett}, \&
  {Esquerdo}}]{brown11}
{Brown}, T.~M., {Latham}, D.~W., {Everett}, M.~E., \& {Esquerdo}, G.~A. 2011,
  \aj, 142, 112

\bibitem[{{Burgasser} {et~al.}(2003){Burgasser}, {Kirkpatrick}, {Reid},
  {Brown}, {Miskey}, \& {Gizis}}]{burgasser03}
{Burgasser}, A.~J., {Kirkpatrick}, J.~D., {Reid}, I.~N., {Brown}, M.~E.,
  {Miskey}, C.~L., \& {Gizis}, J.~E. 2003, \apj, 586, 512

\bibitem[{{Burke} {et~al.}(2014){Burke}, {Bryson}, {Mullally}, {Rowe},
  {Christiansen}, {Thompson}, {Coughlin}, {Haas}, {Batalha}, {Caldwell},
  {Jenkins}, {Still}, {Barclay}, {Borucki}, {Chaplin}, {Ciardi}, {Clarke},
  {Cochran}, {Demory}, {Esquerdo}, {Gautier}, {Gilliland}, {Girouard}, {Havel},
  {Henze}, {Howell}, {Huber}, {Latham}, {Li}, {Morehead}, {Morton}, {Pepper},
  {Quintana}, {Ragozzine}, {Seader}, {Shah}, {Shporer}, {Tenenbaum}, {Twicken},
  \& {Wolfgang}}]{burke14}
{Burke}, C.~J., {Bryson}, S.~T., {Mullally}, F., {Rowe}, J.~F., {Christiansen},
  J.~L., {Thompson}, S.~E., {Coughlin}, J.~L., {Haas}, M.~R., {Batalha}, N.~M.,
  {Caldwell}, D.~A., {Jenkins}, J.~M., {Still}, M., {Barclay}, T., {Borucki},
  W.~J., {Chaplin}, W.~J., {Ciardi}, D.~R., {Clarke}, B.~D., {Cochran}, W.~D.,
  {Demory}, B.-O., {Esquerdo}, G.~A., {Gautier}, III, T.~N., {Gilliland},
  R.~L., {Girouard}, F.~R., {Havel}, M., {Henze}, C.~E., {Howell}, S.~B.,
  {Huber}, D., {Latham}, D.~W., {Li}, J., {Morehead}, R.~C., {Morton}, T.~D.,
  {Pepper}, J., {Quintana}, E., {Ragozzine}, D., {Seader}, S.~E., {Shah}, Y.,
  {Shporer}, A., {Tenenbaum}, P., {Twicken}, J.~D., \& {Wolfgang}, A. 2014,
  \apjs, 210, 19

\bibitem[{{Ciardi} {et~al.}(2015){Ciardi}, {Beichman}, {Horch}, \&
  {Howell}}]{ciardi15}
{Ciardi}, D.~R., {Beichman}, C.~A., {Horch}, E.~P., \& {Howell}, S.~B. 2015,
  \apj, 805, 16

\bibitem[{{Ciardi} {et~al.}(2011){Ciardi}, {von Braun}, {Bryden}, {van Eyken},
  {Howell}, {Kane}, {Plavchan}, {Ram{\'{\i}}rez}, \& {Stauffer}}]{ciardi11}
{Ciardi}, D.~R., {von Braun}, K., {Bryden}, G., {van Eyken}, J., {Howell},
  S.~B., {Kane}, S.~R., {Plavchan}, P., {Ram{\'{\i}}rez}, S.~V., \& {Stauffer},
  J.~R. 2011, \aj, 141, 108

\bibitem[{{Coughlin} {et~al.}(2015){Coughlin}, {Mullally}, {Thompson}, {Rowe},
  {Burke}, {Latham}, {Batalha}, {Ofir}, {Quarles}, {Henze}, {Wolfgang},
  {Caldwell}, {Bryson}, {Shporer}, {Catanzarite}, {Akeson}, {Barclay},
  {Borucki}, {Boyajian}, {Campbell}, {Christiansen}, {Girouard}, {Haas},
  {Howell}, {Huber}, {Jenkins}, {Li}, {Patil-Sabale}, {Quintana}, {Ramirez},
  {Seader}, {Smith}, {Tenenbaum}, {Twicken}, \& {Zamudio}}]{coughlin15}
{Coughlin}, J.~L., {Mullally}, F., {Thompson}, S.~E., {Rowe}, J.~F., {Burke},
  C.~J., {Latham}, D.~W., {Batalha}, N.~M., {Ofir}, A., {Quarles}, B.~L.,
  {Henze}, C.~E., {Wolfgang}, A., {Caldwell}, D.~A., {Bryson}, S.~T.,
  {Shporer}, A., {Catanzarite}, J., {Akeson}, R., {Barclay}, T., {Borucki},
  W.~J., {Boyajian}, T.~S., {Campbell}, J.~R., {Christiansen}, J.~L.,
  {Girouard}, F.~R., {Haas}, M.~R., {Howell}, S.~B., {Huber}, D., {Jenkins},
  J.~M., {Li}, J., {Patil-Sabale}, A., {Quintana}, E.~V., {Ramirez}, S.,
  {Seader}, S., {Smith}, J.~C., {Tenenbaum}, P., {Twicken}, J.~D., \&
  {Zamudio}, K.~A. 2015, ArXiv e-prints

\bibitem[{{D{\'e}sert} {et~al.}(2015){D{\'e}sert}, {Charbonneau}, {Torres},
  {Fressin}, {Ballard}, {Bryson}, {Knutson}, {Batalha}, {Borucki}, {Brown},
  {Deming}, {Ford}, {Fortney}, {Gilliland}, {Latham}, \& {Seager}}]{desert15}
{D{\'e}sert}, J.-M., {Charbonneau}, D., {Torres}, G., {Fressin}, F., {Ballard},
  S., {Bryson}, S.~T., {Knutson}, H.~A., {Batalha}, N.~M., {Borucki}, W.~J.,
  {Brown}, T.~M., {Deming}, D., {Ford}, E.~B., {Fortney}, J.~J., {Gilliland},
  R.~L., {Latham}, D.~W., \& {Seager}, S. 2015, \apj, 804, 59

\bibitem[{{Dressing} {et~al.}(2014){Dressing}, {Adams}, {Dupree}, {Kulesa}, \&
  {McCarthy}}]{dressing14}
{Dressing}, C.~D., {Adams}, E.~R., {Dupree}, A.~K., {Kulesa}, C., \&
  {McCarthy}, D. 2014, \aj, 148, 78

\bibitem[{{Dressing} \& {Charbonneau}(2013)}]{dressing13}
{Dressing}, C.~D. \& {Charbonneau}, D. 2013, \apj, 767, 95

\bibitem[{{Duch{\^e}ne} \& {Kraus}(2013)}]{duchene13}
{Duch{\^e}ne}, G. \& {Kraus}, A. 2013, \araa, 51, 269

\bibitem[{{Duquennoy} \& {Mayor}(1991)}]{duquennoy91}
{Duquennoy}, A. \& {Mayor}, M. 1991, \aap, 248, 485

\bibitem[{{Evans} {et~al.}(2016){Evans}, {Southworth}, {Maxted}, {Skottfelt},
  {Hundertmark}, {J{\o}rgensen}, {Dominik}, {Alsubai}, {Andersen}, {Bozza},
  {Bramich}, {Burgdorf}, {Ciceri}, {D'Ago}, {Figuera Jaimes}, {Gu},
  {Haugb{\o}lle}, {Hinse}, {Juncher}, {Kains}, {Kerins}, {Korhonen},
  {Kuffmeier}, {Peixinho}, {Popovas}, {Rabus}, {Rahvar}, {Schmidt},
  {Snodgrass}, {Starkey}, {Surdej}, {Tronsgaard}, {von Essen}, {Wang}, \&
  {Wertz}}]{evans16}
{Evans}, D.~F., {Southworth}, J., {Maxted}, P.~F.~L., {Skottfelt}, J.,
  {Hundertmark}, M., {J{\o}rgensen}, U.~G., {Dominik}, M., {Alsubai}, K.~A.,
  {Andersen}, M.~I., {Bozza}, V., {Bramich}, D.~M., {Burgdorf}, M.~J.,
  {Ciceri}, S., {D'Ago}, G., {Figuera Jaimes}, R., {Gu}, S.~H., {Haugb{\o}lle},
  T., {Hinse}, T.~C., {Juncher}, D., {Kains}, N., {Kerins}, E., {Korhonen}, H.,
  {Kuffmeier}, M., {Peixinho}, N., {Popovas}, A., {Rabus}, M., {Rahvar}, S.,
  {Schmidt}, R.~W., {Snodgrass}, C., {Starkey}, D., {Surdej}, J., {Tronsgaard},
  R., {von Essen}, C., {Wang}, Y.-B., \& {Wertz}, O. 2016, ArXiv e-prints

\bibitem[{{Everett} {et~al.}(2015){Everett}, {Barclay}, {Ciardi}, {Horch},
  {Howell}, {Crepp}, \& {Silva}}]{everett15}
{Everett}, M.~E., {Barclay}, T., {Ciardi}, D.~R., {Horch}, E.~P., {Howell},
  S.~B., {Crepp}, J.~R., \& {Silva}, D.~R. 2015, \aj, 149, 55

\bibitem[{{Fabrycky} \& {Tremaine}(2007)}]{fabrycky07}
{Fabrycky}, D. \& {Tremaine}, S. 2007, \apj, 669, 1298

\bibitem[{{Fabrycky} {et~al.}(2012){Fabrycky}, {Ford}, {Steffen}, {Rowe},
  {Carter}, {Moorhead}, {Batalha}, {Borucki}, {Bryson}, {Buchhave},
  {Christiansen}, {Ciardi}, {Cochran}, {Endl}, {Fanelli}, {Fischer}, {Fressin},
  {Geary}, {Haas}, {Hall}, {Holman}, {Jenkins}, {Koch}, {Latham}, {Li},
  {Lissauer}, {Lucas}, {Marcy}, {Mazeh}, {McCauliff}, {Quinn}, {Ragozzine},
  {Sasselov}, \& {Shporer}}]{fabrycky12}
{Fabrycky}, D.~C., {Ford}, E.~B., {Steffen}, J.~H., {Rowe}, J.~F., {Carter},
  J.~A., {Moorhead}, A.~V., {Batalha}, N.~M., {Borucki}, W.~J., {Bryson}, S.,
  {Buchhave}, L.~A., {Christiansen}, J.~L., {Ciardi}, D.~R., {Cochran}, W.~D.,
  {Endl}, M., {Fanelli}, M.~N., {Fischer}, D., {Fressin}, F., {Geary}, J.,
  {Haas}, M.~R., {Hall}, J.~R., {Holman}, M.~J., {Jenkins}, J.~M., {Koch},
  D.~G., {Latham}, D.~W., {Li}, J., {Lissauer}, J.~J., {Lucas}, P., {Marcy},
  G.~W., {Mazeh}, T., {McCauliff}, S., {Quinn}, S., {Ragozzine}, D.,
  {Sasselov}, D., \& {Shporer}, A. 2012, \apj, 750, 114

\bibitem[{{Fragner} {et~al.}(2011){Fragner}, {Nelson}, \& {Kley}}]{fragner11}
{Fragner}, M.~M., {Nelson}, R.~P., \& {Kley}, W. 2011, \aap, 528, A40

\bibitem[{{Fressin} {et~al.}(2013){Fressin}, {Torres}, {Charbonneau}, {Bryson},
  {Christiansen}, {Dressing}, {Jenkins}, {Walkowicz}, \& {Batalha}}]{fressin13}
{Fressin}, F., {Torres}, G., {Charbonneau}, D., {Bryson}, S.~T.,
  {Christiansen}, J., {Dressing}, C.~D., {Jenkins}, J.~M., {Walkowicz}, L.~M.,
  \& {Batalha}, N.~M. 2013, \apj, 766, 81

\bibitem[{{Fruchter} \& {Hook}(2002)}]{fruchter02}
{Fruchter}, A.~S. \& {Hook}, R.~N. 2002, \pasp, 114, 144

\bibitem[{{Gilliland} {et~al.}(2015){Gilliland}, {Cartier}, {Adams}, {Ciardi},
  {Kalas}, \& {Wright}}]{gilliland15}
{Gilliland}, R.~L., {Cartier}, K.~M.~S., {Adams}, E.~R., {Ciardi}, D.~R.,
  {Kalas}, P., \& {Wright}, J.~T. 2015, \aj, 149, 24

\bibitem[{{Gross} \& {Vitells}(2010)}]{gross10}
{Gross}, E. \& {Vitells}, O. 2010, European Physical Journal C, 70, 525

\bibitem[{{Haas} {et~al.}(2010){Haas}, {Batalha}, {Bryson}, {Caldwell},
  {Dotson}, {Hall}, {Jenkins}, {Klaus}, {Koch}, {Kolodziejczak}, {Middour},
  {Smith}, {Sobeck}, {Stober}, {Thompson}, \& {Van Cleve}}]{haas10}
{Haas}, M.~R., {Batalha}, N.~M., {Bryson}, S.~T., {Caldwell}, D.~A., {Dotson},
  J.~L., {Hall}, J., {Jenkins}, J.~M., {Klaus}, T.~C., {Koch}, D.~G.,
  {Kolodziejczak}, J., {Middour}, C., {Smith}, M., {Sobeck}, C.~K., {Stober},
  J., {Thompson}, R.~S., \& {Van Cleve}, J.~E. 2010, \apjl, 713, L115

\bibitem[{{Heller}(2012)}]{heller12}
{Heller}, R. 2012, \aap, 545, L8

\bibitem[{{Hodapp} {et~al.}(2003){Hodapp}, {Jensen}, {Irwin}, {Yamada},
  {Chung}, {Fletcher}, {Robertson}, {Hora}, {Simons}, {Mays}, {Nolan}, {Bec},
  {Merrill}, \& {Fowler}}]{hodapp03}
{Hodapp}, K.~W., {Jensen}, J.~B., {Irwin}, E.~M., {Yamada}, H., {Chung}, R.,
  {Fletcher}, K., {Robertson}, L., {Hora}, J.~L., {Simons}, D.~A., {Mays}, W.,
  {Nolan}, R., {Bec}, M., {Merrill}, M., \& {Fowler}, A.~M. 2003, \pasp, 115,
  1388

\bibitem[{{Horch} {et~al.}(2012){Horch}, {Howell}, {Everett}, \&
  {Ciardi}}]{horch12}
{Horch}, E.~P., {Howell}, S.~B., {Everett}, M.~E., \& {Ciardi}, D.~R. 2012,
  \aj, 144, 165

\bibitem[{{Horch} {et~al.}(2014){Horch}, {Howell}, {Everett}, \&
  {Ciardi}}]{horch14}
---. 2014, \apj, 795, 60

\bibitem[{{Howell} {et~al.}(2011){Howell}, {Everett}, {Sherry}, {Horch}, \&
  {Ciardi}}]{howell11}
{Howell}, S.~B., {Everett}, M.~E., {Sherry}, W., {Horch}, E., \& {Ciardi},
  D.~R. 2011, \aj, 142, 19

\bibitem[{{Howell} {et~al.}(2014){Howell}, {Sobeck}, {Haas}, {Still},
  {Barclay}, {Mullally}, {Troeltzsch}, {Aigrain}, {Bryson}, {Caldwell},
  {Chaplin}, {Cochran}, {Huber}, {Marcy}, {Miglio}, {Najita}, {Smith},
  {Twicken}, \& {Fortney}}]{K2}
{Howell}, S.~B., {Sobeck}, C., {Haas}, M., {Still}, M., {Barclay}, T.,
  {Mullally}, F., {Troeltzsch}, J., {Aigrain}, S., {Bryson}, S.~T., {Caldwell},
  D., {Chaplin}, W.~J., {Cochran}, W.~D., {Huber}, D., {Marcy}, G.~W.,
  {Miglio}, A., {Najita}, J.~R., {Smith}, M., {Twicken}, J.~D., \& {Fortney},
  J.~J. 2014, \pasp, 126, 398

\bibitem[{{Kasting} {et~al.}(1993){Kasting}, {Whitmire}, \&
  {Reynolds}}]{kasting93}
{Kasting}, J.~F., {Whitmire}, D.~P., \& {Reynolds}, R.~T. 1993, \icarus, 101,
  108

\bibitem[{{Katz} {et~al.}(2011){Katz}, {Dong}, \& {Malhotra}}]{katz11}
{Katz}, B., {Dong}, S., \& {Malhotra}, R. 2011, Physical Review Letters, 107,
  181101

\bibitem[{{Kolbl} {et~al.}(2015){Kolbl}, {Marcy}, {Isaacson}, \&
  {Howard}}]{kolbl15}
{Kolbl}, R., {Marcy}, G.~W., {Isaacson}, H., \& {Howard}, A.~W. 2015, \aj, 149,
  18

\bibitem[{{Kraus} \& {Hillenbrand}(2007)}]{kraus07}
{Kraus}, A.~L. \& {Hillenbrand}, L.~A. 2007, \aj, 134, 2340

\bibitem[{{Kraus} {et~al.}(2016){Kraus}, {Ireland}, {Huber}, {Mann}, \&
  {Dupuy}}]{kraus16}
{Kraus}, A.~L., {Ireland}, M.~J., {Huber}, D., {Mann}, A.~W., \& {Dupuy}, T.~J.
  2016, \aj, 152, 8

\bibitem[{{Lafreni{\`e}re} {et~al.}(2007){Lafreni{\`e}re}, {Marois}, {Doyon},
  {Nadeau}, \& {Artigau}}]{lafreniere07}
{Lafreni{\`e}re}, D., {Marois}, C., {Doyon}, R., {Nadeau}, D., \& {Artigau},
  {\'E}. 2007, \apj, 660, 770

\bibitem[{{Law} {et~al.}(2015){Law}, {Fors}, {Ratzloff}, {Wulfken},
  {Kavanaugh}, {Sitar}, {Pruett}, {Birchard}, {Barlow}, {Cannon}, {Cenko},
  {Dunlap}, {Kraus}, \& {Maccarone}}]{evryscope}
{Law}, N.~M., {Fors}, O., {Ratzloff}, J., {Wulfken}, P., {Kavanaugh}, D.,
  {Sitar}, D.~J., {Pruett}, Z., {Birchard}, M.~N., {Barlow}, B.~N., {Cannon},
  K., {Cenko}, S.~B., {Dunlap}, B., {Kraus}, A., \& {Maccarone}, T.~J. 2015,
  \pasp, 127, 234

\bibitem[{{Law} {et~al.}(2009){Law}, {Mackay}, {Dekany}, {Ireland}, {Lloyd},
  {Moore}, {Robertson}, {Tuthill}, \& {Woodruff}}]{law09}
{Law}, N.~M., {Mackay}, C.~D., {Dekany}, R.~G., {Ireland}, M., {Lloyd}, J.~P.,
  {Moore}, A.~M., {Robertson}, J.~G., {Tuthill}, P., \& {Woodruff}, H.~C. 2009,
  \apj, 692, 924

\bibitem[{{Law} {et~al.}(2014){Law}, {Morton}, {Baranec}, {Riddle},
  {Ravichandran}, {Ziegler}, {Johnson}, {Tendulkar}, {Bui}, {Burse}, {Das},
  {Dekany}, {Kulkarni}, {Punnadi}, \& {Ramaprakash}}]{law14}
{Law}, N.~M., {Morton}, T., {Baranec}, C., {Riddle}, R., {Ravichandran}, G.,
  {Ziegler}, C., {Johnson}, J.~A., {Tendulkar}, S.~P., {Bui}, K., {Burse},
  M.~P., {Das}, H.~K., {Dekany}, R.~G., {Kulkarni}, S., {Punnadi}, S., \&
  {Ramaprakash}, A.~N. 2014, \apj, 791, 35

\bibitem[{{Lawrence} {et~al.}(2007){Lawrence}, {Warren}, {Almaini}, {Edge},
  {Hambly}, {Jameson}, {Lucas}, {Casali}, {Adamson}, {Dye}, {Emerson},
  {Foucaud}, {Hewett}, {Hirst}, {Hodgkin}, {Irwin}, {Lodieu}, {McMahon},
  {Simpson}, {Smail}, {Mortlock}, \& {Folger}}]{lawrence07}
{Lawrence}, A., {Warren}, S.~J., {Almaini}, O., {Edge}, A.~C., {Hambly}, N.~C.,
  {Jameson}, R.~F., {Lucas}, P., {Casali}, M., {Adamson}, A., {Dye}, S.,
  {Emerson}, J.~P., {Foucaud}, S., {Hewett}, P., {Hirst}, P., {Hodgkin}, S.~T.,
  {Irwin}, M.~J., {Lodieu}, N., {McMahon}, R.~G., {Simpson}, C., {Smail}, I.,
  {Mortlock}, D., \& {Folger}, M. 2007, \mnras, 379, 1599

\bibitem[{{Lillo-Box} {et~al.}(2012){Lillo-Box}, {Barrado}, \&
  {Bouy}}]{lillo12}
{Lillo-Box}, J., {Barrado}, D., \& {Bouy}, H. 2012, \aap, 546, A10

\bibitem[{{Lillo-Box} {et~al.}(2014){Lillo-Box}, {Barrado}, \&
  {Bouy}}]{lillo14}
---. 2014, \aap, 566, A103

\bibitem[{{Lissauer} {et~al.}(2014){Lissauer}, {Dawson}, \&
  {Tremaine}}]{lissauer14}
{Lissauer}, J.~J., {Dawson}, R.~I., \& {Tremaine}, S. 2014, \nat, 513, 336

\bibitem[{{Marcy} {et~al.}(2014){Marcy}, {Isaacson}, {Howard}, {Rowe},
  {Jenkins}, {Bryson}, {Latham}, {Howell}, {Gautier}, {Batalha}, {Rogers},
  {Ciardi}, {Fischer}, {Gilliland}, {Kjeldsen}, {Christensen-Dalsgaard},
  {Huber}, {Chaplin}, {Basu}, {Buchhave}, {Quinn}, {Borucki}, {Koch}, {Hunter},
  {Caldwell}, {Van Cleve}, {Kolbl}, {Weiss}, {Petigura}, {Seager}, {Morton},
  {Johnson}, {Ballard}, {Burke}, {Cochran}, {Endl}, {MacQueen}, {Everett},
  {Lissauer}, {Ford}, {Torres}, {Fressin}, {Brown}, {Steffen}, {Charbonneau},
  {Basri}, {Sasselov}, {Winn}, {Sanchis-Ojeda}, {Christiansen}, {Adams},
  {Henze}, {Dupree}, {Fabrycky}, {Fortney}, {Tarter}, {Holman}, {Tenenbaum},
  {Shporer}, {Lucas}, {Welsh}, {Orosz}, {Bedding}, {Campante}, {Davies},
  {Elsworth}, {Handberg}, {Hekker}, {Karoff}, {Kawaler}, {Lund}, {Lundkvist},
  {Metcalfe}, {Miglio}, {Silva Aguirre}, {Stello}, {White}, {Boss}, {Devore},
  {Gould}, {Prsa}, {Agol}, {Barclay}, {Coughlin}, {Brugamyer}, {Mullally},
  {Quintana}, {Still}, {Thompson}, {Morrison}, {Twicken}, {D{\'e}sert},
  {Carter}, {Crepp}, {H{\'e}brard}, {Santerne}, {Moutou}, {Sobeck}, {Hudgins},
  {Haas}, {Robertson}, {Lillo-Box}, \& {Barrado}}]{marcy14}
{Marcy}, G.~W., {Isaacson}, H., {Howard}, A.~W., {Rowe}, J.~F., {Jenkins},
  J.~M., {Bryson}, S.~T., {Latham}, D.~W., {Howell}, S.~B., {Gautier}, III,
  T.~N., {Batalha}, N.~M., {Rogers}, L., {Ciardi}, D., {Fischer}, D.~A.,
  {Gilliland}, R.~L., {Kjeldsen}, H., {Christensen-Dalsgaard}, J., {Huber}, D.,
  {Chaplin}, W.~J., {Basu}, S., {Buchhave}, L.~A., {Quinn}, S.~N., {Borucki},
  W.~J., {Koch}, D.~G., {Hunter}, R., {Caldwell}, D.~A., {Van Cleve}, J.,
  {Kolbl}, R., {Weiss}, L.~M., {Petigura}, E., {Seager}, S., {Morton}, T.,
  {Johnson}, J.~A., {Ballard}, S., {Burke}, C., {Cochran}, W.~D., {Endl}, M.,
  {MacQueen}, P., {Everett}, M.~E., {Lissauer}, J.~J., {Ford}, E.~B., {Torres},
  G., {Fressin}, F., {Brown}, T.~M., {Steffen}, J.~H., {Charbonneau}, D.,
  {Basri}, G.~S., {Sasselov}, D.~D., {Winn}, J., {Sanchis-Ojeda}, R.,
  {Christiansen}, J., {Adams}, E., {Henze}, C., {Dupree}, A., {Fabrycky},
  D.~C., {Fortney}, J.~J., {Tarter}, J., {Holman}, M.~J., {Tenenbaum}, P.,
  {Shporer}, A., {Lucas}, P.~W., {Welsh}, W.~F., {Orosz}, J.~A., {Bedding},
  T.~R., {Campante}, T.~L., {Davies}, G.~R., {Elsworth}, Y., {Handberg}, R.,
  {Hekker}, S., {Karoff}, C., {Kawaler}, S.~D., {Lund}, M.~N., {Lundkvist}, M.,
  {Metcalfe}, T.~S., {Miglio}, A., {Silva Aguirre}, V., {Stello}, D., {White},
  T.~R., {Boss}, A., {Devore}, E., {Gould}, A., {Prsa}, A., {Agol}, E.,
  {Barclay}, T., {Coughlin}, J., {Brugamyer}, E., {Mullally}, F., {Quintana},
  E.~V., {Still}, M., {Thompson}, S.~E., {Morrison}, D., {Twicken}, J.~D.,
  {D{\'e}sert}, J.-M., {Carter}, J., {Crepp}, J.~R., {H{\'e}brard}, G.,
  {Santerne}, A., {Moutou}, C., {Sobeck}, C., {Hudgins}, D., {Haas}, M.~R.,
  {Robertson}, P., {Lillo-Box}, J., \& {Barrado}, D. 2014, \apjs, 210, 20

\bibitem[{{McCullough} {et~al.}(2005){McCullough}, {Stys}, {Valenti},
  {Fleming}, {Janes}, \& {Heasley}}]{xo}
{McCullough}, P.~R., {Stys}, J.~E., {Valenti}, J.~A., {Fleming}, S.~W.,
  {Janes}, K.~A., \& {Heasley}, J.~N. 2005, \pasp, 117, 783

\bibitem[{Morton {et~al.}(2016)Morton, Bryson, Coughlin, Rowe, Ravichandran,
  Petigura, Haas, \& Batalha}]{morton16}
Morton, T.~D., Bryson, S.~T., Coughlin, J.~L., Rowe, J.~F., Ravichandran, G.,
  Petigura, E.~A., Haas, M.~R., \& Batalha, N.~M. 2016, The Astrophysical
  Journal, 822, 86

\bibitem[{{Morton} \& {Johnson}(2011)}]{morton11}
{Morton}, T.~D. \& {Johnson}, J.~A. 2011, \apj, 738, 170

\bibitem[{{Mullally} {et~al.}(2015){Mullally}, {Coughlin}, {Thompson}, {Rowe},
  {Burke}, {Latham}, {Batalha}, {Bryson}, {Christiansen}, {Henze}, {Ofir},
  {Quarles}, {Shporer}, {Van Eylen}, {Van Laerhoven}, {Shah}, {Wolfgang},
  {Chaplin}, {Xie}, {Akeson}, {Argabright}, {Bachtell}, {Barclay}, {Borucki},
  {Caldwell}, {Campbell}, {Catanzarite}, {Cochran}, {Duren}, {Fleming},
  {Fraquelli}, {Girouard}, {Haas}, {He{\l}miniak}, {Howell}, {Huber}, {Larson},
  {Gautier}, {Jenkins}, {Li}, {Lissauer}, {McArthur}, {Miller}, {Morris},
  {Patil-Sabale}, {Plavchan}, {Putnam}, {Quintana}, {Ramirez}, {Silva Aguirre},
  {Seader}, {Smith}, {Steffen}, {Stewart}, {Stober}, {Still}, {Tenenbaum},
  {Troeltzsch}, {Twicken}, \& {Zamudio}}]{mullally15}
{Mullally}, F., {Coughlin}, J.~L., {Thompson}, S.~E., {Rowe}, J., {Burke}, C.,
  {Latham}, D.~W., {Batalha}, N.~M., {Bryson}, S.~T., {Christiansen}, J.,
  {Henze}, C.~E., {Ofir}, A., {Quarles}, B., {Shporer}, A., {Van Eylen}, V.,
  {Van Laerhoven}, C., {Shah}, Y., {Wolfgang}, A., {Chaplin}, W.~J., {Xie},
  J.-W., {Akeson}, R., {Argabright}, V., {Bachtell}, E., {Barclay}, T.,
  {Borucki}, W.~J., {Caldwell}, D.~A., {Campbell}, J.~R., {Catanzarite}, J.~H.,
  {Cochran}, W.~D., {Duren}, R.~M., {Fleming}, S.~W., {Fraquelli}, D.,
  {Girouard}, F.~R., {Haas}, M.~R., {He{\l}miniak}, K.~G., {Howell}, S.~B.,
  {Huber}, D., {Larson}, K., {Gautier}, III, T.~N., {Jenkins}, J.~M., {Li}, J.,
  {Lissauer}, J.~J., {McArthur}, S., {Miller}, C., {Morris}, R.~L.,
  {Patil-Sabale}, A., {Plavchan}, P., {Putnam}, D., {Quintana}, E.~V.,
  {Ramirez}, S., {Silva Aguirre}, V., {Seader}, S., {Smith}, J.~C., {Steffen},
  J.~H., {Stewart}, C., {Stober}, J., {Still}, M., {Tenenbaum}, P.,
  {Troeltzsch}, J., {Twicken}, J.~D., \& {Zamudio}, K.~A. 2015, \apjs, 217, 31

\bibitem[{{Mullally}(2016)}]{k2fov}
{Mullally}, Fergal;~{Barclay}, T. B.~G. 2016, {K2fov: Field of view software
  for NASA's K2 mission}, Astrophysics Source Code Library

\bibitem[{{Naoz} {et~al.}(2012){Naoz}, {Farr}, \& {Rasio}}]{naox12}
{Naoz}, S., {Farr}, W.~M., \& {Rasio}, F.~A. 2012, \apjl, 754, L36

\bibitem[{{Ngo} {et~al.}(2015){Ngo}, {Knutson}, {Hinkley}, {Crepp}, {Bechter},
  {Batygin}, {Howard}, {Johnson}, {Morton}, \& {Muirhead}}]{ngo15}
{Ngo}, H., {Knutson}, H.~A., {Hinkley}, S., {Crepp}, J.~R., {Bechter}, E.~B.,
  {Batygin}, K., {Howard}, A.~W., {Johnson}, J.~A., {Morton}, T.~D., \&
  {Muirhead}, P.~S. 2015, \apj, 800, 138

\bibitem[{{Nutzman} \& {Charbonneau}(2008)}]{mearth}
{Nutzman}, P. \& {Charbonneau}, D. 2008, \pasp, 120, 317

\bibitem[{{Pepper} {et~al.}(2012){Pepper}, {Kuhn}, {Siverd}, {James}, \&
  {Stassun}}]{kelt2}
{Pepper}, J., {Kuhn}, R.~B., {Siverd}, R., {James}, D., \& {Stassun}, K. 2012,
  \pasp, 124, 230

\bibitem[{{Pepper} {et~al.}(2007){Pepper}, {Pogge}, {DePoy}, {Marshall},
  {Stanek}, {Stutz}, {Poindexter}, {Siverd}, {O'Brien}, {Trueblood}, \&
  {Trueblood}}]{kelt1}
{Pepper}, J., {Pogge}, R.~W., {DePoy}, D.~L., {Marshall}, J.~L., {Stanek},
  K.~Z., {Stutz}, A.~M., {Poindexter}, S., {Siverd}, R., {O'Brien}, T.~P.,
  {Trueblood}, M., \& {Trueblood}, P. 2007, \pasp, 119, 923

\bibitem[{{Pickles}(1998)}]{pickles98}
{Pickles}, A.~J. 1998, \pasp, 110, 863

\bibitem[{{Pollacco} {et~al.}(2006){Pollacco}, {Skillen}, {Collier Cameron},
  {Christian}, {Hellier}, {Irwin}, {Lister}, {Street}, {West}, {Anderson},
  {Clarkson}, {Deeg}, {Enoch}, {Evans}, {Fitzsimmons}, {Haswell}, {Hodgkin},
  {Horne}, {Kane}, {Keenan}, {Maxted}, {Norton}, {Osborne}, {Parley}, {Ryans},
  {Smalley}, {Wheatley}, \& {Wilson}}]{superwasp}
{Pollacco}, D.~L., {Skillen}, I., {Collier Cameron}, A., {Christian}, D.~J.,
  {Hellier}, C., {Irwin}, J., {Lister}, T.~A., {Street}, R.~A., {West}, R.~G.,
  {Anderson}, D.~R., {Clarkson}, W.~I., {Deeg}, H., {Enoch}, B., {Evans}, A.,
  {Fitzsimmons}, A., {Haswell}, C.~A., {Hodgkin}, S., {Horne}, K., {Kane},
  S.~R., {Keenan}, F.~P., {Maxted}, P.~F.~L., {Norton}, A.~J., {Osborne}, J.,
  {Parley}, N.~R., {Ryans}, R.~S.~I., {Smalley}, B., {Wheatley}, P.~J., \&
  {Wilson}, D.~M. 2006, \pasp, 118, 1407

\bibitem[{{Raghavan} {et~al.}(2010){Raghavan}, {McAlister}, {Henry}, {Latham},
  {Marcy}, {Mason}, {Gies}, {White}, \& {ten Brummelaar}}]{raghavan10}
{Raghavan}, D., {McAlister}, H.~A., {Henry}, T.~J., {Latham}, D.~W., {Marcy},
  G.~W., {Mason}, B.~D., {Gies}, D.~R., {White}, R.~J., \& {ten Brummelaar},
  T.~A. 2010, \apjs, 190, 1

\bibitem[{{Rasio} \& {Ford}(1996)}]{rasio96}
{Rasio}, F.~A. \& {Ford}, E.~B. 1996, Science, 274, 954

\bibitem[{{Rauer} {et~al.}(2014){Rauer}, {Catala}, {Aerts}, {Appourchaux},
  {Benz}, {Brandeker}, {Christensen-Dalsgaard}, {Deleuil}, {Gizon}, {Goupil},
  {G{\"u}del}, {Janot-Pacheco}, {Mas-Hesse}, {Pagano}, {Piotto}, {Pollacco},
  {Santos}, {Smith}, {Su{\'a}rez}, {Szab{\'o}}, {Udry}, {Adibekyan}, {Alibert},
  {Almenara}, {Amaro-Seoane}, {Eiff}, {Asplund}, {Antonello}, {Barnes},
  {Baudin}, {Belkacem}, {Bergemann}, {Bihain}, {Birch}, {Bonfils}, {Boisse},
  {Bonomo}, {Borsa}, {Brand{\~a}o}, {Brocato}, {Brun}, {Burleigh}, {Burston},
  {Cabrera}, {Cassisi}, {Chaplin}, {Charpinet}, {Chiappini}, {Church},
  {Csizmadia}, {Cunha}, {Damasso}, {Davies}, {Deeg}, {D{\'{\i}}az}, {Dreizler},
  {Dreyer}, {Eggenberger}, {Ehrenreich}, {Eigm{\"u}ller}, {Erikson}, {Farmer},
  {Feltzing}, {de Oliveira Fialho}, {Figueira}, {Forveille}, {Fridlund},
  {Garc{\'{\i}}a}, {Giommi}, {Giuffrida}, {Godolt}, {Gomes da Silva},
  {Granzer}, {Grenfell}, {Grotsch-Noels}, {G{\"u}nther}, {Haswell}, {Hatzes},
  {H{\'e}brard}, {Hekker}, {Helled}, {Heng}, {Jenkins}, {Johansen},
  {Khodachenko}, {Kislyakova}, {Kley}, {Kolb}, {Krivova}, {Kupka}, {Lammer},
  {Lanza}, {Lebreton}, {Magrin}, {Marcos-Arenal}, {Marrese}, {Marques},
  {Martins}, {Mathis}, {Mathur}, {Messina}, {Miglio}, {Montalban}, {Montalto},
  {Monteiro}, {Moradi}, {Moravveji}, {Mordasini}, {Morel}, {Mortier},
  {Nascimbeni}, {Nelson}, {Nielsen}, {Noack}, {Norton}, {Ofir}, {Oshagh},
  {Ouazzani}, {Papics}, {Parro}, {Petit}, {Plez}, {Poretti}, {Quirrenbach},
  {Ragazzoni}, {Raimondo}, {Rainer}, {Reese}, {Redmer}, {Reffert},
  {Rojas-Ayala}, {Roxburgh}, {Salmon}, {Santerne}, {Schneider}, {Schou},
  {Schuh}, {Schunker}, {Silva-Valio}, {Silvotti}, {Skillen}, {Snellen}, {Sohl},
  {Sousa}, {Sozzetti}, {Stello}, {Strassmeier}, {{\v S}vanda}, {Szab{\'o}},
  {Tkachenko}, {Valencia}, {Van Grootel}, {Vauclair}, {Ventura}, {Wagner},
  {Walton}, {Weingrill}, {Werner}, {Wheatley}, \& {Zwintz}}]{PLATO}
{Rauer}, H., {Catala}, C., {Aerts}, C., {Appourchaux}, T., {Benz}, W.,
  {Brandeker}, A., {Christensen-Dalsgaard}, J., {Deleuil}, M., {Gizon}, L.,
  {Goupil}, M.-J., {G{\"u}del}, M., {Janot-Pacheco}, E., {Mas-Hesse}, M.,
  {Pagano}, I., {Piotto}, G., {Pollacco}, D., {Santos}, {\.C}., {Smith}, A.,
  {Su{\'a}rez}, J.-C., {Szab{\'o}}, R., {Udry}, S., {Adibekyan}, V., {Alibert},
  Y., {Almenara}, J.-M., {Amaro-Seoane}, P., {Eiff}, M.~A.-v., {Asplund}, M.,
  {Antonello}, E., {Barnes}, S., {Baudin}, F., {Belkacem}, K., {Bergemann}, M.,
  {Bihain}, G., {Birch}, A.~C., {Bonfils}, X., {Boisse}, I., {Bonomo}, A.~S.,
  {Borsa}, F., {Brand{\~a}o}, I.~M., {Brocato}, E., {Brun}, S., {Burleigh}, M.,
  {Burston}, R., {Cabrera}, J., {Cassisi}, S., {Chaplin}, W., {Charpinet}, S.,
  {Chiappini}, C., {Church}, R.~P., {Csizmadia}, S., {Cunha}, M., {Damasso},
  M., {Davies}, M.~B., {Deeg}, H.~J., {D{\'{\i}}az}, R.~F., {Dreizler}, S.,
  {Dreyer}, C., {Eggenberger}, P., {Ehrenreich}, D., {Eigm{\"u}ller}, P.,
  {Erikson}, A., {Farmer}, R., {Feltzing}, S., {de Oliveira Fialho}, F.,
  {Figueira}, P., {Forveille}, T., {Fridlund}, M., {Garc{\'{\i}}a}, R.~A.,
  {Giommi}, P., {Giuffrida}, G., {Godolt}, M., {Gomes da Silva}, J., {Granzer},
  T., {Grenfell}, J.~L., {Grotsch-Noels}, A., {G{\"u}nther}, E., {Haswell},
  C.~A., {Hatzes}, A.~P., {H{\'e}brard}, G., {Hekker}, S., {Helled}, R.,
  {Heng}, K., {Jenkins}, J.~M., {Johansen}, A., {Khodachenko}, M.~L.,
  {Kislyakova}, K.~G., {Kley}, W., {Kolb}, U., {Krivova}, N., {Kupka}, F.,
  {Lammer}, H., {Lanza}, A.~F., {Lebreton}, Y., {Magrin}, D., {Marcos-Arenal},
  P., {Marrese}, P.~M., {Marques}, J.~P., {Martins}, J., {Mathis}, S.,
  {Mathur}, S., {Messina}, S., {Miglio}, A., {Montalban}, J., {Montalto}, M.,
  {Monteiro}, M.~J.~P.~F.~G., {Moradi}, H., {Moravveji}, E., {Mordasini}, C.,
  {Morel}, T., {Mortier}, A., {Nascimbeni}, V., {Nelson}, R.~P., {Nielsen},
  M.~B., {Noack}, L., {Norton}, A.~J., {Ofir}, A., {Oshagh}, M., {Ouazzani},
  R.-M., {Papics}, P., {Parro}, V.~C., {Petit}, P., {Plez}, B., {Poretti}, E.,
  {Quirrenbach}, A., {Ragazzoni}, R., {Raimondo}, G., {Rainer}, M., {Reese},
  D.~R., {Redmer}, R., {Reffert}, S., {Rojas-Ayala}, B., {Roxburgh}, I.~W.,
  {Salmon}, S., {Santerne}, A., {Schneider}, J., {Schou}, J., {Schuh}, S.,
  {Schunker}, H., {Silva-Valio}, A., {Silvotti}, R., {Skillen}, I., {Snellen},
  I., {Sohl}, F., {Sousa}, S.~G., {Sozzetti}, A., {Stello}, D., {Strassmeier},
  K.~G., {{\v S}vanda}, M., {Szab{\'o}}, G.~M., {Tkachenko}, A., {Valencia},
  D., {Van Grootel}, V., {Vauclair}, S.~D., {Ventura}, P., {Wagner}, F.~W.,
  {Walton}, N.~A., {Weingrill}, J., {Werner}, S.~C., {Wheatley}, P.~J., \&
  {Zwintz}, K. 2014, Experimental Astronomy, 38, 249

\bibitem[{{Ricker} {et~al.}(2015){Ricker}, {Winn}, {Vanderspek}, {Latham},
  {Bakos}, {Bean}, {Berta-Thompson}, {Brown}, {Buchhave}, {Butler}, {Butler},
  {Chaplin}, {Charbonneau}, {Christensen-Dalsgaard}, {Clampin}, {Deming},
  {Doty}, {De Lee}, {Dressing}, {Dunham}, {Endl}, {Fressin}, {Ge}, {Henning},
  {Holman}, {Howard}, {Ida}, {Jenkins}, {Jernigan}, {Johnson}, {Kaltenegger},
  {Kawai}, {Kjeldsen}, {Laughlin}, {Levine}, {Lin}, {Lissauer}, {MacQueen},
  {Marcy}, {McCullough}, {Morton}, {Narita}, {Paegert}, {Palle}, {Pepe},
  {Pepper}, {Quirrenbach}, {Rinehart}, {Sasselov}, {Sato}, {Seager},
  {Sozzetti}, {Stassun}, {Sullivan}, {Szentgyorgyi}, {Torres}, {Udry}, \&
  {Villasenor}}]{TESS}
{Ricker}, G.~R., {Winn}, J.~N., {Vanderspek}, R., {Latham}, D.~W., {Bakos},
  G.~{\'A}., {Bean}, J.~L., {Berta-Thompson}, Z.~K., {Brown}, T.~M.,
  {Buchhave}, L., {Butler}, N.~R., {Butler}, R.~P., {Chaplin}, W.~J.,
  {Charbonneau}, D., {Christensen-Dalsgaard}, J., {Clampin}, M., {Deming}, D.,
  {Doty}, J., {De Lee}, N., {Dressing}, C., {Dunham}, E.~W., {Endl}, M.,
  {Fressin}, F., {Ge}, J., {Henning}, T., {Holman}, M.~J., {Howard}, A.~W.,
  {Ida}, S., {Jenkins}, J.~M., {Jernigan}, G., {Johnson}, J.~A., {Kaltenegger},
  L., {Kawai}, N., {Kjeldsen}, H., {Laughlin}, G., {Levine}, A.~M., {Lin}, D.,
  {Lissauer}, J.~J., {MacQueen}, P., {Marcy}, G., {McCullough}, P.~R.,
  {Morton}, T.~D., {Narita}, N., {Paegert}, M., {Palle}, E., {Pepe}, F.,
  {Pepper}, J., {Quirrenbach}, A., {Rinehart}, S.~A., {Sasselov}, D., {Sato},
  B., {Seager}, S., {Sozzetti}, A., {Stassun}, K.~G., {Sullivan}, P.,
  {Szentgyorgyi}, A., {Torres}, G., {Udry}, S., \& {Villasenor}, J. 2015,
  Journal of Astronomical Telescopes, Instruments, and Systems, 1, 014003

\bibitem[{{Riddle} {et~al.}(2012){Riddle}, {Burse}, {Law}, {Tendulkar},
  {Baranec}, {Rudy}, {Sitt}, {Arya}, {Papadopoulos}, {Ramaprakash}, \&
  {Dekany}}]{riddle12}
{Riddle}, R.~L., {Burse}, M.~P., {Law}, N.~M., {Tendulkar}, S.~P., {Baranec},
  C., {Rudy}, A.~R., {Sitt}, M., {Arya}, A., {Papadopoulos}, A., {Ramaprakash},
  A.~N., \& {Dekany}, R.~G. 2012, in Society of Photo-Optical Instrumentation
  Engineers (SPIE) Conference Series, Vol. 8447, Society of Photo-Optical
  Instrumentation Engineers (SPIE) Conference Series, 2

\bibitem[{{Roberts} {et~al.}(2015){Roberts}, {Tokovinin}, {Mason}, {Riddle},
  {Hartkopf}, {Law}, \& {Baranec}}]{roberts15}
{Roberts}, Jr., L.~C., {Tokovinin}, A., {Mason}, B.~D., {Riddle}, R.~L.,
  {Hartkopf}, W.~I., {Law}, N.~M., \& {Baranec}, C. 2015, \aj, 149, 118

\bibitem[{{Roell} {et~al.}(2012){Roell}, {Neuh{\"a}user}, {Seifahrt}, \&
  {Mugrauer}}]{roell12}
{Roell}, T., {Neuh{\"a}user}, R., {Seifahrt}, A., \& {Mugrauer}, M. 2012, \aap,
  542, A92

\bibitem[{{Rogers}(2015)}]{rogers15}
{Rogers}, L.~A. 2015, \apj, 801, 41

\bibitem[{{Rowe} {et~al.}(2014){Rowe}, {Bryson}, {Marcy}, {Lissauer},
  {Jontof-Hutter}, {Mullally}, {Gilliland}, {Issacson}, {Ford}, {Howell},
  {Borucki}, {Haas}, {Huber}, {Steffen}, {Thompson}, {Quintana}, {Barclay},
  {Still}, {Fortney}, {Gautier}, {Hunter}, {Caldwell}, {Ciardi}, {Devore},
  {Cochran}, {Jenkins}, {Agol}, {Carter}, \& {Geary}}]{rowe14}
{Rowe}, J.~F., {Bryson}, S.~T., {Marcy}, G.~W., {Lissauer}, J.~J.,
  {Jontof-Hutter}, D., {Mullally}, F., {Gilliland}, R.~L., {Issacson}, H.,
  {Ford}, E., {Howell}, S.~B., {Borucki}, W.~J., {Haas}, M., {Huber}, D.,
  {Steffen}, J.~H., {Thompson}, S.~E., {Quintana}, E., {Barclay}, T., {Still},
  M., {Fortney}, J., {Gautier}, III, T.~N., {Hunter}, R., {Caldwell}, D.~A.,
  {Ciardi}, D.~R., {Devore}, E., {Cochran}, W., {Jenkins}, J., {Agol}, E.,
  {Carter}, J.~A., \& {Geary}, J. 2014, \apj, 784, 45

\bibitem[{{Santerne} {et~al.}(2012){Santerne}, {D{\'{\i}}az}, {Moutou},
  {Bouchy}, {H{\'e}brard}, {Almenara}, {Bonomo}, {Deleuil}, \&
  {Santos}}]{santerne12}
{Santerne}, A., {D{\'{\i}}az}, R.~F., {Moutou}, C., {Bouchy}, F.,
  {H{\'e}brard}, G., {Almenara}, J.-M., {Bonomo}, A.~S., {Deleuil}, M., \&
  {Santos}, N.~C. 2012, \aap, 545, A76

\bibitem[{{Santerne} {et~al.}(2013){Santerne}, {Fressin}, {D{\'{\i}}az},
  {Figueira}, {Almenara}, \& {Santos}}]{santerne13}
{Santerne}, A., {Fressin}, F., {D{\'{\i}}az}, R.~F., {Figueira}, P.,
  {Almenara}, J.-M., \& {Santos}, N.~C. 2013, \aap, 557, A139

\bibitem[{{Santerne} {et~al.}(2015){Santerne}, {Moutou}, {Tsantaki}, {Bouchy},
  {H{\'e}brard}, {Adibekyan}, {Almenara}, {Amard}, {Barros}, {Boisse},
  {Bonomo}, {Bruno}, {Courcol}, {Deleuil}, {Demangeon}, {D{\'{\i}}az},
  {Guillot}, {Havel}, {Montagnier}, {Rajpurohit}, {Rey}, \&
  {Santos}}]{santerne15}
{Santerne}, A., {Moutou}, C., {Tsantaki}, M., {Bouchy}, F., {H{\'e}brard}, G.,
  {Adibekyan}, V., {Almenara}, J.-M., {Amard}, L., {Barros}, S.~C.~C.,
  {Boisse}, I., {Bonomo}, A.~S., {Bruno}, G., {Courcol}, B., {Deleuil}, M.,
  {Demangeon}, O., {D{\'{\i}}az}, R.~F., {Guillot}, T., {Havel}, M.,
  {Montagnier}, G., {Rajpurohit}, A.~S., {Rey}, J., \& {Santos}, N.~C. 2015,
  ArXiv e-prints

\bibitem[{{Schwamb} {et~al.}(2013){Schwamb}, {Orosz}, {Carter}, {Welsh},
  {Fischer}, {Torres}, {Howard}, {Crepp}, {Keel}, {Lintott}, {Kaib}, {Terrell},
  {Gagliano}, {Jek}, {Parrish}, {Smith}, {Lynn}, {Simpson}, {Giguere}, \&
  {Schawinski}}]{schwamb13}
{Schwamb}, M.~E., {Orosz}, J.~A., {Carter}, J.~A., {Welsh}, W.~F., {Fischer},
  D.~A., {Torres}, G., {Howard}, A.~W., {Crepp}, J.~R., {Keel}, W.~C.,
  {Lintott}, C.~J., {Kaib}, N.~A., {Terrell}, D., {Gagliano}, R., {Jek}, K.~J.,
  {Parrish}, M., {Smith}, A.~M., {Lynn}, S., {Simpson}, R.~J., {Giguere},
  M.~J., \& {Schawinski}, K. 2013, \apj, 768, 127

\bibitem[{{Seager}(2013)}]{seager13}
{Seager}, S. 2013, Science, 340, 577

\bibitem[{{Selsis} {et~al.}(2007){Selsis}, {Kasting}, {Levrard}, {Paillet},
  {Ribas}, \& {Delfosse}}]{selis07}
{Selsis}, F., {Kasting}, J.~F., {Levrard}, B., {Paillet}, J., {Ribas}, I., \&
  {Delfosse}, X. 2007, \aap, 476, 1373

\bibitem[{{Skrutskie} {et~al.}(2006){Skrutskie}, {Cutri}, {Stiening},
  {Weinberg}, {Schneider}, {Carpenter}, {Beichman}, {Capps}, {Chester},
  {Elias}, {Huchra}, {Liebert}, {Lonsdale}, {Monet}, {Price}, {Seitzer},
  {Jarrett}, {Kirkpatrick}, {Gizis}, {Howard}, {Evans}, {Fowler}, {Fullmer},
  {Hurt}, {Light}, {Kopan}, {Marsh}, {McCallon}, {Tam}, {Van Dyk}, \&
  {Wheelock}}]{skrutskie06}
{Skrutskie}, M.~F., {Cutri}, R.~M., {Stiening}, R., {Weinberg}, M.~D.,
  {Schneider}, S., {Carpenter}, J.~M., {Beichman}, C., {Capps}, R., {Chester},
  T., {Elias}, J., {Huchra}, J., {Liebert}, J., {Lonsdale}, C., {Monet}, D.~G.,
  {Price}, S., {Seitzer}, P., {Jarrett}, T., {Kirkpatrick}, J.~D., {Gizis},
  J.~E., {Howard}, E., {Evans}, T., {Fowler}, J., {Fullmer}, L., {Hurt}, R.,
  {Light}, R., {Kopan}, E.~L., {Marsh}, K.~A., {McCallon}, H.~L., {Tam}, R.,
  {Van Dyk}, S., \& {Wheelock}, S. 2006, \aj, 131, 1163

\bibitem[{{Torres} {et~al.}(2015){Torres}, {Kipping}, {Fressin}, {Caldwell},
  {Twicken}, {Ballard}, {Batalha}, {Bryson}, {Ciardi}, {Henze}, {Howell},
  {Isaacson}, {Jenkins}, {Muirhead}, {Newton}, {Petigura}, {Barclay},
  {Borucki}, {Crepp}, {Everett}, {Horch}, {Howard}, {Kolbl}, {Marcy},
  {McCauliff}, \& {Quintana}}]{torres15}
{Torres}, G., {Kipping}, D.~M., {Fressin}, F., {Caldwell}, D.~A., {Twicken},
  J.~D., {Ballard}, S., {Batalha}, N.~M., {Bryson}, S.~T., {Ciardi}, D.~R.,
  {Henze}, C.~E., {Howell}, S.~B., {Isaacson}, H.~T., {Jenkins}, J.~M.,
  {Muirhead}, P.~S., {Newton}, E.~R., {Petigura}, E.~A., {Barclay}, T.,
  {Borucki}, W.~J., {Crepp}, J.~R., {Everett}, M.~E., {Horch}, E.~P., {Howard},
  A.~W., {Kolbl}, R., {Marcy}, G.~W., {McCauliff}, S., \& {Quintana}, E.~V.
  2015, \apj, 800, 99

\bibitem[{{van Dam} {et~al.}(2006){van Dam}, {Bouchez}, {Le Mignant},
  {Johansson}, {Wizinowich}, {Campbell}, {Chin}, {Hartman}, {Lafon}, {Stomski},
  \& {Summers}}]{KeckLGS2}
{van Dam}, M.~A., {Bouchez}, A.~H., {Le Mignant}, D., {Johansson}, E.~M.,
  {Wizinowich}, P.~L., {Campbell}, R.~D., {Chin}, J.~C.~Y., {Hartman}, S.~K.,
  {Lafon}, R.~E., {Stomski}, Jr., P.~J., \& {Summers}, D.~M. 2006, \pasp, 118,
  310

\bibitem[{{Wang} {et~al.}(2015{\natexlab{a}}){Wang}, {Fischer}, {Horch}, \&
  {Xie}}]{wang15a}
{Wang}, J., {Fischer}, D.~A., {Horch}, E.~P., \& {Xie}, J.-W.
  2015{\natexlab{a}}, \apj, 806, 248

\bibitem[{{Wang} {et~al.}(2014){Wang}, {Fischer}, {Xie}, \& {Ciardi}}]{wang14}
{Wang}, J., {Fischer}, D.~A., {Xie}, J.-W., \& {Ciardi}, D.~R. 2014, \apj, 791,
  111

\bibitem[{{Wang} {et~al.}(2015{\natexlab{b}}){Wang}, {Fischer}, {Xie}, \&
  {Ciardi}}]{wang15b}
---. 2015{\natexlab{b}}, ArXiv e-prints

\bibitem[{{Wheatley} {et~al.}(2013){Wheatley}, {Pollacco}, {Queloz}, {Rauer},
  {Watson}, {West}, {Chazelas}, {Louden}, {Walker}, {Bannister}, {Bento},
  {Burleigh}, {Cabrera}, {Eigm{\"u}ller}, {Erikson}, {Genolet}, {Goad},
  {Grange}, {Jord{\'a}n}, {Lawrie}, {McCormac}, \& {Neveu}}]{ngts}
{Wheatley}, P.~J., {Pollacco}, D.~L., {Queloz}, D., {Rauer}, H., {Watson},
  C.~A., {West}, R.~G., {Chazelas}, B., {Louden}, T.~M., {Walker}, S.,
  {Bannister}, N., {Bento}, J., {Burleigh}, M., {Cabrera}, J., {Eigm{\"u}ller},
  P., {Erikson}, A., {Genolet}, L., {Goad}, M., {Grange}, A., {Jord{\'a}n}, A.,
  {Lawrie}, K., {McCormac}, J., \& {Neveu}, M. 2013, in European Physical
  Journal Web of Conferences, Vol.~47, European Physical Journal Web of
  Conferences, 13002

\bibitem[{{Wizinowich} {et~al.}(2000){Wizinowich}, {Acton}, {Shelton},
  {Stomski}, {Gathright}, {Ho}, {Lupton}, {Tsubota}, {Lai}, {Max}, {Brase},
  {An}, {Avicola}, {Olivier}, {Gavel}, {Macintosh}, {Ghez}, \&
  {Larkin}}]{wizinowich00}
{Wizinowich}, P., {Acton}, D.~S., {Shelton}, C., {Stomski}, P., {Gathright},
  J., {Ho}, K., {Lupton}, W., {Tsubota}, K., {Lai}, O., {Max}, C., {Brase}, J.,
  {An}, J., {Avicola}, K., {Olivier}, S., {Gavel}, D., {Macintosh}, B., {Ghez},
  A., \& {Larkin}, J. 2000, \pasp, 112, 315

\bibitem[{{Wizinowich} {et~al.}(2006){Wizinowich}, {Le Mignant}, {Bouchez},
  {Campbell}, {Chin}, {Contos}, {van Dam}, {Hartman}, {Johansson}, {Lafon},
  {Lewis}, {Stomski}, {Summers}, {Brown}, {Danforth}, {Max}, \&
  {Pennington}}]{KeckLGS1}
{Wizinowich}, P.~L., {Le Mignant}, D., {Bouchez}, A.~H., {Campbell}, R.~D.,
  {Chin}, J.~C.~Y., {Contos}, A.~R., {van Dam}, M.~A., {Hartman}, S.~K.,
  {Johansson}, E.~M., {Lafon}, R.~E., {Lewis}, H., {Stomski}, P.~J., {Summers},
  D.~M., {Brown}, C.~G., {Danforth}, P.~M., {Max}, C.~E., \& {Pennington},
  D.~M. 2006, \pasp, 118, 297

\bibitem[{{Xie} {et~al.}(2014){Xie}, {Wu}, \& {Lithwick}}]{xie14}
{Xie}, J.-W., {Wu}, Y., \& {Lithwick}, Y. 2014, \apj, 789, 165

\bibitem[{{Yelda} {et~al.}(2010){Yelda}, {Lu}, {Ghez}, {Clarkson}, {Anderson},
  {Do}, \& {Matthews}}]{Yelda10}
{Yelda}, S., {Lu}, J.~R., {Ghez}, A.~M., {Clarkson}, W., {Anderson}, J., {Do},
  T., \& {Matthews}, K. 2010, \apj, 725, 331

\bibitem[{{Ziegler} {et~al.}(2015){Ziegler}, {Law}, {Baranec}, {Riddle}, \&
  {Fuchs}}]{ziegler15}
{Ziegler}, C., {Law}, N.~M., {Baranec}, C., {Riddle}, R.~L., \& {Fuchs}, J.~T.
  2015, \apj, 804, 30

\bibitem[{{Zsom} {et~al.}(2013){Zsom}, {Seager}, {de Wit}, \&
  {Stamenkovi{\'c}}}]{zsom13}
{Zsom}, A., {Seager}, S., {de Wit}, J., \& {Stamenkovi{\'c}}, V. 2013, \apj,
  778, 109

\end{thebibliography}

\clearpage

\clearpage

\section{Appendix}

In Table$~\ref{tab:whitelist}$, we list our Robo-AO observed KOIs, including date the target was observed, observation quality (as described in Section \ref{sec:imageperf}), the estimated latest detectable companion spectral type (as described in Section \ref{sec:spectraltypes}), and the presence of detected companions.

{\footnotesize
\tabcolsep=0.12cm
\begin{longtable}{ccccccc}
\caption{Full Robo-AO Observation List \label{tab:whitelist}}\\
\hline
\hline
\noalign{\vskip 3pt}  
\text{KOI} & m\textit{$_i$} & \text{ObsID} & \text{Filter} & \text{Obs. Qual.} & \text{Latest Det.} & \text{Comp.}\\ [0.1ex]
 & (mags) & &  &  & \text{Comp. SpT} & \text{Det.?} \\ [0.1ex]
\hline
\noalign{\vskip 3pt}  
\endfirsthead
\multicolumn{6}{c}
{\tablename\ \thetable\ -- \textit{Continued}} \\
\hline \hline
\noalign{\vskip 3pt} 
\text{KOI} & m\textit{$_i$} & \text{ObsID} & \text{Filter} & \text{Obs. Qual.} & \text{Latest Det.} & \text{Comp.}\\ [0.1ex]
 & (mags) & &  &  & \text{Comp. SpT} & \text{Det.?} \\ [0.1ex]
\hline
\noalign{\vskip 3pt}  
\endhead
\endfoot
\hline
\endlastfoot
K020 & 13.29 & 2014 Aug 22 & \textit{i}\textsuperscript{$\prime$} & medium & M4 &  \\ 
K051 & 14.48 & 2013 Jul 25 & LP600 & medium & M4 & yes \\ 
K076 & 10.05 & 2014 Aug 22 & \textit{i}\textsuperscript{$\prime$} & high & M4 &  \\ 
K0104 & 12.78 & 2014 Aug 22 & \textit{i}\textsuperscript{$\prime$} & medium & M6 &  \\ 
K0126 & 13.11 & 2015 Jun 08 & LP600 & high & M5 & yes \\ 
K0134 & 15.02 & 2012 Jul 16 & \textit{i}\textsuperscript{$\prime$} & medium & M4 &  \\ 
K0135 & 13.8 & 2012 Jul 17 & \textit{i}\textsuperscript{$\prime$} & low & M1 &  \\ 
K0163 & 13.3 & 2012 Jul 18 & LP600 & high & M6 & yes \\
\end{longtable}}

\end{document}